\RequirePackage{fix-cm}
\makeatletter
\def\cl@chapter{\@elt {theorem}}
\makeatother

\documentclass[smallextended,envcountsame,nospthms]{svjour3}
\journalname{}
\smartqed 
\makeatletter \def\@citecolor{blue}%
\def\@urlcolor{blue}%
\def\@linkcolor{blue}%

\def\orcidID#1{\href{http://orcid.org/#1}{\smash{\protect\raisebox{-1.25pt}{\protect\includegraphics{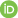}}}}}
\makeatother
\usepackage{bbm}
  \usepackage{amsmath, amssymb, amsthm}
\usepackage{thmtools}
\usepackage{thm-restate}
\PassOptionsToPackage{hyphens}{url}
\usepackage[bookmarksopen,bookmarksdepth=3]{hyperref}
\usepackage{lineno}
\usepackage{tikz}
\usetikzlibrary{shapes,calc,arrows,automata,decorations.pathmorphing}
\usepackage{xifthen}
\usepackage{xspace}
\usepackage[ruled, vlined, linesnumbered]{algorithm2e}
\usepackage[capitalize,nameinlink]{cleveref}
\usepackage{mymatrix}
\usepackage{subcaption}

\SetKwInOut{Input}{input}
\SetKwInOut{Output}{output}
\newcommand*{\return}{\textbf{return}\,\,}
\theoremstyle{plain}
\newtheorem{theorem}{Theorem}
\numberwithin{theorem}{section}

\numberwithin{notation}{section}

\numberwithin{conjecture}{section}
\newtheorem{lemma}[theorem]{Lemma}
\numberwithin{lemma}{section}
\newtheorem{corollary}[theorem]{Corollary}
\numberwithin{corollary}{section}
\theoremstyle{definition}

\numberwithin{remark}{section}
\newtheorem{definition}[theorem]{Definition}
\numberwithin{definition}{section}
\theoremstyle{example}
\newtheoremstyle{example} 
{\topsep} 
{\topsep} 
{\slshape\fontfamily{ptm}\selectfont} 
{} 
{\fontfamily{ptm}\selectfont\scshape\color{blue}} 
{:} 
{.5em} 
{} 

\newtheorem{example}[theorem]{Example}
\numberwithin{example}{section}
\usepackage{stackengine}
\usepackage[utf8]{inputenc}
\usepackage[shortlabels]{enumitem}
\setlist[enumerate,1]{label=(\alph*), wide=0pt, leftmargin=*}
\setlist[itemize,1]{label=\textbullet}

\renewcommand{\emptyset}{\varnothing}

\usepackage{mathtools}

\usepackage{stmaryrd}

\makeatletter \let\xx@thm\@thm \AtBeginDocument{\let\@thm\xx@thm}
\makeatother

\usepackage[nottoc]{tocbibind}
\usepackage{cite}
\usepackage[numbers]{natbib}
\usepackage{doi}

\hypersetup{
  pdftitle={Targeting Completeness: Automated Complexity Analysis of Integer Programs}, colorlinks=true, linkcolor=blue, citecolor=olive, filecolor=magenta, urlcolor=cyan
}

 \usepackage[location=appendix,prependtoappendix=true,hideproofs=false]{moveproofs}

\newcommand{\tool}[1]{\textsf{#1}}
\newcommand{\KoAT}[0]{\tool{KoAT}}
\newcommand{\braced}[1]{\left\lbrace #1 \right\rbrace}
\newcommand{\var}{\normalfont\texttt}
\newcommand{\wildcard}{\underline{\hspace{0.15cm}}}

\newcommand{\paragraphProof}[1]{\newline
  \begingroup\raggedright\textit{#1}\;\endgroup
}

\newcommand{\NN}{\mathbb{N}}
\newcommand{\ZZ}{\mathbb{Z}}
\newcommand{\QQ}{\mathbb{Q}}
\newcommand{\RR}{\mathbb{R}}
\renewcommand{\AA}{\mathbb{A}}
\newcommand{\NNC}{\overline{\mathbb{N}}}

\newcommand{\RRC}{\overline{\mathbb{R}}}
\renewcommand{\emptyset}{\varnothing}
\newcommand{\identity}{{\normalfont\textsf{id}}}
\newcommand{\abs}[1]{\lvert #1 \rvert}

\newcommand{\im}{\mathrm{i}}
\newcommand{\compconj}[1]{%
  \overline{#1}%
}
\newcommand{\myvec}[1]{\ensuremath{
    \begin{sbmatrix}
      #1
    \end{sbmatrix}
  }}
\newcommand{\xvec}{\vec{x}}
\newcommand{\pvec}[1]{\vec{#1}\mkern2mu\vphantom{#1}}
\newcommand{\sgn}{\mathrm{sgn}}

\DeclareMathOperator{\sign}{sign}

\newcommand{\AtomSet}{\mathcal{A}}

\newcommand{\FormulaSet}{\mathcal{F}}
\newcommand{\true}{\var{true}}
\newcommand{\false}{\var{false}}

\newcommand{\initial}{\sigma_0}
\newcommand{\valuation}{\sigma}
\newcommand{\Valuation}{\Sigma}

\newcommand{\VSet}{\mathcal{V}}
\newcommand{\TVSet}{\mathcal{TV}}
\newcommand{\indv}{d}
\newcommand{\guard}{\varphi}
\newcommand{\update}{\eta}
\newcommand{\Program}{\mathcal{P}}
\newcommand{\TSet}{\mathcal{T}}
\newcommand{\LSet}{\mathcal{L}}

\newcommand{\location}{\ell}
\newcommand{\IntProgram}{(\VSet,\TVSet,\LSet,\location_0,\TSet)}
\DeclareMathOperator{\rc}{rc}
\newcommand{\entry}{\mathcal{E}}
\newcommand{\chain}{\star}

\newcommand{\pret}{r}

\newcommand{\prestate}{\valuation}
\newcommand{\actstate}{{\tilde{\valuation}}}
\newcommand{\actl}{{\tilde{\location}}}
\newcommand{\prel}{\location}

\newcommand{\IntLoop}{(\guard,\update)}
\newcommand{\WhileLoop}[2]{\normalfont\textbf{while } #1 \textbf{ do } #2}
\newcommand{\cl}[1]{\normalfont{\texttt{cl}^{#1}}}
\newcommand{\PPEE}{\mathbb{PE}}
\newcommand{\SSet}{\mathcal{S}}
\newcommand{\clExp}[2]{\normalfont{\texttt{cl}^{#1}_{#2}}}

\newcommand{\sth}{\operatorname{sth}}

\newcommand{\lex}{>_{\mathrm{lex}}}
\newcommand{\ud}{a_1}
\newcommand{\ub}{b_1}
\newcommand{\ld}{a_2}
\newcommand{\lb}{b_2}
\newcommand{\auto}{\vartheta}

\newcommand{\BoundSet}{\mathcal{B}}
\newcommand{\Size}{{\mathcal{SB}}}
\newcommand{\size}{\normalfont\texttt{sb}}
\newcommand{\run}{\normalfont\texttt{rb}}
\newcommand{\bound}{\normalfont{b}}

\newcommand{\locs}[3]{{\mathcal{RB}_{\normalfont\text{loc}}^{#1}}( \to_{#3} #2)}
\newcommand{\locsold}[3]{{\mathcal{RB}_{\normalfont\text{loc}}^{#1,#2}}(#3)}
\newcommand{\locsoldtwo}[2]{{\mathcal{RB}_{\normalfont\text{loc}}^{#1,#2}}}
\newcommand{\locsize}[1]{{\mathcal{SB}_{\normalfont\text{loc}}^{#1}}}
\newcommand{\glo}{{\mathcal{RB}_{\normalfont\text{glo}}}}
\newcommand{\glopr}{{\mathcal{RB}'_{\normalfont\text{glo}}}}

\newcommand{\MRFs}{\text{M}\Phi\text{RFs}}

\crefname{definition}{Def.}{Def.}
\crefname{example}{Ex.}{Ex.}
\crefname{counterexample}{Counterex.}{Counterex.}
\crefname{appendix}{App.}{App.}
\crefname{ex}{Ex.}{Ex.}
\crefname{theorem}{Thm.}{Thm.}
\crefname{lemma}{Lemma}{Lemmas}
\crefname{remark}{Rem.}{Rem.}
\crefname{section}{Sect.}{Sect.}
\crefname{subsection}{Sect.}{Sect.}
\crefname{subsubsection}{Sect.}{Sect.}
\crefname{line}{Line}{Lines}
\crefname{corollary}{Cor.}{Cor.}
\crefname{figure}{Fig.}{Fig.}
\crefname{enumi}{}{}
\crefname{algorithm}{Alg.}{Alg.}

\renewenvironment{proof}{}{}
\newenvironment{myproof}{
  \medskip
  \noindent{\it Proof.}
}{
  \medskip}

\allowdisplaybreaks

\begin{document}
\title{Targeting Completeness: Automated Complexity Analysis of Integer Programs \thanks{funded by the DFG Research Training Group 2236 UnRAVeL and supported by a fellowship within the IFI program of the German Academic Exchange Service (DAAD).}}

\author{\mbox{Nils Lommen\orcidID{0000-0003-3187-9217} \and Éléanore Meyer\orcidID{0000-0003-1038-4944} \and Jürgen Giesl\orcidID{0000-0003-0283-8520}}}

\authorrunning{Nils Lommen et al.}

\institute{Nils Lommen \and Éléanore Meyer \and Jürgen Giesl \\
  RWTH Aachen University, Germany}

\date{}

\maketitle

\begin{abstract}
  There exist several approaches to infer runtime or resource bounds for integer programs automatically.
  In this paper, we study the subclass of \emph{periodic rational solvable loops (prs-loops)}, where questions regarding the runtime and the size of variable values are decidable and where we can therefore obtain techniques that are ``complete'' for such subclasses.
  We show how to use these results for the complexity analysis of arbitrary general integer programs.
  To this end, we present a modular approach which computes local runtime and size bounds for subprograms which correspond to \emph{prs}-loops.
  These local bounds\linebreak
  are then lifted to global runtime and size bounds for the whole integer program.
  Furthermore, we introduce several techniques to transform larger programs\linebreak
  into \emph{prs}-loops to increase the scope of the approach.
  The power of the procedure is shown by our implementation in the complexity analysis tool \tool{KoAT}.
\end{abstract}

\keywords{Automated Complexity Analysis \and Runtime Bounds \and Size Bounds \and Decidability \and Integer Programs}

\section{Introduction}
\label{sect:introduction}

There are numerous automatic approaches to infer bounds on the runtime or on other resources for integer programs.
Most of these approaches are based on incomplete techniques like ranking functions (see, e.g., \cite{ben-amram2017MultiphaseLinearRankingFunctions,albert2019ResourceAnalysisDriven,sinn2017ComplexityResourceBound,brockschmidt2016AnalyzingRuntimeSize,flores-montoya2016UpperLowerAmortized,giesl2022ImprovingAutomaticComplexity,hoffmann2017AutomaticResourceBound,lopez18IntervalBasedResource,carbonneaux2015CompositionalCertifiedResource,albert2012CostAnalysisObjectoriented,lommen2024ControlFlowRefinementProbabilistic,pham2024RobustResourceBounds,HoffmannJ22}).
However, there also exist many techniques to decide termination, analyze runtime complexity, or study memory consumption that are complete for certain (syntactically characterized) classes of programs, e.g., \cite{tiwari04,braverman06,frohn2019TerminationTriangularInteger,ben-amram2019TightWorstCaseBounds,hosseini2019TerminationLinearLoops,frohn2020TerminationPolynomialLoops,hark2020PolynomialLoopsTermination,ben-amram2016FlowchartProgramsRegular,ben-amram2008LinearPolynomialExponential,xuSymbolicTerminationAnalysis2013,neumann2020RankingFunctionSynthesis}.

In this paper, we show that complete techniques for subclasses of programs can be integrated into a general incomplete approach for complexity analysis of arbitrary integer programs.
Our results indicate that such complete techniques are not only theoretically interesting, but they have an important practical value, since their integration increases the power of the general incomplete approach substantially.

The current paper is based on our previous two conference papers \cite{lommen2022AutomaticComplexityAnalysis,lommen2023TargetingCompletenessUsing}.
Here, \cite{lommen2022AutomaticComplexityAnalysis} shows how to embed complete procedures into the analysis of runtime complexity, and \cite{lommen2023TargetingCompletenessUsing} focuses on bounds on the possible sizes of variable values.
The current paper presents the results of these two papers in a unified framework, and extends them considerably.

We regard a subclass of integer programs, i.e., \emph{periodic rational solvable loops (prs-loops)}, for which questions concerning termination, runtime, and size bounds are decidable.
For example, consider the following \emph{prs}-loop (which contains non-linear arithmetic in both its guard and its update):
\begin{equation}
  \label{WhileExample}
  \WhileLoop{x_2^2 - x_3^5 < x_1 \land x_2 \neq 0}{\myvec{x_1 \\ x_2 \\ x_3 \\ x_4 \\ x_5}
    \leftarrow
    \begin{bmatrix}
      3 & 0  & 0 & 0  & 0  \\
      0 & -2 & 0 & 0  & 0  \\
      0 & 0  & 1 & 0  & 0  \\
      0 & 0  & 0 & 3  & 2  \\
      0 & 0  & 0 & -5 & -3
    \end{bmatrix}
    \myvec{x_1\\ x_2 \\ x_3 \\ x_4 \\ x_5}
    +
    \myvec{x_3^2 \vspace{0.075cm}\\ 0 \\ 0 \\ 0 \\ 0}
  }
\end{equation}
The advantage of \emph{prs}-loops is that one can compute \emph{closed forms} which correspond to applying the loop's update $n$ times.
For example, the closed form for $x_2$ in Loop \eqref{WhileExample} is $x_2 \cdot (-2)^n$, i.e., this expression describes the value of $x_2$ after $n$ loop iterations.

These closed forms are particularly suitable for automatic termination and complexity analysis.
Based on a technique for termination analysis of the smaller class of triangular weakly-non linear loops (\emph{twn}-loops) \cite{frohn2020TerminationPolynomialLoops}, the question of termination of a \emph{prs}-loop can be reduced to an existential arithmetic formula (whose validity is decidable for linear integer arithmetic and where SMT solvers often also prove (in)validity for non-linear integer arithmetic).
This implies that termination of linear \emph{prs}-loops (i.e., where the loop has only linear arithmetic in the guard and the update) is decidable and that non-termination of \emph{prs}-loops over $\ZZ$ (possibly with non-linear arithmetic) is semi-decidable \cite{frohn2020TerminationPolynomialLoops}.

Closed forms can also be used for complexity analysis.
To this end, \cite{hark2020PolynomialLoopsTermination} presented a ``\emph{complete}'' complexity analysis technique for \emph{runtime bounds}.
More precisely, for every \emph{twn}-loop, it infers a polynomial which is an upper bound on the runtime for all those inputs where the loop terminates.

In the current paper, we lift this result to \emph{prs}-loops.
Furthermore, in contrast to the previous conference papers \cite{lommen2022AutomaticComplexityAnalysis,lommen2023TargetingCompletenessUsing}, we improve the technique in order to also infer \emph{logarithmic} instead of just polynomial runtime bounds for \emph{prs}-loops like \eqref{WhileExample}.
The reason is that in the guard $x_2^2 - x_3^5 < x_1$, $x_2^2$ grows faster than $x_1$ (since the closed form for $x_2^2$ is $(x_2 \cdot (-2)^n)^2 = x_2^2 \cdot 4^n$ whereas the closed form for $x_1$ is $\left( \frac{1}{2}
  \cdot x_3^2 + x_1 \right)\cdot 3^n - \frac{1}{2}\cdot x_3^2$, see \Cref{ex:closedFormEx}).

Moreover, we present a new approach to use closed forms in order to compute \emph{size bounds} which indicate how large the absolute value of an integer variable may become.
In contrast to other complete procedures for the inference of size bounds which are based on fixpoint computations \cite{ben-amram2019TightWorstCaseBounds,ben-amram2008LinearPolynomialExponential}, our technique can also handle (possibly negative) constants and exponential size bounds.
As mentioned, in Loop \eqref{WhileExample}, $x_2$ has the closed form $x_2 \cdot (-2)^n$.
We over-approximate closed forms to obtain non-negative, weakly monotonically increasing functions.
The (absolute value of this) closed form can be over-approximated by $x_2 \cdot 2^n$, which is monotonically increasing in all variables when instantiating them with non-negative numbers.
Finally, $n$ is substituted by a runtime bound for the loop.
While the techniques from our previous conference papers \cite{lommen2022AutomaticComplexityAnalysis,lommen2023TargetingCompletenessUsing} would yield an exponential size bound for $x_2$, when substituting $n$ by the logarithmic runtime bound we obtain a \emph{polynomial}
bound on the size of $x_2$.
Due to the restriction to weakly monotonically increasing over-approximations, we can plug in any over-approximation of the runtime and do not necessarily need exact bounds.

In this paper, we embed the complete procedures for runtime and size bounds of \emph{prs}-loops into an incomplete approach for general integer programs \cite{brockschmidt2016AnalyzingRuntimeSize,giesl2022ImprovingAutomaticComplexity}.
This incomplete technique for complexity analysis uses individual ranking functions to infer runtime bounds for different subprograms.
Based on this, we introduce a modular approach to automatically infer runtime and size bounds for programs possibly consisting of multiple consecutive or nested loops by handling some subprograms as \emph{prs}-loops and by using ranking functions and the size bound technique of \cite{brockschmidt2016AnalyzingRuntimeSize} for others.
In order to compute runtime and size bounds, we analyze subprograms in topological order, i.e., in case of multiple consecutive loops, we start with the first loop and propagate knowledge about the resulting values of variables to subsequent loops.
By handling one subprogram after the other, in the end we obtain a bound on the runtime complexity of the whole program.
The modular approach uses size bounds to infer runtime bounds and vice versa.
Hence, runtime and size bound computations are alternated until all bounds are finite or no bound can be improved further.

So for the first time, ``complete'' complexity analysis techniques like \cite{hark2020PolynomialLoopsTermination} for subclasses of programs are combined with incomplete techniques based on (linear) ranking functions like \cite{brockschmidt2016AnalyzingRuntimeSize,giesl2022ImprovingAutomaticComplexity}.
In this way, the power of the incomplete approach is increased significantly, in particular for programs with non-linear arithmetic (in fact, existing tools based on incomplete techniques for complexity analysis often fail for such programs and cannot handle loops like \eqref{WhileExample}).

\paragraph{Structure of the Paper:}
In \Cref{sect:loops}, we consider classes of loops where there exist complete techniques for the computation of bounds.
We show how to compute closed forms for such loops in \Cref{sect:closed_forms}.
Afterwards, we study runtime and size bounds in \Cref{sect:Runtime Bounds for PRS-Loops,sect:loops_size_bounds}, respectively, and summarize the completeness results for loops in \Cref{sect:completenessLoops}.

In \Cref{sect:global_integer_programs}, we introduce (Turing-complete) integer programs in \Cref{sect:integer_programs} and define the concept of runtime and size bounds for integer programs in \Cref{sect:global_bounds}.
Next in \Cref{sect:Runtime Bounds for Integer Programs,sect:Size Bounds for Integer Programs}, we present a modular approach to lift runtime and size bounds of subprograms to bounds for the full program.
Finally, we summarize our completeness results for integer programs in \Cref{sect:Completeness}.

We conclude in \Cref{sect:conclusion} by evaluating the implementation of the approach in the complexity analysis tool \tool{KoAT} and show that \tool{KoAT} can now also successfully analyze the runtime of programs containing non-linear arithmetic.

\paragraph{New Contributions Compared to \cite{lommen2022AutomaticComplexityAnalysis,lommen2023TargetingCompletenessUsing}:}

As mentioned, the current paper is based on our previous conference papers \cite{lommen2022AutomaticComplexityAnalysis,lommen2023TargetingCompletenessUsing}, and it extends them by the following new contributions:
\begin{itemize}
  \item We develop a unified framework to integrate complete methods for both runtime and size bounds into a general approach for integer programs.
        In particular, in \Cref{sect:completenessLoops}
        we also present a new algorithm to compute both runtime and size bounds for loops.
  \item We develop numerous improvements concerning the formalization and presentation of our contributions throughout the paper.
  \item As defined in \Cref{sect:Runtime Bounds for PRS-Loops}, we now also consider bounds with logarithms and maxima, which allows us to express tighter runtime and size bounds.
  \item In addition to polynomial bounds, we can now also compute \emph{logarithmic} runtime bounds for \emph{prs}-loops (see \Cref{lem:complexity_logarithmic}).
        As sketched in the introductory example, this also leads to improved size bounds (see \Cref{exa:sizeboundLoop} in \Cref{sect:loops_size_bounds}).
  \item In \Cref{sect:Runtime Bounds for PRS-Loops}, we extend the procedure for the computation of runtime bounds to \emph{unsolvable loops} \cite{amrollahi2022SolvingInvariantGeneration}.
        For these loops, it is often not clear how to infer closed forms for each variable individually.
        However, for the runtime bound procedure it is enough to find closed forms for the guard of the loop.
        We introduce a preprocessing step which makes use of this observation in order to handle certain unsolvable loops as well.
  \item In \Cref{sect:Size Bounds for Integer Programs}, we extend the complete approach for size bounds to handle multiple commuting loops, i.e., loops for which the order of execution is irrelevant.
  \item We give the proofs of central contributions (\Cref{lem:correctness_chaining}, \ref{lem:complexity}, and \ref{lem:complexity_logarithmic}, and \Cref{thm:completeness,thm:completeness_integer_programs}) in the paper, and all remaining proofs can be found in the appendix.
  \item We extend and improve the implementation of the complexity analysis tool \tool{KoAT} (see \Cref{sect:conclusion}).
\end{itemize}

\paragraph{Main Completeness Results:}
In the following, we present our main completeness results.
\Cref{thm:completeness} captures our core completeness results for \emph{prs}-loops.
For all linear \emph{prs}-loops, termination is decidable.
Moreover, for all terminating \emph{prs}-loops, one can derive meaningful bounds on their runtime and on the sizes of variable values.
Additional syntactic restrictions -- such as unity (i.e., all eigenvalues of the update matrix are inside the closed unit disk) or strictness of the updates (which is a condition on the loop's eigenvalues and its guard) -- yield even smaller complexity
\pagebreak[3]
bounds.
\setcounter{section}{2}
\setcounter{theorem}{44}

\begin{theorem}[Completeness Results for Loops]
  \hspace*{0.1cm}
  \begin{itemize}
    \item Termination is decidable for all linear prs-loops.
    \item Polynomial runtime bounds and finite size bounds are computable for all terminating prs-loops.
    \item For terminating \emph{unit} prs-loops, these size bounds are polynomial as well.
    \item For terminating \emph{strict} prs-loops, the runtime bounds are logarithmic and the size bounds are polynomial.
  \end{itemize}
\end{theorem}

Building on these results for \emph{prs}-loops, we show how to handle general integer programs.
Here, we have to restrict ourselves to \emph{simple}
programs in order to obtain completeness results.
In particular, for a simple program we require that every loop is reachable from every initial program configuration.
Then the integer program is terminating iff each of its loops is terminating.
The following theorem for integer programs then corresponds to the previous theorem for loops.
\setcounter{section}{3}
\setcounter{theorem}{32}
\begin{theorem}[Completeness Results for Integer Programs]
  \hspace*{0.1cm}
  \begin{enumerate}
    \item Termination is decidable for all simple \emph{linear} integer programs where after chaining, all simple cycles correspond to prs-loops.
    \item Finite runtime and size bounds are computable for all simple integer programs where after chaining, all simple cycles correspond to \emph{terminating}
          prs-loops.
    \item If in addition to (b), all simple cycles correspond to \emph{unit} prs-loops, then the runtime and size bounds are \emph{polynomial}.
    \item If in addition to (b), all simple cycles correspond to \emph{unit} and \emph{strict} prs-loops, then the runtime bounds are logarithmic.
  \end{enumerate}
\end{theorem}

However, our results are not restricted to complete techniques.
An important contribution of our work is a modular framework which can be used to combine various (complete and incomplete) techniques in order to handle general (possibly non-simple) integer programs.

In this way, our contributions give rise to a powerful approach to analyze complexity in practice.
This can be seen in the evaluation of \Cref{sect:conclusion}
where we demonstrate the strengths of our implementation in the tool \tool{KoAT}.

\setcounter{section}{1}

\section{Runtime and Size Bounds for Solvable Loops}
\label{sect:loops}
Before we analyze the Turing-complete language of integer programs in \Cref{sect:global_integer_programs}, we consider runtime and size bounds of single-path loops like the following one:
\[
  \WhileLoop{x_2^2 - x_3^5 < x_1 \land x_2 \neq 0}{\myvec{x_1 \\ x_2 \\ x_3 \\ x_4 \\ x_5}
    \leftarrow \begin{bmatrix}
      3 & 0  & 0 & 0  & 0  \\
      0 & -2 & 0 & 0  & 0  \\
      0 & 0  & 1 & 0  & 0  \\
      0 & 0  & 0 & 3  & 2  \\
      0 & 0  & 0 & -5 & -3
    \end{bmatrix}
    \myvec{x_1\\ x_2 \\ x_3 \\ x_4 \\ x_5}
    +
    \myvec{x_3^2 \vspace{0.075cm}\\ 0 \\ 0 \\ 0 \\ 0}
  } \hspace*{\fill} \eqref{WhileExample}
\]
These results are then used in a modular way within the approach for general integer programs in \Cref{sect:global_integer_programs}.

Our approach is based on \emph{solvable loops} \cite{rodriguez2004AutomaticGenerationPolynomial,xuSymbolicTerminationAnalysis2013,kincaidClosedFormsNumerical2019,frohn2020TerminationPolynomialLoops,kovacsReasoningAlgebraicallyPSolvable2008,humenberger2018InvariantGenerationMulti}.
These loops admit \emph{closed forms} which correspond to applying the loop's update $n$ times.
As shown in \cite{frohn2020TerminationPolynomialLoops} for the subclass of \emph{twn}-loops, such closed forms can be used to reduce the question of termination to an existential formula over $\ZZ$.
Furthermore, we will also use these closed forms to compute runtime and size bounds in \Cref{sect:Runtime Bounds for PRS-Loops,sect:loops_size_bounds}.

For a finite set of variables $\VSet$, as usual, $\ZZ[\VSet]$ denotes the polynomial ring over the variables $\VSet$ with integer coefficients.

\begin{definition}[Atoms and Formulas]
  \label{def:formulas}
  The set of \emph{atoms} $\AtomSet(\VSet)$ consists of all inequations $p_1 < p_2$ for polynomials $p_1,p_2\in\ZZ[\VSet]$.
  $\FormulaSet(\VSet)$ is the set of all propositional \emph{formulas} built from atoms $\AtomSet(\VSet)$, $\land$, and $\lor$.
\end{definition}
In addition to ``$<$'', we also use ``$\geq$'', ``$=$'', ``$\neq$'', etc., and negations ``$\neg$'' which can be simulated by formulas (e.g., $p_1 \geq p_2$ is equivalent to $p_2 < p_1 + 1$ for integers).
We often write $\alpha\in\varphi$ if the atom $\alpha$ occurs in the formula $\varphi$.

When regarding loops, we consider a set of variables $\VSet = \braced{x_1, \ldots, x_\indv}$ where $\vec{x} = (x_1,\ldots,x_d)$.
A \emph{loop} $\IntLoop$ consists of a \emph{guard} $\guard\in\FormulaSet(\VSet)$ and an \emph{update}
$\update: \VSet\to\ZZ[\VSet]$ which maps variables to polynomials.
A loop is \emph{linear} if all these polynomials and also the guard only contain linear arithmetic.
A loop $(\guard,\update)$ is \emph{solvable} if $\update$ is a \emph{solvable update} (see \Cref{def:solvable} below for a formal definition), which partitions $\VSet$ into blocks $\SSet_1,\ldots, \SSet_m$.
Each block allows updates with \emph{cyclic dependencies} between its variables and \emph{non-linear} dependencies on variables in blocks with higher indices.

\begin{definition}[Solvable Loops \cite{rodriguez2004AutomaticGenerationPolynomial,xuSymbolicTerminationAnalysis2013,kincaidClosedFormsNumerical2019,frohn2020TerminationPolynomialLoops,kovacsReasoningAlgebraicallyPSolvable2008,humenberger2018InvariantGenerationMulti}]
  \label{def:solvable}
  An update $\update:\VSet \to \ZZ[\VSet]$ is \emph{solvable}
  if there exists a partition $\SSet_1, \ldots, \SSet_m$ of $\braced{x_1, \ldots, x_d}$ such that for all $1 \leq i \leq m$ we have $\vec{\update}_{\SSet_i} = A_{\SSet_i} \cdot \vec{x}_{\SSet_i} + \vec{p}_{\SSet_i}$ for a matrix $A_{\SSet_i}\in\ZZ^{|\SSet_i|\times |\SSet_i|}$ and a vector $\vec{p}_{\SSet_i}$ of polynomials over the variables $\SSet_{i+1} \cup \ldots \cup \SSet_{m}$, i.e., $\vec{p}_{\SSet_i} \in(\ZZ[\SSet_{i + 1}
      \cup \ldots \cup \SSet_m])^{|\SSet_i|}$.
  Here, $\vec{\update}_{\SSet_i}$ is the vector of all $\update(x_j)$ and $\vec{x}_{\SSet_i}$ is the vector of all $x_j$ with $j\in\SSet_i$.
  The eigenvalues of a solvable loop are defined as the union of the eigenvalues of all matrices $A_{\SSet_i}$.
\end{definition}
\begin{example}
  \label{exa:solvable}
  The loop \eqref{WhileExample} is an example for a solvable loop using the partition $\SSet_1 = \braced{x_1}$, $\SSet_2 = \braced{x_2}$, $\SSet_3 = \braced{x_3}$, and $\SSet_4 = \braced{x_4,x_5}$.
  Here, the updates $\update(x_4) = 3 \cdot x_4 + 2 \cdot x_5$ and $\update(x_5) = -5 \cdot x_4 - 3 \cdot x_5$ of $x_4$ and $x_5$ both contain the variables $x_4$ and $x_5$, i.e., they have a cyclic dependency.
  The update $\update(x_1) = 3 \cdot x_1 + x_3^2$ of $x_1$ depends non-linearly on $x_3$, but only linearly on the previous value of $x_1$.
\end{example}

The crucial idea for the results in \Cref{sect:closed_forms,sect:Runtime Bounds for PRS-Loops} is to reduce the problem of finding closed forms and runtime bounds from solvable loops to \emph{triangular weakly non-linear} loops (\emph{twn-loops}) \cite{frohn2019TerminationTriangularInteger,frohn2020TerminationPolynomialLoops,hark2020PolynomialLoopsTermination}.
A \emph{twn-update} is a solvable update where each block $\SSet_i$ has cardinality one.
Thus, a \emph{twn}-update is \emph{triangular}, i.e., the update of a variable does not depend on variables with smaller indices.
Furthermore, the update is \emph{weakly non-linear}, i.e.,
a variable does not occur non-linear in its own update.
We are mainly interested in loops over $\ZZ$,
but to handle solvable updates, we will transform them into \emph{twn}-updates with coefficients from the algebraic numbers $\AA$.
Here, $\AA$ is the field of all (possibly complex) roots of polynomials in $\ZZ[x]$.
\begin{definition}[\emph{twn}-Loops, \emph{tnn}-Loops \cite{frohn2020TerminationPolynomialLoops,hark2020PolynomialLoopsTermination,frohn2019TerminationTriangularInteger}]
  \label{def:twn}
  An update $\update: \VSet\rightarrow\AA[\VSet]$ is \emph{twn} if for all $1 \leq i \leq d$ we have $\update(x_i) = c_i\cdot x_i + p_i$ for some $c_i \in \AA$ and some polynomial $p_i\in\AA[x_{i + 1}, \ldots,x_d]$.
  A \emph{twn}-update is called \emph{non-negative} (or a \emph{tnn-update}) if $c_i \in \NN$.
  A loop with a \emph{twn}-update (resp.\ \emph{tnn}-update) is called a \emph{twn-loop} (resp.\ \emph{tnn-loop}).
\end{definition}
\noindent
Thus, \eqref{WhileExample} is not a \emph{twn}-loop due to the cyclic dependency between $x_4$ and $x_5$.
However, if one removes the variables $x_4$ and $x_5$ (which both do not influence the termination behavior of \eqref{WhileExample}) then one obtains a \emph{twn}-loop.

In \Cref{sect:closed_forms}, we explain how to compute \emph{closed forms} for solvable loops (via a transformation to \emph{twn}-loops).
Afterwards, we consider the subclass of \emph{periodic solvable loops} (\emph{prs}-loops) which encompasses \emph{twn}-loops, and introduce techniques for the inference of runtime and size bounds for \emph{prs}-loops in \Cref{sect:Runtime Bounds for PRS-Loops,sect:loops_size_bounds}.
In \Cref{sect:completenessLoops}, we show that these techniques are \emph{complete} for terminating \emph{prs}-loops.

\subsection{Closed Forms}
\label{sect:closed_forms}
A \emph{closed form} $\cl{x_i}$ (formally defined in \Cref{def:closed_form} below) is an expression in the distinguished variable $n$ and in the (initial values of the) variables $x_1,\ldots,x_\indv$ which corresponds to the value of $x_i$ after iterating the loop $n$ times.
For our purpose, we only need closed forms which hold for all $n \geq n_0$ for some fixed $n_0\in\NN$.
Moreover, we restrict ourselves to closed forms which are \emph{poly-exponential expressions}, as in \cite{kincaidClosedFormsNumerical2019}.
Such closed forms are needed for our technique to compute runtime bounds, whereas our technique for size bounds would work for any closed form expression with a finite number of arithmetic operations (i.e., the number of operations must be independent of $n$), which is weakly monotonically increasing in $n$.
\begin{definition}[Poly-Exponential Expression]
  \label{def:Poly-Exponential Expression}
  The set of \emph{poly-exponential expressions} is $$\PPEE = \braced{ \sum_{j=1}^\ell p_j \cdot n^{a_j} \cdot b_j^n \mathrel{\Big|}\ell, a_j\in \NN, \; b_j\in\AA, \; p_j\in\AA[\VSet]}.$$ We write $\PPEE(\QQ,\NN)$ (resp.\
  $\PPEE(\ZZ,\NN)$) for poly-exponential expressions where $p_j\in\QQ[\VSet]$ (resp.\
  $p_j\in\ZZ[\VSet]$) and $b_j\in\NN$ holds for all $1 \leq j \leq \ell$.
\end{definition}
Note that while we only consider loops with integer coefficients,
their closed forms can be poly-exponential expressions with non-integer and even complex algebraic numbers $b_j\in\AA$ and polynomials $p_j\in\AA[\VSet]$ (see, e.g., \Cref{exa:transform_solvable}).
More precisely, poly-exponential expressions with complex numbers $b_j$ and polynomials $p_j$ with complex coefficients can result from the transformation from solvable loops to \emph{twn}-loops, because the transformation can yield \emph{twn}-loops over $\AA$.
If a \emph{twn}-loop has only rational coefficients and integer eigenvalues, then it has closed forms in $\PPEE(\QQ,\ZZ)$ and for a \emph{tnn}-loop with rational coefficients and integer (and thus, natural) eigenvalues the closed forms are in $\PPEE(\QQ,\NN)$.
Such a poly-exponential expression has the advantage that its asymptotic behavior is determined by its unique dominating summand.

For the computation of runtime bounds, we will need closed forms in $\PPEE(\QQ,\NN)$.
For that reason, we will consider \emph{prs}-loops.
These are solvable loops that can be transformed into \emph{twn}-loops and further into \emph{tnn}-loops with rational coefficients and natural eigenvalues such that their closed forms are indeed in $\PPEE(\QQ,\NN)$.
Then we can infer runtime bounds for these \emph{tnn}-loops which also yield runtime bounds for the original \emph{prs}-loops, see \Cref{sect:Runtime Bounds for PRS-Loops}.

A mapping $\valuation : \VSet \to \ZZ$ is called a \emph{state} and
$\Valuation$ denotes the set of all states.
We extend the application of functions like $\valuation : \VSet \to \ZZ$ and $\update: \VSet \rightarrow \ZZ[\VSet]$ also to polynomials, vectors, and formulas, etc., by replacing each variable $v$ in the expression by $\update(v)$.
So, e.g., for a polynomial $p$, $\update(p)$ results from replacing every variable $v$ in $p$ by $\update(v)$.
In particular, $(\update_1 \circ \update_2)(x) = \update_1(\update_2(x))$ stands for the polynomial $\update_2(x)$ in which every variable $v$ is replaced by $\update_1(v)$.
So if $\update_1(x) = x+1$ and $\update_2(x) = 2 \cdot x$, then $\update_1(\update_2(x)) = \update_1(2 \cdot x) = (2 \cdot x) [x / (x+1)] = 2 \cdot (x+1) = 2 \cdot x + 2$.
Here, ``$(2 \cdot x) [x / (x+1)]$'' means that every occurrence of $x$ in the polynomial $2 \cdot x$ is replaced by $x+1$.

Moreover, $\update^n$ denotes the $n$-fold application of $\update$.
For example, if $\update(x_4) = 3 \cdot x_4 + 2 \cdot x_5$ and $\update(x_5) = -5 \cdot x_4 - 3 \cdot x_5$, then $\update^2(x_4) = \update(\update(x_4)) = \update(3 \cdot x_4 + 2 \cdot x_5) = 3 \cdot (3 \cdot x_4 + 2 \cdot x_5) + 2 \cdot (-5 \cdot x_4 - 3 \cdot x_5) = - x_4$.

\begin{definition}[Closed Forms]
  \label{def:closed_form}
  For a loop $L = (\guard,\update)$, an arithmetic expression $\cl{x_i}$ is a \emph{closed form} for $x_i$ with \emph{start value} $n_0 \in \NN$ if $\cl{x_i}\in\PPEE$ and for all $\valuation:\VSet\cup \{n\}\to \ZZ$ with $\valuation(n) \geq n_0$ we have $\valuation(\cl{x_i}) = \valuation(\update^n(x_i))$.
  Similarly, we call $\cl{\vec{x}} = (\cl{x_1},\ldots,\cl{x_\indv})$ a \emph{closed form} of the update $\update$ (resp.\ for the loop $L$) with start value $n_0$ if for all $1 \leq i \leq d$, $\cl{x_i}$ are closed forms for $x_i$ with start value $n_0$.
\end{definition}

\begin{example}
  \label{ex:closedFormEx}
  For the loop \eqref{WhileExample}, in \Cref{exa:closed_form_solvable} we will compute the closed form $\normalfont{\cl{x_4}} = \tfrac{1}{2}\cdot\alpha\cdot (-\im)^n + \tfrac{1}{2}\cdot\compconj{\alpha}\cdot \im^n$ for $x_4$ (with start value 0) where $\alpha = (1 + 3\im)\cdot x_4 + 2\im\cdot x_5$.
  Here, $\compconj{\alpha}$ denotes the complex conjugate of $\alpha$, i.e., the sign of those monomials is flipped where the coefficient is a real multiple of the imaginary unit $\im$.
  A closed form for $x_1$ (also with start value 0) is $\cl{x_1} = \left( \frac{1}{2}
    \cdot x_3^2 + x_1 \right)\cdot 3^n - \frac{1}{2}\cdot x_3^2$.
\end{example}

\begin{example}
  \label{ex:StartValue}
  The reason for using start values in the definition of closed forms is that there are loops where a start value $n_0 > 0$ is needed in order to obtain closed forms in poly-exponential form (see \Cref{Closed Forms for TWN Updates}).
  As an example, consider the update $
    \begin{sbmatrix}
      x\\
      y\\
      z
    \end{sbmatrix}
    \leftarrow
    \begin{sbmatrix}
      y\\
      z\\
      42
    \end{sbmatrix}
  $.
  The closed form {\normalfont{$\texttt{cl}^x$}} of $x$ is just $42$ for $n \geq 2$.
  However, for $n = 0$ the value of $x$ is the initial value of $x$ and for $n = 1$ the variable $x$ has the initial value of $y$.
  Thus, a closed form for $x$ with start value $n_0 = 0$ would have to perform a case analysis for the cases $n = 0$, $n = 1$, and $n \geq 2$.
\end{example}

\Cref{lem:transform_solvable} (which extends \cite[Thm.\ 5.15]{frohn2020TerminationPolynomialLoops}
from solvable updates with real eigenvalues to arbitrary solvable updates) illustrates that one can transform any \underline{s}olvable update $\update_s$ into a \emph{\underline{t}wn}-update $\update_t$ by an automorphism $\auto$.
Here, $\auto$ is induced by the change-of-basis matrix of the Jordan normal form of each block of $\update_s$.
Note that the Jordan normal form is always computable in polynomial time (see \cite{cai1994ComputingJordanNormal}).
\begin{lemma}
  [Transforming Solvable Updates (see {\cite[Thm.\ 5.15]{frohn2020TerminationPolynomialLoops}})] \label{lem:transform_solvable}
  Let $\update_s$ be a solvable update.
  Then $\auto: \VSet\rightarrow\AA[\VSet]$ is an automorphism, where $\auto$ is defined by $\auto(\SSet) = P\cdot \xvec_\SSet$ for each block $\SSet$, and where $J(A_{\SSet}) = P\cdot A_{\SSet}\cdot P^{-1}$ is the Jordan normal form of $A_\SSet$.
  Furthermore, $\update_t = \auto^{-1}\circ\update_s\circ \auto$ is a twn-update.
\end{lemma}
\begin{example}
  \label{exa:transform_solvable}
  To illustrate \Cref{lem:transform_solvable}, we transform the solvable update $\update_s$ of \eqref{WhileExample} into a twn-update $\update_t$.
  As the blocks $\SSet_1 = \braced{x_1}$, $\SSet_2 = \braced{x_2}$, and $\SSet_3 = \braced{x_3}$ have cardinality one, we only have to consider $\SSet_4 = \braced{x_4,x_5}$.
  The restriction of $\update_s$ to $\SSet_4$ is $\myvec{x_4\\
      x_5} \leftarrow A_{\SSet_4} \cdot \myvec{x_4\\
      x_5}$ with $A_{\SSet_4} =
    \begin{sbmatrix}
      3 & 2 \\
      -5 & -3
    \end{sbmatrix}
  $.
  Thus, we get the Jordan normal form $J(A_{\SSet_4}) = P\cdot A_{\SSet_4} \cdot P^{-1} =
    \begin{sbmatrix}
      -\!\im & 0 \\
      0 & \im
    \end{sbmatrix}
  $ where $P =
    \begin{sbmatrix}
      - \tfrac{5}{2}\im 							 & \;\tfrac{1}{2}(1\!-\!3\im) \\
      \phantom{-} \;\, \tfrac{5}{2}\im & \;\tfrac{1}{2}(1\!+\!3\im)
    \end{sbmatrix}
  $ and $P^{-1} =
    \begin{sbmatrix}
      \tfrac{1}{5}(\im\!-\!3) & \;-\tfrac{1}{5}(\im\!+\!3) \\
      1 										 & \;1
    \end{sbmatrix}
  $.
  Hence, we have the following automorphism $\vartheta$ and its inverse $\vartheta^{-1}$:
  \[
    \mbox{\small
      $
        \begin{array}{rl@{\qquad}lrll}
          \vartheta\myvec{x_1                                                         \\ x_2 \\ x_3}
           & = \myvec{x_1                                                             \\ x_2 \\ x_3} \vspace*{0.1cm} &
           & \vartheta\myvec{x_4                                                      \\ x_5}
           & = P \cdot \myvec{x_4                                                     \\ x_5}
           & = \myvec{- \tfrac{5}{2}\im\cdot x_4 + \tfrac{1}{2}(1-3\im)\cdot x_5      \\ \phantom{-} \;\,	 \tfrac{5}{2}\im\cdot x_4 + \tfrac{1}{2}(1 + 3\im)\cdot x_5 } \\
          \vartheta^{-1}\myvec{x_1                                                    \\ x_2 \\ x_3}
           & = \myvec{x_1                                                             \\ x_2 \\ x_3} &
           & \vartheta^{-1}\myvec{x_4                                                 \\ x_5}
           & = P^{-1} \cdot \myvec{x_4                                                \\ x_5}
           & = \myvec{\tfrac{1}{5}(\im - 3)\cdot x_4 - \tfrac{1}{5}(\im + 3)\cdot x_5 \\ x_4 + x_5}
        \end{array}
      $}
  \]
  Thus, $\update_t = \vartheta^{-1} \circ \update_s \circ \vartheta$ is the following twn-update:
  \[\begin{array}{l@{\qquad}l@{\qquad}l}
      \update_t(x_1) = 3\cdot x_1 + x_3^2 & \update_t(x_2) = -2\cdot x_2  & \update_t(x_3) = x_3 \\
      \update_t(x_4) = -\im\cdot x_4      & \update_t(x_5) = \im\cdot x_5 &
    \end{array}\]
\end{example}

The reason for transforming solvable updates to \emph{twn}-updates is that for the latter, we can re-use the algorithm from \cite{frohn2019TerminationTriangularInteger} to compute poly-exponential closed forms.
While \cite{frohn2019TerminationTriangularInteger} only considered updates with linear arithmetic over $\ZZ$, it can directly be extended to \emph{twn}-updates over $\AA$.

\begin{lemma}[Closed Forms for \emph{twn}-Updates (see \cite{frohn2019TerminationTriangularInteger})]
  \label{Closed Forms for TWN Updates}
  Let $\update$ be a twn-update.
  Then a (poly-exponential) closed form is computable for $\update$.
\end{lemma}
\begin{example}
  \label{exa:closed_form_twn}
  For $\update_t$ from \Cref{exa:transform_solvable}, we obtain the following closed form (with start value $0$):
  $$\clExp{\vec{x}}{t} = \left( (x_1 + \tfrac{1}{2}\cdot x_3^2)\cdot 3^n - \tfrac{1}{2}\cdot x_3^2, \; x_2 \cdot (-2)^n, \; x_3, \; x_4 \cdot (-\im)^n, \; x_5 \cdot \im^n\right)$$
\end{example}

The following theorem shows that poly-exponential closed forms can be computed for all solvable updates $\update_s$.
To this end, we first transform $\update_s$ into a \emph{twn}-update $\update_t$ via \Cref{lem:transform_solvable}, and then compute the closed form $\clExp{\vec{x}}{t}$ of $\update_t$ (\Cref{Closed Forms for TWN Updates}).
Finally, we obtain a closed form for $\update_s$ from $\clExp{\vec{x}}{t}$.
\begin{restatable}[Closed Forms for Solvable Updates]{theorem}{closedformsSolvableUpdates}
  \label{thm:closed_form_solvable}
  Let $\update_s$ be a solvable update and $\auto$ be an automorphism as in \Cref{lem:transform_solvable}
  such that $\update_t = \auto^{-1}\circ\update_s\circ \auto$ is a twn-update.
  If $\clExp{\vec{x}}{t}$ is a closed form of $\update_t$ with start value $n_0$, then $\clExp{\vec{x}}{s} = \auto\circ\clExp{\vec{x}}{t}\circ\auto^{-1}$ is a closed form of $\update_s$ with start value $n_0$.
\end{restatable}
\makeproof{thm:closed_form_solvable}{
  \closedformsSolvableUpdates*
  \begin{myproof}
    Let $\clExp{\vec{x}}{t}$ be a closed form for $\update_t$ with start value $n_0$.
    We have to prove that $\auto\circ\normalfont{\clExp{\vec{x}}{t}}\circ\auto^{-1}$ is a closed form for $\update_s$ with start value $n_0$.
    Let $\valuation: \VSet\rightarrow\ZZ$ be an arbitrary state and $v\in\VSet$.
    Note that $\update_t = \auto^{-1}\circ\update_s\circ \auto$ implies $\update_s = \auto\circ\update_t\circ \auto^{-1}$.

    Thus, for all $v \in \VSet$ and all $n \geq n_0$ we have
    \begin{align*}
      \valuation(\update_s^n(v)) & = \valuation((\update_s\circ\ldots\circ\update_s)(v)) \\
                                 & = \valuation(((\auto \circ \update_t \circ \auto^{-1})\circ\ldots\circ(\auto \circ \update_t \circ \auto^{-1}))(v)) \\
                                 & = \valuation((\auto \circ (\update_t \circ \ldots\circ \update_t) \circ \auto^{-1})(v)) \\
                                 & = \valuation((\auto \circ \update_t^n \circ \auto^{-1})(v)) \\
                                 & = \valuation((\auto \circ \clExp{\vec{x}}{t} \circ
      \auto^{-1})(v)).
    \end{align*} \qed
  \end{myproof}
}

\begin{example}
  \label{exa:closed_form_solvable}
  In \cref{exa:transform_solvable} we transformed $\update_s$ into the twn-update $\update_t$ via an automorphism $\vartheta$ and in \cref{exa:closed_form_twn}, we gave the closed form $\normalfont{\clExp{\vec{x}}{t}}$ of $\update_t$.
  Thus, by \cref{thm:closed_form_solvable}, we can infer a closed form $\normalfont{\clExp{\vec{x}}{s}}= \vartheta\circ\normalfont{\clExp{\vec{x}}{t}}\circ\vartheta^{-1}$ of $\update_s$.
  For example, in this way we compute the following closed form for $x_4$ with start value $0$ (the remaining closed forms can be inferred in a similar way).
  Here, ``$v/\clExp{v}{t}$'' means that the variable $v$ is substituted by the expression $\clExp{v}{t}$, and ``$\left[v/\clExp{v}{t} \mid v\in\VSet \right]$'' stands for $[x_1/\clExp{x_1}{t}, \ldots, x_d/\clExp{x_d}{t}]$.
  \begin{align*}
    \clExp{x_4}{s}
     & = \auto^{-1}(x_4) \; \left[v/\clExp{v}{t} \mid v\in\VSet \right] \; \left[v/\vartheta(v) \mid v\in\VSet \right] \\
     & = \left(\tfrac{1}{5}(\im - 3)\cdot x_4 - \tfrac{1}{5}(\im + 3)\cdot x_5\right)\; \left[v/\clExp{v}{t} \mid v\in\VSet \right] \; \left[v/\vartheta(v) \mid v\in\VSet \right] \\
     & = \left(\tfrac{1}{5}(\im - 3)\cdot(-\im)^n \cdot x_4 - \tfrac{1}{5}(\im + 3)\cdot\im^n \cdot x_5\right) \; \left[v/\vartheta(v) \mid v\in\VSet \right] \\
     & = \tfrac{1}{2}(\underbrace{(1 + 3\im)\cdot x_4 + 2\im\cdot x_5}_{\alpha}) \cdot(-\im)^n + \tfrac{1}{2}(\underbrace{(1 - 3\im)\cdot x_4 - 2\im\cdot x_5}_{\compconj{\alpha}}) \cdot\im^n.
  \end{align*}
\end{example}

\subsection{Runtime Bounds for \emph{prs}-Loops}
\label{sect:Runtime Bounds for PRS-Loops}

In this section, we show that for the subclass of terminating \emph{periodic rational}
solvable (\emph{prs}) loops, one can always compute polynomial runtime bounds (and under certain conditions even logarithmic runtime bounds).
To this end, we first show in \Cref{sect:Reducing Runtime Bounds of PRS-Loops to TNN-Loops}
that when transforming a \emph{prs}-loop into a \emph{tnn}-loop, every runtime bound for the resulting \emph{tnn}-loop can be transformed into a runtime bound for the original \emph{prs}-loop.
Afterwards, we explain how to compute runtime bounds for \emph{tnn}-loops in \Cref{sect:Computing Runtime Bounds for TNN-Loops}.

The results on the transformation of \emph{prs}- to \emph{twn}- or \emph{tnn}-loops also help us to show termination of \emph{prs}-loops, because while \Cref{lem:transform_solvable} only ensures that every solvable loop can be transformed into a \emph{twn}-loop (whose update may contain complex algebraic coefficients), we now show that \emph{prs}-loops can be transformed into \emph{twn}-loops whose updates only contain rational coefficients.
Then, as shown in \cite{frohn2020TerminationPolynomialLoops}, the question of termination of such \emph{twn}-loops can be reduced to (in)validity of an existential formula.

On the one hand, we use runtime bounds of \emph{prs}-loops to compute size bounds for loops in \Cref{sect:loops_size_bounds}.
On the other hand, runtime bounds of \emph{prs}-loops play a crucial role in the analysis of Turing-complete programs in \Cref{sect:global_integer_programs}.
In \Cref{sect:global_integer_programs}, we investigate \emph{subprograms} which correspond to \emph{prs}-loops and use \emph{local} runtime bounds for these loops in order to obtain runtime bounds for the \emph{full program}.

We start with defining the notion of \emph{bounds}.
We only consider bounds which correspond to functions $f$ that are weakly monotonically increasing in all arguments, i.e., where $x \leq y$ implies $f(\ldots x \ldots) \leq f(\ldots y \ldots)$.
In this way, if $f$ and $g$ are both upper bounds, then $f\circ g$ is also an upper bound, i.e., bounds can be ``composed'' easily.
In principle, every weakly monotonically increasing function could be used as a bound in our framework.
However, here we restrict ourselves to bounds which are easy to represent and to compute with, and which cover the most prominent complexity classes.
In contrast to our earlier papers \cite{brockschmidt2016AnalyzingRuntimeSize,giesl2022ImprovingAutomaticComplexity,lommen2022AutomaticComplexityAnalysis,lommen2023TargetingCompletenessUsing}, we now extend this approach to also compute \emph{logarithmic} bounds.
Furthermore, bounds contain $\RRC_{\geq 0} = \RR_{\geq 0} \cup \braced{\omega}$ instead of $\NNC = \NN \cup \braced{\omega} $ as the logarithm yields real numbers.
Moreover, in contrast to our conference papers \cite{lommen2022AutomaticComplexityAnalysis,lommen2023TargetingCompletenessUsing}, we also use expressions with ``maximum'' in bounds.
These improvements allow us to express and infer tighter bounds.
In particular, as already sketched in \Cref{sect:introduction} and shown in detail in \Cref{exa:sizeboundLoop} in \Cref{sect:loops_size_bounds}, logarithmic runtime bounds can be used to ``cancel out'' exponential size bounds, such that in the end one again obtains polynomial bounds.

\begin{definition}[Bounds]
  \label{def:bounds}
  The set of \emph{bounds} $\BoundSet$ is the smallest set with $\RRC_{\geq 0} \subseteq \BoundSet$, $\VSet\subseteq\BoundSet$, and $\braced{\bound_1 + \bound_2, \bound_1\cdot \bound_2, k^{\bound_1}, \log_k(\bound_1), \max \braced{\bound_1,\ldots,\bound_m}}\subseteq\BoundSet$ for all $\bound_1,\ldots,\bound_m\in\BoundSet$ with $m \geq 2$ and $k\in\RR_{> 1}$.\footnote{\label{logFunction}
    More precisely, instead of $\log_k(\bound_1)$ we use the function $\log_k(\max\braced{1,\bound_1})$ to ensure that bounds are well defined, non-negative, and weakly monotonically increasing.}
\end{definition}

For $\bound \in \BoundSet$, let $\bound[v/n \mid v \in \VSet]$ denote the result of replacing all variables in $\bound$ by $n$.
For $\mathit{deg} \in \NN$, let $\textsf{PB}(n^\mathit{deg}) = \{ \bound \in \BoundSet \mid \bound[v/n \mid v \in \VSet] \in \mathcal{O}(n^\mathit{deg}) \}$ denote the class of all \emph{polynomial bounds} of degree at most $\mathit{deg}$, where $\textsf{PB} = \bigcup_{\mathit{deg}\in\NN}\textsf{PB}(n^\mathit{deg})$ is the set of all polynomial bounds.
The bounds in $\textsf{PB}(n)$ are called \emph{linear}.
Similarly, the set of \emph{logarithmic bounds} is $\textsf{LB} = \{ \bound \in \BoundSet \mid \bound[v/n \mid v \in \VSet] \in \mathcal{O}(\log(n)) \, \}$.
So for example, $4 + 3\cdot\log_{2}(2\cdot x_1 + x_3^2 + 2\cdot x_3^5)$ is a logarithmic (and thus, also linear) bound, whereas the bound $x_1 + \max(\braced{x_3 + x_4,5})$ is only linear but not logarithmic.
In these bounds, ``$x_i$'' always stands for the initial (absolute) value of the variable $x_i$.

Note that the transformation from solvable to \emph{twn}-loops via an automorphism $\auto : \VSet \to \AA[\VSet]$ in \Cref{lem:transform_solvable} may yield \emph{twn}-loops with non-integer coefficients.
While the complete approach for the computation of runtime bounds cannot be used for \emph{twn}-loops resulting from arbitrary automorphisms, it can be used when restricting ourselves to $\QQ$-automorphisms.
Thus, we now define the \emph{runtime complexity} and \emph{runtime bounds} for loops w.r.t.
some automorphism $\auto : \VSet \to \QQ[\VSet]$.
Then the states of the loop have the form $\valuation\in\auto(\Valuation) = \braced{\valuation' \circ \auto \mid \valuation'\in\Valuation }$, i.e., for all $v \in \VSet$ we have $\valuation(v) = \valuation'(\auto(v))$ for some $\valuation': \VSet \to \ZZ$.
In other words, instead of executing the loop on $\ZZ^\indv$, we now execute it on $\auto(\ZZ^\indv) = \braced{\auto(\vec{x}) [\vec{x}/\vec{e}]\mid \vec{e}\in\ZZ^d}$ where $\auto(\vec{x}) [\vec{x}/\vec{e}]$ stands for \[(\auto(x_1)[x_1/e_1, \ldots, x_d/e_d], \ldots, \auto(x_d)[x_1/e_1, \ldots, x_d/e_d]).\]
Here, $\auto(x_i)[x_1/e_1, \ldots, x_d/e_d]$ means that all variables in the polynomial $\auto(x_i)$ are instantiated according to the substitution $[x_1/e_1, \ldots, x_d/e_d]$.
\begin{example}
  For the update with $\update(x_1) = x_1 + 2\cdot x_2$ and $\update(x_2) = 2\cdot x_1 + x_2$ and the automorphism with $\auto(x_1) = \tfrac{1}{2}(x_2 - x_1)$ and $\auto(x_2) = \tfrac{1}{2}(x_1 + x_2)$, we obtain $\auto(\ZZ^2) = \braced{(\tfrac{1}{2}(e_2 - e_1),\tfrac{1}{2}(e_1 + e_2)) \mid e_1, e_2\in\ZZ}\subseteq\QQ^2$.
\end{example}
The \emph{runtime complexity} of a loop $L$ for an initial state $\valuation\in\auto(\Valuation)$ -- denoted as $\rc_L^\auto(\valuation)$ -- is the (possibly infinite) number of iterations of $L$ when starting in $\valuation$.
Our aim is to infer a \emph{runtime bound} which over-approximates the length of all possible executions of $L$.
We measure the sizes of variables by their absolute values, i.e., we want to infer a bound on the number of iterations of a loop, where this bound depends on the absolute initial values of the variables.
For any $\valuation \in \Valuation$, let $|\valuation|$ be the state with $|\valuation|(v) = |\valuation(v)|$ for all $v \in \VSet$.

\begin{definition}[Runtime Complexity and Runtime Bounds for Loops]
  \label{def:runtimeLoop}
  For an automorphism $\auto:\VSet\to\QQ[\VSet]$, the \emph{runtime complexity} of a loop $L = (\guard,\update)$ is $\rc_L^\auto : \auto(\Valuation) \to \NNC$ with $\rc_L^\auto(\valuation) = \min\braced{n\in\NN\mid \valuation(\update^n( \neg\guard))}$, where $\min \emptyset	= \omega$.
  An expression $\run \in \BoundSet$ is a \emph{runtime bound} for $L$ on $\auto(\ZZ^\indv)$ if $|\valuation|(\run) \geq \rc_L^\auto(\valuation)$ for all $\valuation\in\auto(\Valuation)$.
\end{definition}
If the loop is clear from the context, then we often omit the index $L$ in the following.
Similarly, if we want to consider the runtime complexity over $\ZZ$, i.e., $\auto$ is the identity $\identity$, then we omit the automorphism.
\begin{example}
  \label{run example}
  When denoting states $\valuation\in\Valuation$ as tuples $(\valuation(x_1),\ldots,\valuation(x_5)) \in \ZZ^5$, then a run of the loop \eqref{WhileExample} that starts in $(10,1,2,10,10)$ has the form $(10,1,2,10,10) \to (12\cdot 3 - 2,-2,2,50,-80) \to \dots \to (12\cdot 3^8 - 2,2^8,2,10,10) \to (12\cdot 3^9 - 2,(-2)^9,2,50,-80)$.
  Note that $(12\cdot 3^9 - 2,(-2)^9,2,50,-80)$ is the first state which does not fulfill the guard and it is reached after $9$ evaluation steps.
  Hence, the runtime complexity of \eqref{WhileExample} is $\rc(\valuation) = 9$.
  In \Cref{runtimeBoundWhileExample}, we will show that $\run = 4 + 3\cdot\log_{2}(2\cdot x_1 + x_3^2 + 2\cdot x_3^5)$ is a runtime bound for \eqref{WhileExample}.
  As mentioned, in these bounds ``$x_i$'' always stands for the initial (absolute) value of the variable $x_i$.
  Thus, we have $|\valuation|(4 + 3\cdot\log_{2}(2\cdot x_1 + x_3^2 + 2\cdot x_3^5)) \geq \rc(\valuation)$ for all states $\valuation\in\Valuation$.
  For example, we obtain $|\valuation|(4 + 3\cdot\log_{2}(2\cdot x_1 + x_3^2 + 2\cdot x_3^5)) \approx 23.38 > 9$ for $\valuation = (10,1,2,10,10)$.
\end{example}

\subsubsection{Reducing Runtime Bounds of prs-Loops to tnn-Loops}
\label{sect:Reducing Runtime Bounds of PRS-Loops to TNN-Loops}
In \Cref{sect:closed_forms}, we discussed how to compute closed forms for solvable updates (by transforming them to \emph{twn}-updates).
In \cite{hark2020PolynomialLoopsTermination} it was shown that (polynomial) runtime bounds can always be computed for terminating \emph{twn}-loops over the integers.
However, in general, transforming solvable loops via \Cref{lem:transform_solvable} yields \emph{twn}-updates which may contain algebraic (complex) numbers.
Nevertheless, we now show that for the subclass of terminating \emph{periodic rational} solvable loops, (finite) polynomial runtime bounds can always be computed as well.

\begin{definition}[\emph{prs}-Loops]
  \label{def:prs-loops}
  A number $\lambda\in\AA$ is \emph{periodic rational} \cite{kincaidClosedFormsNumerical2019} if $\lambda^p\in\QQ$ for some $p\in\NN$ with $p > 0$.
  The \emph{period} of $\lambda$ is the smallest such $p$ with $\lambda^p\in\QQ$.
  A solvable loop is \emph{periodic rational} (i.e., it is a \emph{prs-loop}) with period $p$ if all its eigenvalues $\lambda$ are periodic rational and $p$ is the least common multiple of all their periods.
\end{definition}
So $\im$, $-\im$, and $\sqrt{2} \cdot \im$ are periodic rational with period 2, while $\sqrt{2} + \im$ is not periodic rational.
The following lemma from \cite{kincaidClosedFormsNumerical2019} gives a bound on the period of \emph{prs}-loops and thus yields an algorithm to detect \emph{prs}-loops and to compute their period.
\begin{lemma}[Bound on the Period \cite{kincaidClosedFormsNumerical2019}]
  \label{Bound on the Period}
  Let $A\in\ZZ^{n\times n}$.
  If $\lambda$ is a periodic rational eigenvalue of $A$ with period $p$, then $p \leq n^3$.
\end{lemma}

Now we show that by \emph{chaining}
(i.e., by performing $p$ iterations of a \emph{prs}-loop with period $p$ in a single step), one can transform any \emph{prs}-loop into a solvable loop with only integer eigenvalues.
In \Cref{sect:Computing Runtime Bounds for TNN-Loops}, we will show that based on the transformation to \emph{twn}-loops from \Cref{lem:transform_solvable}, one can then infer runtime bounds for the original \emph{prs}-loop.

Note that although the period of each eigenvalue is polynomially bounded by \Cref{Bound on the Period}, their least common multiple can still be exponential.
For future work, it may be interesting to explore ways to circumvent our explicit construction which may add exponentially many conjuncts to the loop guard.

\begin{definition}[Chaining Loops]
  \label{def:chaining}
  Let $L = \IntLoop$ be a loop and $p \in \NN \setminus \{0\}$.
  Then $L_p = (\guard_p,\update_p)$ results from iterating $L$ $p$ times, i.e., $\guard_p = \guard \, \land \, \update(\guard) \, \land \, \update(\update(\guard)) \, \land \, \ldots \,\, \land \, \update^{p-1}(\guard) \; \text{ and } \; \update_p(v) = \update^p(v) \text{ for all } v\in\VSet.$
\end{definition}
\noindent
We often denote $p$ chaining steps of the loop $L$ by $\underbrace{L\chain\dots\chain L}_{p \, \mathrm{times}}$.

Note that here, we only define the concept of chaining a loop with itself.
Of course, one could also define chaining in order to combine two arbitrary loops.
We will introduce such a more general definition of chaining for transitions of general integer programs in \Cref{Chaining Transitions}.

\begin{example}
  \label{exa:chaining}
  The update matrix of the loop \eqref{WhileExample} has the eigenvalues $\pm\im, 1, -2$, and $3$.
  The eigenvalues $\{1,-2,3\}$ have period 1 whereas the eigenvalues $\pm\im$ have period 2.
  Hence, \eqref{WhileExample} is a \emph{prs}-loop with period 2.
  Chaining of $L = (\guard, \update)$ yields $L \chain L = (\guard\wedge\update(\guard), \update^2)$.
  To ease the presentation, when chaining \eqref{WhileExample}
  we will keep the guard $\guard$ instead of using $\guard\wedge\update(\guard)$ (ignoring $\update(\guard)$ in the conjunction of the guard does not decrease the runtime complexity).
  Thus, $\eqref{WhileExample} \chain \eqref{WhileExample}$ is
  \begin{equation}
    \label{WhileExampleChained}
    \WhileLoop{x_2^2 - x_3^5 < x_1 \land x_2 \neq 0}{\myvec{x_1 \\ x_2 \\ x_3 \\ x_4 \\ x_5} \gets
      \myvec{\,9\cdot x_1 \\ 4\cdot x_2 \\ x_3 \\ -x_4 \\ - x_5 }
      +
      \myvec{4\cdot x_3^2 \\ 0 \\ 0 \\ 0 \\ 0}
    }.
  \end{equation}
\end{example}

In \cite{hark2020PolynomialLoopsTermination}, a complete method for the inference of runtime bounds was presented for \emph{twn}-loops over the integers.
As mentioned in \cite{hark2020PolynomialLoopsTermination}, this method can be extended to loops that can be transformed to \emph{twn}-loops via a $\QQ$-automorphism.
However, this observation does not allow us to extend the method to arbitrary solvable loops, because while we can transform every solvable update into a \emph{twn}-update by a (linear) automorphism $\auto$ due to \cref{lem:transform_solvable}, in general the automorphism $\auto$ may use (complex) algebraic numbers.
However, for solvable loops with integer eigenvalues, $\QQ$-automorphisms suffice for the transformation to \emph{twn}-loops.
Hence, such automorphisms also suffice for \emph{prs}-loops $L$.
The reason is that we first chain the \emph{prs}-loop $L$.
The resulting chained loop $L_p$ has only integer eigenvalues, and therefore, the automorphism $\auto$ that transforms $L_p$ into a \emph{twn}-loop $L_t$ via \Cref{lem:transform_solvable} is a $\QQ$-automorphism.
By multiplying all inequations with the least common multiple of all occurring denominators, the guard of the transformed loop $L_t$ again only contains integer coefficients, i.e., all atoms are of the form $p_1 < p_2$ with $p_1, p_2 \in \ZZ[\VSet]$.

The loop $L_t\chain L_t$ is a \emph{tnn}-loop, i.e., the update has the form $\update(x_i) = c_i \cdot x_i + p_i$ with $c_i\in\NN$ (instead of $c_i\in\ZZ$ as for \emph{twn}-loops over the integers).
Then we can infer a runtime bound for $L_t\chain L_t$ on $\auto(\ZZ^d)$ as shown in \Cref{sect:Computing Runtime Bounds for TNN-Loops}.
Afterwards, the runtime bound for $L_t \chain L_t$ on $\auto(\ZZ^d)$ can be lifted to a runtime bound for the original loop $L$ on $\ZZ^\indv$ by reconsidering the automorphism $\auto$.

Similarly, in order to prove termination of the \emph{prs}-loop $L$, we analyze termination of $L_t \chain L_t$ on $\auto(\ZZ^d)$.
By \cite{frohn2020TerminationPolynomialLoops}, termination of $L_t \chain L_t$ on $\auto(\ZZ^d)$ is reducible to invalidity of a formula $\exists \vec{x}\in\QQ^d . \psi_{\auto(\ZZ^d)}\wedge \xi_{L_t\chain L_t}$.
Here, $\psi_{\auto(\ZZ^d)}$ holds iff $\vec{x}\in\auto(\ZZ^d)$ and $\xi_{L_t\chain L_t}$ holds iff $L_t \chain L_t$ does not terminate on $\vec{x}$.
As shown in \cite{frohn2020TerminationPolynomialLoops}, non-termination of linear \emph{twn}-loops with integer eigenvalues is \textsf{NP}-complete and it is semi-decidable for \emph{twn}-loops with non-linear arithmetic.
Thus, in order to analyze termination of \emph{prs}-loops, we use SMT solvers and try to prove (in)validity of the existential formula above.

In the following lemma, $||\cdot||$ is used to transform any polynomial into a (weakly monotonic) bound from $\BoundSet$.
For any $\auto:\VSet\rightarrow\QQ[\VSet]$, $||\auto||:\VSet\rightarrow\QQ_{\geq 0}[\VSet]$ is defined as $||\auto||(v) = ||\auto(v)||$, where for any polynomial $p = \sum_i c_i \cdot \beta_i$ with $c_i \in \AA$ and normalized monomials $\beta_i$ of the form $x_1^{e_1}\cdot \ldots \cdot x_d^{e_d}$, $||p||$ denotes $\sum_i |c_i| \cdot \beta_i$.

\begin{restatable}[Runtime Bounds for \emph{prs}-Loops]{lemma}{prsLoops}
  \label{lem:correctness_chaining}
  Let $L$ be a \emph{prs}-loop with period $p$ and let $L_p = (\guard_p,\update_p)$ result from chaining as in \Cref{def:chaining}.
  From $\update_p$, one can compute a linear automorphism $\auto:\VSet\rightarrow\QQ[\VSet]$ as in \cref{lem:transform_solvable} such that:
  \begin{enumerate}
    \item $L_p$ is solvable and only has integer eigenvalues.
    \item $\update_t = (\auto^{-1}\circ\update_p\circ\auto) : \VSet \to \QQ[\VSet]$ is a twn-update as in \Cref{def:twn} where $\update_t(x_i) = c_i \cdot x_i + p_i$ with $c_i\in\ZZ$ and $p_i \in \QQ[x_{i+1},\dots,x_d]$.
          Furthermore, $\update_t \circ \update_t$ is a tnn-update.
    \item Let $L_t = (\guard_t, \update_t)$ where $\guard_t$ results from $\auto^{-1}(\guard_p)$ by multiplying all inequations with the least common multiple of all occurring denominators.
          Then, $L_t \chain L_t$ is a tnn-loop with updates of the form $\update_t^2 (x_i) = c_i' \cdot x_i + p_i'$ with $c_i'\in\NN$ and $p_i' \in \QQ[x_{i+1},\dots,x_d]$, and the following statements are equivalent: \vspace*{0.1cm}
          \begin{itemize}
            \item $L$ terminates on $\ZZ^d$
            \item $L_p$ terminates on $\ZZ^d$
            \item $L_t$ terminates on $\auto(\ZZ^d) = \braced{\auto(\vec{x})[\vec{x}/\vec{e}]\mid \vec{e}\in\ZZ^d}$
            \item $L_t \chain L_t$ terminates on $\auto(\ZZ^d) = \braced{\auto(\vec{x})[\vec{x}/\vec{e}]\mid \vec{e}\in\ZZ^d}$
          \end{itemize}
          \vspace*{0.1cm}
    \item If $\run$ is a runtime bound for $L_t\chain L_t$ on $\auto(\ZZ^d)$, then $2\cdot p\cdot ||\auto||(\run) + 2\cdot p - 1$ is a runtime bound for $L$ on $\ZZ^d$.
          If $L_t$ already has a tnn-update and $\run$ is a runtime bound for $L_t$ on $\auto(\ZZ^d)$, then $p\cdot ||\auto||(\run) + p - 1$ is a runtime bound for $L$ on $\ZZ^d$.
  \end{enumerate}
\end{restatable}
\begin{myproof}
  \begin{enumerate}
    \item Consider a block $\SSet_i$ of $L$'s original update $\update$.
          Then we have $\vec{\update}_{\SSet_i} = A_{\SSet_i} \cdot \vec{x}_{\SSet_i}
            + \vec{p}_{\SSet_i}$ for an $A_{\SSet_i}\in\ZZ^{|\SSet_i|\times |\SSet_i|}$ and a $\vec{p}_{\SSet_i}\in(\ZZ[\SSet_{i+1} \cup \ldots \cup \SSet_m])^{|\SSet_i|}$, as in \Cref{def:solvable}.
          For any $k \geq 1$, let $\vec{\update}^{\,k}(\vec{x}_{\SSet_i})$ denote the vector of $\update^k(x_j)$ for all $j \in \SSet_i$.
          By induction on $k$, we show that $\vec{\update}^{\,k}(\vec{x}_{\SSet_i}) = A_{\SSet_i}^k \cdot \vec{x}_{\SSet_i} + \pvec{p}_{\!\SSet_i}'$ for some $\pvec{p}'_{\!\SSet_i}\in(\ZZ[\SSet_{i + 1} \cup \ldots \cup \SSet_m])^{|\SSet_i|}$.
          For $k = 1$ the claim trivially holds, because $\vec{\update}(\vec{x}_{\SSet_i}) = \vec{\update}_{\SSet_i}$.
          For $k > 1$ we get
          \begin{align*}
            \vec{\update}^{\,k + 1}(\vec{x}_{\SSet_i}) = \update(\vec{\update}^{\,k}(\vec{x}_{\SSet_i})) & \stackrel{\text{IH}}{=}
            \update( A_{\SSet_i}^k \cdot \vec{x}_{\SSet_i} + \pvec{p}_{\!\SSet_i}') \\
                                                                                                         & = A_{\SSet_i}^k \cdot \update(\vec{x}_{\SSet_i})+ \update(\pvec{p}_{\!\SSet_i}') \\
                                                                                                         & = A_{\SSet_i}^k \cdot ( A_{\SSet_i} \cdot \vec{x}_{\SSet_i} + \vec{p}_{\SSet_i}
            ) + \update(\pvec{p}_{\!\SSet_i}') \\
                                                                                                         &
            = A^{k + 1}_{\SSet_i} \cdot \vec{x}_{\SSet_i}
            + \underbrace{A_{\SSet_i}^k \cdot \vec{p}_{\SSet_i}
              + \update(\pvec{p}_{\!\SSet_i}')}_{\in(\ZZ[\SSet_{i + 1} \cup \ldots \cup \SSet_m])^{|\SSet_i|}}.
          \end{align*}
          The claim implies that $\update_p$ still is solvable (with the same partition), since $\update_p(v) = \update^p(v)$ for all $v \in \VSet$ by the definition of chaining in \Cref{def:chaining}.

          We now show that $A_{\SSet_i}^p$ only has integer eigenvalues.
          The characteristic polynomial $\chi_{A_{\SSet_i}^p}$ of $A_{\SSet_i}^p$ is monic, since
          \[
            \chi_{A_{\SSet_i}^p} = \det(\lambda I - A_{\SSet_i}^p) = \sum_{\pi\in S}\sgn(\pi)\prod_{j = 1}^{|\SSet_i|}
            (\lambda\delta_{j,\pi(j)}-(A_{\SSet_i}^p)_{j,\pi(j)})
            \in \ZZ[\lambda]
          \]
          for the symmetric group $S$ of degree $|\SSet_i|$ and the Kronecker delta function $\delta_{i,j}$, where $\lambda^{|\SSet_i|}$ only results from the multiplication of the factors $(\lambda\delta_{j,\pi(j)}- \ldots)$ for $\pi = \text{id}$ and thus, $\sgn(\pi) = 1$.
          Therefore, we have $r_2 =1$ for every root $\frac{r_1}{r_2}\in\QQ$ of $\chi_{A_{\SSet_i}^p}$ by the rational root theorem (see e.g., \cite{bunt1988historical}).
          The matrix $A_{\SSet_i}^p\in\ZZ^{|\SSet_i| \times |\SSet_i|}$ clearly only has eigenvalues in $\QQ$ by the definition of the period $p$.
          Thus, every root of $\chi_{A_{\SSet_i}^p}$ and hence every eigenvalue is integer.
    \item When using the automorphism $\auto$ induced by the Jordan normal form, one obtains the \emph{twn}-update $\update_t = \auto^{-1}\circ\update_p\circ\auto$ by \cref{lem:transform_solvable}.
          Here, all constants $c_i$ are in $\ZZ$ as they are the (integer) eigenvalues of $L_p$.
          Similarly, $\update_t\circ\update_t$ is a \emph{tnn}-update with $c_i\in \NN$, since the $c_i$'s correspond to the eigenvalues of $A^p\cdot A^p$, where $A$ is the update matrix of $L$.

          Now we show that $\auto(v)\in\QQ[\VSet]$ for all $v\in\VSet$.
          According to the definition of $\auto$ in \Cref{lem:transform_solvable}, to this end we have to show that the change-of-basis matrix to transform $A^p_{\SSet_i}$ into Jordan normal form contains only rational numbers.
          Let $\lambda\in\ZZ$ be an eigenvalue of $A^p_{\SSet_i}\in\ZZ^{|\SSet_i|\times |\SSet_i|}$ for the block $\SSet_i$.
          In order to compute the change-of-basis matrix, for all Jordan blocks of the eigenvalue $\lambda$, we have to construct the respective Jordan chains $(\lambda I - A^p_{\SSet_i})^{\rho - 1} \vec{u}, \, \ldots, \, (\lambda I - A^p_{\SSet_i}) \vec{u}, \, \vec{u}$ for suitable generalized eigenvectors $\vec{u}\in\ker((\lambda I - A^p_{\SSet_i})^{\rho})$ of some dimension $\rho$.
          As $\vec{u}\in\ker((\lambda I - A_{\SSet_i}^p)^{\rho})$ iff $(\lambda I - A^p_{\SSet_i})^{\rho}\vec{u} = 0$, we have $\vec{u}\in\QQ^{|\SSet_i|}$.
          Thus, all columns of the change-of-basis matrix -- generated by the eigenvalue $\lambda$ -- are also in $\QQ^{|\SSet_i|}$, because the change-of-basis matrix consists of these Jordan chains (for suitable choices of $\rho$).
          Hence, for all eigenvalues $\lambda$, the corresponding columns are always in $\QQ^{|\SSet_i|}$ and thus the whole change-of-basis matrix is rational.
    \item $L$ terminates on $\ZZ^d$ iff $L_p$ terminates on $\ZZ^d$ because chaining does not change the termination behavior \cite{frohn2019TerminationTriangularInteger}.
          In \cite[Cor.\ 17]{frohn2020TerminationPolynomialLoops} it is shown that $L_p$ terminates on $\ZZ^d$ iff $L_t$ terminates on $\auto(\ZZ^d)$, i.e., the transformation via automorphisms also does not change the termination behavior.
          Furthermore, $L_t$ terminates on $\auto(\ZZ^d)$ iff $L_t \chain L_t$ terminates on $\auto(\ZZ^d)$ again as chaining does not change the termination behavior.
    \item In the following, we only show the second part of the statement, i.e., that $p\cdot ||\auto||(\run) + p - 1$ is a runtime bound for $L$ resulting from the runtime bound $\run$ for $L_t$.
          The first part of the statement can be shown analogously by performing $(2\cdot p)$ chaining steps (instead of $p$ chaining steps) and directly transforming $L$ into $L_t\chain L_t$.
          For any state $\valuation: \VSet \to \ZZ$, we define the state $\auto(\valuation)$ by $(\auto(\valuation))(v) = \valuation(\auto(v))$ for all $v \in \VSet$.
          Similar to \cite[Cor.\ 17]{frohn2020TerminationPolynomialLoops}, we now show that for any state $\valuation$ and any $n \in \NN$ we have $(\auto(\valuation)) (\update_t^n(\guard_t)) \Leftrightarrow \valuation(\update_p^n(\guard_p))$, i.e., a run of $L_p$ starting in the state $\valuation$ corresponds to a run of $L_t$ starting in the state $\auto(\valuation)$.
          We have
          \begin{align*}
             & (\auto(\valuation)) (\update_t^n(\guard_t)) \\
             & \Leftrightarrow (\auto(\valuation)) ((\auto^{-1}\circ\update_p\circ\auto)^n(\guard_t)) \\
             & \Leftrightarrow (\auto(\valuation)) ((\auto^{-1}\circ\update_p^n\circ\auto)(\guard_t)) \\
             & \Leftrightarrow (\auto(\valuation)) ((\auto^{-1}\circ\update_p^n\circ\auto\circ\auto^{-1})(\guard_p)) \\
             & \Leftrightarrow (\auto(\valuation)) ((\auto^{-1}\circ\update_p^n)(\guard_p)) \\
             & \Leftrightarrow \valuation ((\auto\circ\auto^{-1}\circ\update_p^n)(\guard_p)) \\
             & \Leftrightarrow \valuation (\update_p^n(\guard_p)).
          \end{align*}
          Thus, $\min\braced{n\in\NN\mid \valuation(\update_p^n(\neg\guard_p))} = \min\braced{n\in\NN\mid (\auto(\valuation))(\update_t^n(\neg\guard_t))}$ for all $\valuation \in \Sigma$.
          Hence, if we have a runtime bound $\run$ on the runtime complexity of $L_t$, i.e., $|\valuation|(\run) \geq \min\braced{n\in\NN\mid \valuation(\update_t^n(\neg \guard_t))}$ for all $\valuation : \VSet \to \auto(\ZZ^d)$, then applying $\auto$ to $\run$ yields a runtime bound for $L_p$.
          In other words, we have $\rc_{L_p}^\identity(\valuation) = \rc_{L_t}^\auto(\auto(\valuation)) = \min\braced{n\in\NN\mid (\auto(\valuation))(\update_t^n(\neg\guard_t))} \leq |\auto(\valuation)|(\run) = |\valuation(\auto(\run))| \leq |\valuation|(||\auto(\run)||)$ for all states $\valuation\in\Valuation$.
          Hence, $||\auto(\run)||$ is a runtime bound for $L_p$.

          Similar as in \cite[Lemma 18]{hark2020PolynomialLoopsTermination}, we now show that $\rc_L(\valuation) \leq p \cdot \rc_{L_p}(\valuation) +p -1$ holds for all $\valuation\in\Valuation$, where $\rc_L$ and $\rc_{L_p}$ denote the runtime complexities of $L$ and $L_p$, respectively.
          \[
            \begin{array}{ll}
                                  & \rc_{L_p}(\valuation) = \min\braced{n\in\NN\mid \valuation(\update_p^n( \neg\guard_p))}                                                               \\
              \Longleftrightarrow & \forall n\!<\!\rc_{L_p}(\valuation). \; \valuation(\update_p^n(\guard_p) \land \neg \update_p^{\rc_{L_p}(\valuation)}(\guard_p))                      \\
              \Longleftrightarrow & \forall n\!<\!\rc_{L_p}(\valuation). \; \valuation(\update^{p\cdot n}(\guard) \land \ldots \land \update^{p\cdot n + p -1}(\guard) \land              \\
                                  & \hspace*{2cm} \neg ( \update^{p\cdot \rc_{L_p}(\valuation)}(\guard_p) \land \ldots \land \update^{p\cdot \rc_{L_p}(\valuation) + p -1}(\guard_p) ) )  \\
              \Longleftrightarrow & \forall n\!<\!\rc_{L_p}(\valuation). \; \valuation(\update^{p\cdot n}(\guard) \land \ldots \land \update^{p\cdot n + p -1}(\guard) \land              \\
                                  & \hspace*{2cm}
              ( \neg \update^{p\cdot \rc_{L_p}(\valuation)}(\guard_p) \lor \ldots \lor \neg \update^{p\cdot \rc_{L_p}(\valuation) + p -1}(\guard_p) ) )                                   \\
              \Longleftrightarrow & \forall
              n\!<\!p \cdot \rc_{L_p}(\valuation). \; \valuation(\update^{n}(\guard) \land                                                                                                \\
                                  & \hspace*{2.5cm} ( \neg \update^{p\cdot \rc_{L_p}(\valuation)}(\guard_p) \lor ... \lor \neg \update^{p\cdot \rc_{L_p}(\valuation) + p -1}(\guard_p) )) \\
              \Longleftrightarrow & p \cdot \rc_{L_p}(\valuation) \; \leq \; \rc_{L}(\valuation) \; \leq \; p\cdot \rc_{L_p}(\valuation) + p -1
            \end{array}
          \]
          Thus, the runtime bound $||\auto(\run)||$ for $L_p$ yields the runtime bound $p\cdot||\auto(\run)|| + p -1$ for $L$.
          \qed
  \end{enumerate}
\end{myproof}

\begin{example}
  \label{runtimeBoundWhileExample}
  For the loop $L$ from \eqref{WhileExample}, we computed $L \chain L = L_p$ for $p=2$ in \eqref{WhileExampleChained}, see \Cref{exa:chaining}.
  In practice, we omit variables that obviously do not influence the runtime complexity.\footnote{To determine these variables, we perform a fixpoint iteration.
    All variables in the loop guard may influence the runtime.
    In the next step, we add all variables which occur in an update of these variables.
    We repeat this step until we reach a fixpoint.
    The remaining variables do not influence the runtime complexity and can be removed.}
  For example, neither $x_4$ nor $x_5$ influence the runtime complexity of $L_p$.
  If we omit these variables, then $L_p$ already is a tnn-loop and we can use the technique of \Cref{sect:Computing Runtime Bounds for TNN-Loops}
  (see \Cref{exa:tnn_final}) to obtain the runtime bound $\run = \tfrac{3}{2} + \tfrac{3}{2}\cdot \log_{2}(2\cdot x_1 + x_3^2 + 2\cdot x_3^5)$ for $L_p$.
  As this\linebreak
  simplified version of $L_p$ already is a twn-loop and even a tnn-loop, the automorphism $\auto$ is just the identity.
  Hence, \Cref{lem:correctness_chaining} yields the runtime bound $p\cdot ||\auto||(\run) + p - 1 = 4 + 3\cdot\log_{2}(2\cdot x_1 + x_3^2 + 2\cdot x_3^5)$ for the original loop \eqref{WhileExample}.
\end{example}

\paragraph{Unsolvable Loops:}
Closed forms play an important role in the automatic analysis of loops.
For example, to infer complexity bounds in \Cref{sect:Computing Runtime Bounds for TNN-Loops}, we will insert closed forms of variables into the guards of loops to determine the truth values of the guards after $n$ iterations.
However, there are loops that do not permit closed forms which are in poly-exponential form, e.g., the update $x \gets x^2$ has the closed form $x\cdot 2^{(2^n)}$.
In this paragraph, we show that the approach to infer runtime bounds for \emph{prs}-loops may (sometimes) still be used even if a poly-exponential closed form cannot be found.

In \cite{amrollahi2022SolvingInvariantGeneration}, the authors study \emph{unsolvable loops}, i.e., loops which are not solvable,
and compute closed forms for polynomial combinations of the loop variables in a sophisticated way which are then used to infer invariants.

Instead, we now introduce a preprocessing step to transform certain unsolvable loops into solvable loops such that the runtime complexity is preserved.
This allows us to analyze the obtained solvable loops and use these results to infer runtime bounds for the original unsolvable loops.

More precisely, the transformation (formally defined below in \Cref{def:transform_unsolvable}) replaces all occurrences of some polynomial $q\in\ZZ[\VSet]$ in the guard $\guard$ by a fresh variable $x$, denoted by $\guard[q/x]$.
In practice, we try to choose $q$ such that all variables that destroy the ``solvability'' of the loop (we discuss this in more detail after \Cref{def:transform_unsolvable}) are removed from the guard.
If these variables do not influence the runtime complexity anymore, then we can simply remove them from the loop.
Of course, this is not always possible, but if it succeeds, then this leads to solvable loops which can be analyzed instead of the original unsolvable loops.
To ensure that the fresh variable $x$ always has the value of $q$, we extend the update accordingly such that $x$ is updated to $\update(q) \, [q/x]$.
Thus, $x$ is updated to $q$ where first every variable $v$ in $q$ is replaced by its update $\update(v)$ and afterwards every occurrence of $q$ is replaced by $x$.
\begin{definition}[Transforming Unsolvable Loops]
  \label{def:transform_unsolvable}
  Let $L = \IntLoop$ be a loop and $q\in\ZZ[\VSet]$ a polynomial.
  Let $x\not\in\VSet$ be a fresh variable and $\update_x: \VSet\cup\braced{x}\to\ZZ[\VSet\cup\braced{x}]$ where $\update_x(v) = \update(v)$ for all $v\in\VSet$ and $\update_x(x) = \update(q) \, [q/x]$.
  Then we define the loop $L_x = (\guard[q/x],\update_x)$.
\end{definition}
In \cite{amrollahi2022SolvingInvariantGeneration}, unsolvable loops are characterized by the occurrence of \emph{defective variables}.
Essentially, a defective variable is a variable that depends (transitively) \emph{non-linearly} on itself or on another defective variable.
As indicated above, we try to choose $q$ such that all defective variables are eliminated from the guard.

\begin{example}
  \label{exa:unsolvable}
  Consider the following unsolvable loop (which is not solvable as $y_1$ depends non-linearly on itself):
  \begin{equation}
    \label{unsolvable}
    \WhileLoop{x_2^2 - x_3^5 < 2\cdot y_1 - y_2 \land x_2 \neq 0}{\!\!\!\myvec{y_1 \\ y_2 \\ x_2 \\ x_3 \\ x_4 \\ x_5} \leftarrow \myvec{y_1 + y_1^2 + x_3^2\vspace{0.075cm} \\ - 4\cdot y_1 + 2\cdot y_1^2 + 3 \cdot y_2 + x_3^2 \vspace{0.075cm} \\ -2\cdot x_2 \\ x_3 \\ \;\;\,3\cdot x_4 + 2 \cdot x_5 \\ -5\cdot x_4 - 3\cdot x_5}}
  \end{equation}
  We transform this unsolvable loop \eqref{unsolvable} into the solvable loop \eqref{WhileExample} by applying \Cref{def:transform_unsolvable}.
  The defective variables of \eqref{unsolvable} are $y_1$ and $y_2$.
  We now show how to choose $q$ such that both $y_1$ and $y_2$ can be removed.
  These variables only occur in the polynomial $2\cdot y_1 - y_2$ in the guard $x_2^2 - x_3^5 < 2\cdot y_1 - y_2 \land x_2 \neq 0$.
  Thus, by choosing $q$ to be the polynomial $2\cdot y_1 - y_2$, \Cref{def:transform_unsolvable} allows us to replace all defective variables by a fresh variable $x_1$.
  \Cref{def:transform_unsolvable} yields the following update for $x_1$:
  \begin{align*}
         & \update(2\cdot y_1 - y_2) \; [(2\cdot y_1 - y_2) / x_1] \\
    = \; & (2\cdot \update(y_1) - \update(y_2)) \;[(2\cdot y_1 - y_2) / x_1] \\
    = \; & (2\cdot(y_1 + y_1^2 + x_3^2) - (- 4\cdot y_1 + 2\cdot y_1^2 + 3 \cdot y_2 +
    x_3^2)) \; [(2\cdot y_1 - y_2) / x_1] \\
    = \; & (6\cdot y_1 - 3\cdot y_2 + x_3^2) \; [(2\cdot y_1 - y_2) / x_1] \\
    = \; & 3\cdot x_1 + x_3^2
  \end{align*}
  Furthermore, the polynomial $2\cdot y_1 - y_2$ is replaced by $x_1$ in the guard $x_2^2 - x_3^5 < 2\cdot y_1 - y_2 \land x_2 \neq 0$.
  This results in the new guard $x_2^2 - x_3^5 < x_1 \land x_2 \neq 0$.
  The variables $y_1$ and $y_2$ do no longer have any influence on the runtime complexity of this new loop.
  As mentioned, this transformation of course does not always yield solvable loops, since in general, defective variables might still occur in the new update.
  However, in the example we can just remove the defective variables $y_1$ and $y_2$, and obtain the solvable loop \eqref{WhileExample}.
  In the future, it would be interesting to improve this approach.
  For example, it might be possible to over-approximate certain variables in such a way that the runtime complexity is not decreased, thereby enabling the transformation of more unsolvable loops into solvable loops.
\end{example}
The following theorem shows the soundness of \Cref{def:transform_unsolvable}.
Let $L_x$ result from applying \Cref{def:transform_unsolvable} on the loop $L$ and the polynomial $q\in\ZZ[\VSet]$.
Then by \Cref{lem:unsolvable}, both loops $L$ and $L_x$ have the same runtime complexity, i.e., $\rc_L(\valuation) = \rc_{L_x}(\valuation_x)$ for every $\valuation\in\Valuation$ where $\valuation_x$ is the extension of $\valuation$ by $x$ such that $\valuation_x(x) = \valuation(q)$.
Furthermore, if $\run$ is a runtime bound for $L_x$, then $\run\, [\,x/||q||\,]$ is a runtime bound for $L$.

\begin{restatable}[Soundness of \Cref{def:transform_unsolvable}]{theorem}{unsolvable}
  \label{lem:unsolvable}
  Let $L = \IntLoop$ be a loop and $L_x = (\guard[q/x],\update_x)$ be the result of \Cref{def:transform_unsolvable} for $L$ and $q\in\ZZ[\VSet]$.
  Then for all $\valuation \in \Valuation$, we have $\rc_L(\valuation) = \rc_{L_x}(\valuation_x)$ where $\valuation_x(v) = \valuation(v)$ for all $v\in\VSet$ and $\valuation_x(x) = \valuation(q)$.
  If $\run$ is a runtime bound for $L_x$, then $\run \, [\, x/||q||\, ]$ is a runtime bound for $L$.
\end{restatable}
\makeproof{lem:unsolvable}{
  \unsolvable*
  \begin{myproof}
    Let $\valuation\in\Valuation$ and $\valuation_x$ be as in \Cref{lem:unsolvable}.
    Let $\update^n(\valuation)$ and $\update_x^n(\valuation_x)$ denote the state which results from $n$ applications of $\update$ resp.
    $\update_x$ starting in $\valuation$ resp.
    $\valuation_x$.
    We show by induction that $$\left(\update^n(\valuation)\right)(q) = \left(\update_x^n(\valuation_x)\right)(x)$$ holds for all $n\in\NN$.
    \paragraphProof{Induction Base:}
    For $n = 0$ we have $\left(\update^0(\valuation)\right)(q) = \valuation(q) = \valuation_x(x) = \left(\update_x^0(\valuation_x)\right)(x)$.
    \paragraphProof{Induction Step:}
    For $n > 0$ we have
    \begin{align*}
       & \left(\update^{n + 1}(\valuation)\right)(q) \\
       & = \left(\update^n(\valuation)\right)(\update(q)) \\
       & = \update(q) \; \underbrace{[v / \left(\update^n(\valuation)\right)(v) \mid v\in\VSet]}_{\text{evaluating at $\update^n(\valuation)$}} \\
       & =\update(q) \;
      \underbrace{[q/\left(\update^n(\valuation)\right)(q)]\;[v / \left(\update^n(\valuation)\right)(v) \mid v\in\VSet]}_{\text{evaluate $q$ separately beforehand}} \\
       & =\update(q) \;
      [q/\left(\update^n(\valuation)\right)(q)]
      \;[v / \left(\update_x^n(\valuation_x)\right)(v) \mid v\in\VSet] \tag{since $\left(\update_x^n(\valuation_x)\right)(v) = \left(\update^n(\valuation)\right)(v)$ for all $v\in\VSet$} \\
       & \stackrel{\text{IH}}{=}
      \update(q) \;
      \; [q/\left(\update_x^n(\valuation_x)\right)(x)]
      \; [v
      / \left(\update_x^n(\valuation_x)\right)(v) \mid v\in\VSet] \tag{by the induction
        hypothesis $\left(\update^n(\valuation)\right)(q)
      = \left(\update_x^n(\valuation_x)\right)(x)$} \\
       & =\update(q) \; [q/x] \;
      [v
      / \left(\update_x^n(\valuation_x)\right)(v) \mid v\in\VSet \cup \{x\}] \\
       & = \left(\update_x^n(\valuation_x)\right)
      \; \left( \update(q) \; [q/x] \right) \\
       & = \left(\update_x^n(\valuation_x)\right) \; \left(\update_x(x)\right) \tag{definition of $\update_x(x)$} \\
       & = \left(\update_x^{n + 1}(\valuation_x)\right) \, (x)
    \end{align*}

    Hence, we have $\valuation(\update^n(\neg\guard)) = \valuation_x(\update_x^n(\neg\guard[q/x]))$ for all $n\in\NN$.
    This implies $\rc_L(\valuation) = \min\braced{n\in\NN\mid \valuation(\update^n( \neg\guard))} = \min\braced{n\in\NN\mid \valuation_x(\update_x^n( \neg\guard[q/x]))} = \rc_{L_x}(\valuation_x)$ which proves the first part of \Cref{lem:unsolvable}.
    If $\run$ is a runtime bound for $L_x$, then we have $$\rc_L(\valuation) = \rc_{L_x}(\valuation_x) \leq \abs{\valuation_x}(\run) \leq \abs{\valuation}(\run[x/||q||])$$ as bounds are weakly monotonic and $\abs{\valuation_x}(x) \leq \abs{\valuation}(\,||q||\,)$.
    \qed
  \end{myproof}
}
\begin{example}
  Reconsider \Cref{exa:unsolvable} where the transformation of \Cref{def:transform_unsolvable} yields the prs-loop \eqref{WhileExample}.
  We have already seen that $4 + 3\cdot\log_{2}(2\cdot x_1 + x_3^2 + 2\cdot x_3^5)$ is a runtime bound for the loop \eqref{WhileExample}.
  In \Cref{exa:unsolvable}, we have chosen the polynomial $q = 2\cdot y_1 - y_2$, i.e., $||q|| = 2\cdot y_1 + y_2$.
  Hence, $(4 + 3\cdot\log_{2}(2\cdot x_1 + x_3^2 + 2\cdot x_3^5))\; [\, x_5/||q|| \, ] = 4 + 3\cdot\log_{2}(x_3^2 + 2\cdot x_3^5 + 4\cdot y_1 + 2\cdot y_2)$ is a runtime bound for \Cref{exa:unsolvable} by \Cref{lem:unsolvable}.
\end{example}

\subsubsection{Computing Runtime Bounds for tnn-Loops on $\auto(\ZZ^d)$}
\label{sect:Computing Runtime Bounds for TNN-Loops}
Now we show how to compute runtime bounds for terminating \emph{tnn}-loops on $\auto(\ZZ^d)$ for a $\QQ$-automorphism $\auto: \VSet \to \QQ[\VSet]$, where the update has the form $\update(x_i) = c_i \cdot x_i + p_i$ with $c_i\in\NN$ and $p_i$ is a polynomial with rational coefficients.
As observed in \cite{hark2020PolynomialLoopsTermination,frohn2020TerminationPolynomialLoops}, since the closed forms for \emph{tnn}-loops are poly-exponential expressions from $\PPEE(\QQ,\NN)$ that are weakly monotonically increasing in $n$, every \emph{tnn}-loop $\IntLoop$ \emph{stabilizes} for each input $\vec{e} \in \auto(\ZZ^d)$.
So there is a number of loop iterations (a \emph{stabilization threshold} $\sth^\auto_{\IntLoop}(\vec{e})$), such that the truth value of the loop guard $\guard$ does not change anymore when performing further loop iterations.
Hence, the runtime of every terminating \emph{tnn}-loop is bounded by its stabilization threshold.
In the following, we first define the stabilization threshold for poly-exponential expressions from $\PPEE(\ZZ,\NN)$.
Afterwards, we use these results for those poly-exponential expressions which result from instantiating all variables in the loop guard by their closed forms.
Note that we only consider $\PPEE(\ZZ,\NN)$ instead of $\PPEE(\QQ,\NN)$ (since \Cref{lem:complexity,lem:complexity_logarithmic} will require poly-exponential expressions with integer coefficients).
However, this is not a restriction in our setting, because we can always multiply all atoms in the loop guard with the least common multiple of all denominators occurring in the resulting poly-exponential expressions.
\begin{definition}[Stabilization Threshold \cite{hark2020PolynomialLoopsTermination}]
  \label{def:Stabilization Threshold}
  Let $pe\in\PPEE(\ZZ,\NN)$ be a poly-exponential expression.
  Then the \emph{stabilization threshold}
  $\sth^\auto_{pe}: \auto(\Valuation) \to \NN$ of $pe$ is the smallest number such that $\sign(\valuation(pe)) = \sign(\valuation(pe[n / \sth^\auto_{pe}(\valuation)]))$ holds for all $n \geq \sth^\auto_{pe}(\valuation)$ and all $\valuation\in\auto(\Valuation)$.
\end{definition}
\begin{example}
  In the examples of this subsection, we always consider the identity automorphism $\auto(v) = v$ and the following poly-exponential expression:
  \begin{equation}
    pe \; = \; (-2\cdot x_2^2) \cdot 16^n + (2\cdot x_1 + x_3^2) \cdot 9^n - x_3^2 +
    2\cdot x_3^5 \; \in\PPEE(\ZZ,\NN) \label{exa:pe}
  \end{equation}
  For example, the stabilization threshold $\sth^\auto_{pe}(\valuation)$ of the state $\valuation(x_1) = 10, \valuation(x_2)\linebreak
    = 1$, and $\valuation(x_3) = 0$ is $5$ as $\valuation(pe) > 0$ for $n\in\braced{0,\dots,4}$ and $\valuation(pe) < 0$ for all $n > 4$.
  In the first part of this subsection, we will show how to compute a bound on the stabilization threshold of poly-exponential expressions like $pe$.
\end{example}

As mentioned, we always assume that $pe = \sum_{j=1}^\ell p_j \cdot n^{a_j} \cdot b_j^n\in\PPEE(\ZZ,\NN)$.
This allows us to order the summands of $pe$ according to their growth rate w.r.t.\ $n$.
Here, a summand with $(b_1,a_1)$ has a higher growth rate than a summand with $(b_2,a_2)$ iff the pair $(b_1,a_1)$ is lexicographically greater than $(b_2,a_2)$.
To compute upper bounds on the stabilization threshold of poly-exponential expressions like \eqref{exa:pe}, we now present a construction based on \emph{monotonicity thresholds}, which are computable \cite[Lemma 12]{hark2020PolynomialLoopsTermination}.
The 1-monotonicity threshold of two pairs of numbers $(b_1,a_1)$ and $(b_2,a_2)$ is the smallest $n_0\in\NN$ such that $n^{\ud}\cdot \ub^n$ is greater than $n^{\ld} \cdot \lb^n$ for all $n \geq n_0$.
Analogously, the $k$-monotonicity threshold can also be defined for arbitrary $k \in \NN_{\geq 1}$ by comparing $n^{\ud}\cdot \ub^n$ and $k\cdot n^{\ld} \cdot \lb^n$.
Hence, monotonicity thresholds characterize for which values of $n$ a summand of $pe$ dominates another summand.
\begin{definition}[Monotonicity Threshold \cite{hark2020PolynomialLoopsTermination}]
  \label{def:monotonicity_threshold}
  Let\footnote{In contrast to our paper \cite{lommen2022AutomaticComplexityAnalysis}, we consider $\RR_{\geq 0}$ instead of $\NN$ as the construction of logarithmic bounds in \Cref{lem:complexity_logarithmic} uses monotonicity thresholds with $\ub,\lb\in\RR_{\geq 0}$.} $(\ub,\ud), (\lb,\ld)\in\RR_{\geq 0}^2$ such that $(\ub,\ud)\lex(\lb,\ld)$ (i.e., $\ub > \lb$ or both $\ub = \lb$ and $\ud > \ld$).
  For any $k\in\NN_{\geq 1}$, the $k$-\emph{monotonicity threshold} of $(\ub,\ud)$ and $(\lb,\ld)$ is the smallest $n_0\in\NN$ such that for all $n\geq n_0$ we have $n^{\ud}\cdot \ub^n > k\cdot n^{\ld} \cdot \lb^n$.
\end{definition}
For example, the $1$-monotonicity threshold of $(4,0)$ and $(3,1)$ is $7$ as the largest root of $f(n) = 4^n - n \cdot 3^n$ is approximately $6.5139$.

To obtain bounds that are weakly monotonically increasing, the corresponding technique of \cite[Lemma 21]{hark2020PolynomialLoopsTermination}
over-approximated the polynomials $p_j$ in the poly-exponential expressions $\sum_{j=1}^\ell p_j \cdot n^{a_j} \cdot b_j^n$ by a polynomial that did not distinguish the effects of the different variables $x_1,\ldots,x_{\indv}$.
Such an over-approximation is only useful for a direct asymptotic bound on the runtime of the \emph{tnn}-loop, but it is too coarse for a useful modular runtime bound within the complexity analysis of a larger program, e.g., integer programs like the one that we will regard in \Cref{fig:ITS} in \Cref{sect:global_integer_programs}.
Here, $x_3$ is set to the constant value $2$.
Therefore, it is crucial to obtain a local runtime bound like $\max \{1, 4\cdot x_1 + 2\cdot x_3^2 + 4\cdot x_3^5\}$ for the loop as a subprogram, which indicates that only the variable $x_3$ may influence the runtime with an exponent of $2$ or $5$.
Hence, this might yield a substantially smaller global runtime bound for the whole program in the end if the constant value of $x_3$ is considered, see \Cref{exa:SBLifting}.

Thus, we improve the existing results on complexity analysis of \emph{twn}-loops \cite{hark2020PolynomialLoopsTermination,frohn2020TerminationPolynomialLoops,frohn2019TerminationTriangularInteger} such that they yield concrete polynomial bounds.
To this end, we over-approximate the polynomials $p_j$ in the closed form by the polynomial $\sqcup \{ p_1, \ldots, p_{\ell}\}$ which contains every normalized monomial $x_1^{e_1} \cdot \ldots \cdot x_\indv^{e_\indv}$ of $\{p_1,\ldots, p_{\ell}\}$, using the absolute value of the largest coefficient with which the monomial occurs in $\{p_1, \ldots, p_{\ell}\}$.\footnote{In contrast, \cite{hark2020PolynomialLoopsTermination} used the coarser over-approximation $|k_\text{max}| \cdot \sum_{i = 0}^{d_\text{max}}(x_1 + \dots + x_d)^i$ as a bound where $k_\text{max}$ is the coefficient with the largest absolute value and $d_\text{max}$ is the maximum of the degrees of all $p_j$.}
Thus, $\sqcup \{ p_1, \ldots, p_{\ell}\}$ is always a weakly monotonically increasing bound from $\BoundSet$.
For example, $\sqcup \{ 2\cdot x_1 + x_3^2, - x_3^2 + 2\cdot x_3^5\} = 2\cdot x_1 + x_3^2 + 2\cdot x_3^5$.
In the following, for $\vec{e} = (e_1,\ldots,e_d) \in \NN^d$, let $\vec{x}^{\vec{e}}$ denote $x_1^{e_1} \cdot \ldots \cdot x_\indv^{e_\indv}$.

\begin{definition}[Over-Approximation of Polynomials]
  \label{def:overappr_poly}
  Let $p_1, \ldots, p_\ell \in \ZZ[\vec{x}]$, and for all $1 \leq j \leq \ell$, let $\mathcal{I}_j \subseteq (\ZZ \setminus \{0\}) \times \NN^\indv$ be the \emph{index set} of the polynomial $p_j$ where $p_j = \sum_{(c,\vec{e})\in \mathcal{I}_j} c \cdot\vec{x}^{\vec{e}}$ and there are no $c \neq c'$ with $(c,\vec{e}), (c',\vec{e})\in \mathcal{I}_j$.
  For all $\vec{e} \in \NN^d$ we define $c_{\vec{e}} \in \NN$ with $c_{\vec{e}}	= \max \{ |c| \mid (c,\vec{e}) \in \mathcal{I}_1 \cup \ldots \cup \mathcal{I}_\ell \}$, where $\max \emptyset = 0$.
  Then the \emph{over-approximation} of $p_1, \ldots, p_\ell$ is $\sqcup \{p_1, \ldots, p_{\ell}\} = \sum_{\vec{e} \in \NN^d} c_{\vec{e}} \cdot \vec{x}^{\vec{e}}$.
\end{definition}
Clearly, $\sqcup \{ p_1, \ldots, p_{\ell}\}$ indeed over-approximates the absolute value of each $p_j$.\footnote{The degree of the polynomial $\sqcup \{ p_1, \ldots, p_{\ell}\}$ can be reduced further by taking invariants into account, see \cite{lommen2022AutomaticComplexityAnalysis}.
  We omitted this refinement here for the sake of readability.}

\begin{corollary}[Soundness of $\sqcup \{ p_1, \ldots, p_{\ell}\}$]
  \label{cor:SoundnessOverAppr}
  For all $\valuation\in\auto(\Valuation)$ and all $1 \leq j \leq \ell$, we have $|\valuation|(\sqcup \{ p_1, \ldots, p_{\ell}\}) \geq |\valuation(p_j)|$.
\end{corollary}

The following lemma yields a polynomial upper bound on the stabilization threshold of a poly-exponential expression.
The main idea is to consider an ordered poly-exponential expression $pe = \sum_{j=1}^{\ell} p_j \cdot n^{a_j} \cdot b_j^n$ with $(b_\ell,a_\ell) \lex \ldots\lex (b_1,a_1)$.
For a state $\valuation \in \auto(\Valuation)$, we consider the largest summand $p_{j'} \cdot n^{a_{j'}} \cdot b_{j'}^n$ such that $\valuation(p_{j'}) \neq 0$.
In \Cref{lem:complexity}, we compute a bound for $n$ after which $n^{a_{j'}} \cdot b_{j'}^n$ is always larger than $\abs{\sum_{j=1}^{j' - 1} \valuation(p_j) \cdot n^{a_j} \cdot b_j^n}$.
For all larger values of $n$, $p_{j'} \cdot n^{a_{j'}} \cdot b_{j'}^n$ dominates the expression $pe$, i.e., $\sign(\valuation(p_{j'}))$ determines the sign of the whole expression.
The construction of this bound is based on monotonicity thresholds as in \Cref{def:monotonicity_threshold}.
Note that \Cref{lem:complexity} (as well as \Cref{lem:complexity_logarithmic}) requires that $\valuation(p_j) \in \ZZ$ for all $\valuation\in\auto(\Valuation)$.
We will discuss how to ensure this when explaining how to obtain poly-exponential expressions $pe_\alpha$ for the atoms $\alpha$ in loop guards and how to obtain stabilization thresholds for loops (see \Cref{def:Stabilization Threshold for Loops}).

\begin{restatable}[Polynomial Bound on Stabilization Threshold]{lemma}{boundSTH}
  \label{lem:complexity}
  Let $pe = \sum_{j=1}^{\ell} p_j \cdot n^{a_j} \cdot b_j^n\in\PPEE(\ZZ,\NN)$ be a poly-exponential expression with $p_j \neq 0$ for all $1 \leq j \leq \ell$ and $(b_\ell,a_\ell) \lex \ldots\lex (b_1,a_1)$ such that $\valuation(p_j) \in \ZZ$ for all $\valuation\in\auto(\Valuation)$.

  Let $C = \max\{1, M_2, N_2, \ldots, M_\ell, N_\ell \}$, where we have:
  \[
    \mbox{\small $M_j=\left\{
        \begin{array}{ll}
          0,                                                     & \text{if $b_j = b_{j-1}$} \\
          \text{1-monotonicity threshold of}                     &                           \\
          \text{$\;\;\;(b_j,a_j)$ and $(b_{j-1}, a_{j-1} + 1)$}, & \text{if $b_j > b_{j-1}$}
        \end{array}
        \right. \quad\;\; N_j=\left\{
        \begin{array}{ll}
          1,              & \text{if $j = 2$} \\
          mt',            & \text{if $j = 3$} \\
          \max\{mt,mt'\}, & \text{if $j > 3$}
        \end{array}
        \right.$}
  \]
  Here, $mt'$ is the $(j-2)$-monotonicity threshold of $(b_{j-1},a_{j-1})$ and $(b_{j-2},a_{j-2})$.
  Moreover, $mt= \max\{1\text{-monotonicity threshold of } (b_{j-2},a_{j-2}) \text{ and } (b_i,a_i) \mid 1 \leq i \leq j-3\}$.
  Let $Pol = \braced{p_1,\ldots, p_{\ell-1}}$.
  Then $\max\braced{C,2 \cdot \sqcup Pol}$ is a bound on the stabilization threshold of $pe$, i.e., $\max\{ C, 2 \cdot \abs{\valuation}(\sqcup Pol) \} \; \geq \; \sth^\auto_{pe}(\valuation)$ for all $\valuation \in \auto(\Valuation)$.
\end{restatable}
\begin{myproof}
  Let $\valuation \in \auto(\Valuation)$.
  We prove that
  \begin{equation}
    \max\{ C, 2 \cdot \abs{\valuation}(\sqcup Pol) \} \; \geq \; \sth^\auto_{pe}(\valuation)
  \end{equation}
  holds for the poly-exponential expression $pe = \sum_{j=1}^{\ell} p_j \cdot n^{a_j} \cdot b_j^n$.

  If $\valuation(p_j) = 0$ for all $1 \leq j \leq \ell$, then the claim is trivial as $\sth^\auto_{pe}(\valuation) = 0$.
  The reason is that $\sign(\valuation(pe)) = 0$ for all $n \geq 0$.

  Otherwise, there exists a maximal index $1 \leq j \leq \ell$ where $\valuation(p_j) \neq 0$.
  If $j = 1$, then for every $n \geq 1$, the sign of $\valuation(pe) = \valuation(p_1) \cdot n^{a_1}\cdot b_1^n$ is $\sign(\valuation(p_1))$.
  Hence, $\sth^\auto_{pe}(\valuation) = 1$ for all $n \geq 1$.
  Clearly, we have $\max\{ C, 2 \cdot \abs{\valuation}(\sqcup Pol) \} \geq C \geq 1$.

  Otherwise we have $2 \leq j \leq \ell$.
  Then we obtain the following for all $n \geq N_j$:
  \begin{align}
            & \abs{\sum_{i=1}^{j-1} \valuation(p_i) \cdot n^{a_i} \cdot b_i^n} \nonumber \\
    \leq \; & \sum_{i=1}^{j-1}\abs{ \valuation(p_i)} \cdot n^{a_i} \cdot b_i^n\nonumber \\
    \leq \; & \sum_{i=1}^{j-1} |\valuation|(\sqcup \{p_1,\ldots,p_{j-1} \}) \cdot n^{a_i} \cdot b_i^n\tag{by \cref{cor:SoundnessOverAppr}}\nonumber \\
    = \;    & |\valuation|(\sqcup \{p_1,\ldots,p_{j-1} \})\cdot \Big(n^{a_{j-1}}\cdot b_{j-1}^n + \sum_{i=1}^{j-2} n^{a_i} \cdot b_i^n\Big)\nonumber \\
    \leq \; & |\valuation|(\sqcup \{p_1,\ldots,p_{j-1} \})\cdot \Big(n^{a_{j-1}}\cdot b_{j-1}^n + \sum_{i=1}^{j-2} n^{a_{j-2}}\cdot b_{j-2}^n\Big) \tag{as $n \geq N_j \geq mt$ for $j > 3$} \\
    = \;    & |\valuation|(\sqcup \{p_1,\ldots,p_{j-1} \})\cdot \Big(n^{a_{j-1}}\cdot b_{j-1}^n + (j-2)\cdot n^{a_{j-2}}\cdot b_{j-2}^n\Big)\nonumber \\
    \leq \; & 2 \cdot |\valuation|(\sqcup \{p_1,\ldots,p_{j-1} \})\cdot n^{a_{j-1}}\cdot b_{j-1}^n\tag{as $n \geq N_j \geq mt'$ for $j \geq 3$}
  \end{align}
  Note that if $|\valuation|(\sqcup \{p_1,\ldots,p_{j-1} \}) \neq 0$, then the last inequation is strict.

  Clearly, $(b_j, a_j) \lex (b_{j-1},a_{j-1})$ implies $b_j > b_{j-1}$ or both $b_j = b_{j-1}$ and $a_j \geq a_{j-1} + 1$.
  If $b_j = b_{j-1}$ and $a_j \geq a_{j-1} + 1$, we have that
  \begin{align*}
    2 \cdot |\valuation|(\sqcup \{p_1,\ldots,p_{j-1} \})\cdot n^{a_{j-1}}\cdot b_{j-1}^n & = 2 \cdot |\valuation|(\sqcup \{p_1,\ldots,p_{j-1} \})\cdot n^{a_{j-1}}\cdot b_{j}^n \\
                                                                                         & \leq n^{a_{j-1} + 1}\cdot b_{j}^n \\
                                                                                         & \leq n^{a_j}\cdot b_{j}^n
  \end{align*}
  holds for all $n \geq 2 \cdot |\valuation|(\sqcup \{p_1,\ldots,p_{j-1}\})$,
  where the second-to-last inequation is strict if $|\valuation|(\sqcup \{p_1,\ldots, p_{j-1} \}) = 0$ and $n \geq 1$.

  In the second case $b_j > b_{j-1}$, we can derive $$ 2 \cdot |\valuation|(\sqcup \{p_1,\ldots,p_{j-1} \})\cdot n^{a_{j-1}}\cdot b_{j-1}^n \leq n^{a_{j-1} + 1}\cdot b_{j-1}^n < n^{a_j}\cdot b_{j}^n$$ for all $n \geq \max\{M_j,2 \cdot |\valuation|(\sqcup \{p_1,\ldots,p_{j-1} \})\}$, as $M_j$ is the 1-monotonicity threshold of $(b_j,a_j)$ and $(b_{j-1}, a_{j-1} + 1)$.
  Thus, in total we have shown that for all $n \geq \max\{N_j,M_j,2\cdot|\valuation|(\sqcup \{p_1,\ldots,p_{j-1} \})\}$, we have
  \begin{equation}
    \abs{\sum_{i=1}^{j-1} \valuation(p_i) \cdot n^{a_i} \cdot b_i^n} < n^{a_j}\cdot b_j^n.\label{proof:eqSTH}
  \end{equation}

  Now we prove that either always $\valuation(pe) < 0$ (if $\valuation(p_j) < 0$) or $\valuation(pe) > 0$ (if $\valuation(p_j) > 0$) hold for all $n \geq \max\{M_j,N_j,2\cdot|\valuation|(\sqcup\{p_1,\ldots,p_{j-1} \})\}$.
  In the following, we prove the case $\valuation(p_j) < 0$.
  \begin{align*}
            & \valuation(pe) \\
    = \;    & \valuation(p_j)\cdot n^{a_j}\cdot b_j^n + \sum_{i=1}^{j - 1} \valuation(p_i) \cdot n^{a_i} \cdot b_i^n \nonumber \\
    \leq \; & \valuation(p_j)\cdot n^{a_j}\cdot b_j^n + \abs{\sum_{i=1}^{j-1} \valuation(p_i) \cdot n^{a_i} \cdot b_i^n}\tag{as $x + y \leq x + \abs{y}$ holds for all $x,y \in \ZZ$}\nonumber \\
    < \;    & \valuation(p_j)\cdot n^{a_j}\cdot b_j^n + n^{a_j}\cdot b_j^n\tag{by \eqref{proof:eqSTH}}\nonumber \\
    = \;    & (\valuation(p_j) + 1) \cdot n^{a_j}\cdot b_j^n \\
    \leq \; & 0 \tag{since $\valuation(p_j) \in \ZZ$ and thus $\valuation(p_j) < 0$ implies $\valuation(p_j) + 1 \leq 0$}
  \end{align*}
  The other case can be proven similarly by using $\forall x,y\in\ZZ.\, x + y \geq x - \abs{y}$ and $\valuation(p_j) - 1 \geq 0$.

  Note that we have $C \geq M_j$ and $C \geq N_j$ for all $2 \leq j \leq \ell$.
  Moreover, since $Pol \supseteq \{p_1,\ldots,p_{j-1}\}$, we have $|\valuation|(\sqcup Pol) \geq |\valuation|(\sqcup \{p_1,\ldots,p_{j-1}\})$.
  Hence, for all $n \geq \max\{ C, 2 \cdot |\valuation|( \sqcup Pol) \} \geq \max\{ M_j, N_j, 2 \cdot |\valuation|(\sqcup\{p_1,\ldots,p_{j-1} \}) \}$, we have always $\valuation(pe) < 0$ or always $\valuation(pe) > 0$.
  Thus, $\max\{ C, 2 \cdot |\valuation|( \sqcup Pol) \} \geq \sth^\auto_{pe}(\valuation)$.
  \qed
\end{myproof}

\begin{example}
  \label{exa:sthPoly}
  We want to apply \Cref{lem:complexity} on $(-2\cdot x_2^2) \cdot 16^n + (2\cdot x_1 + x_3^2) \cdot 9^n - x_3^2 + 2\cdot x_3^5$, see \eqref{exa:pe}.
  Here, we have $p_1 = - x_3^2 + 2\cdot x_3^5$, $p_2 = 2\cdot x_1 + x_3^2$, and $p_3 = -2\cdot x_2^2$.
  Furthermore, we have $b_1 = 1$, $b_2 = 9$, $b_3 = 16$, and $a_j = 0$ for $j\in\braced{1,\dots,3}$.
  We obtain
  \[
    \begin{array}{rcll}
      M_2                      & = & 0, & \text{as 0 is the 1-monotonicity threshold of $(9,0)$ and $(1,1)$}  \\
      M_3                      & = & 0, & \text{as 0 is the 1-monotonicity threshold of $(16,0)$ and $(9,1)$} \\
      N_2 = 1 \text{ and } N_3 & = & 1, & \text{as 1 is the 1-monotonicity threshold of $(9,0)$ and $(1,0)$}. \\
    \end{array}
  \]
  Hence, we get $C = \max\{1, M_2, N_2, M_3, N_3\} = 1$.
  So by \Cref{lem:complexity}, $$\max\braced{C,2\cdot\sqcup\braced{- x_3^2 + 2\cdot x_3^5, 2\cdot x_1 + x_3^2}} = \max\braced{1,4\cdot x_1 + 2\cdot x_3^2 + 4\cdot x_3^5}$$ is a bound on the stabilization threshold of \eqref{exa:pe}.
\end{example}

Finally, we present an improved version of \Cref{lem:complexity} for logarithmic bounds.
If the addends of the poly-exponential expression satisfy $b_\ell >\!\ldots\!> b_1$ instead of just $(b_\ell,a_\ell) \lex \ldots\lex (b_1,a_1)$, then \Cref{lem:complexity_logarithmic} yields logarithmic (sub-linear) bounds instead of polynomial bounds.
The reason is that then the summand $p_j \cdot n^{a_j} \cdot b_j^n$ grows exponentially faster than all summands $p_i \cdot n^{a_i} \cdot b_i^n$ for $i < j$.

\begin{restatable}[Logarithmic Bound on Stabilization Threshold]{lemma}{boundSTHLog}
  \label{lem:complexity_logarithmic}
  Let $pe = \sum_{j=1}^{\ell} p_j \cdot n^{a_j} \cdot b_j^n\in\PPEE(\ZZ,\NN)$ be a poly-exponential expression with $p_j \neq 0$ for all $1 \leq j \leq \ell$ and $b_\ell > \ldots > b_1$ such that $\valuation(p_j) \in \ZZ$ for all $\valuation\in\auto(\Valuation)$.
  Let $C = \max\{1, M_2', N_2, \ldots, M_\ell', N_\ell \}$, where $N_j$ is as in \Cref{thm:time-bound} and
  \[
    \mbox{$M_j'$ is the 1-monotonicity threshold of $(b_{j - 1} + \varepsilon_j,a_j)$ and $(b_{j - 1}, a_{j-1})$}
  \]
  for $0 < \varepsilon_j < b_j - b_{j - 1}$.
  Let\footnote{Here, $\log_{\frac{b_j}{b_{j - 1}
        + \varepsilon_j}}(2\cdot \sqcup \braced{p_1,\ldots, p_{j-1}})$ is always a bound, where $\log_k(b_1)$ stands for $\log_k (\max(\braced{1,b_1}))$, see \Cref{def:bounds}.
  Hence, this logarithmic bound is also well defined and non-negative for
  bounds $2\cdot \sqcup \braced{p_1,\ldots, p_{j-1}}$ which evaluate to numbers from $[0,1)$.} \[Log = \max\braced{\log_{\frac{b_j}{b_{j - 1}
          + \varepsilon_j}}(2\cdot \sqcup \braced{p_1,\ldots, p_{j-1}})\mid 2 \leq
      j \leq \ell}.\]
  Then $\max\braced{C, Log}$ is a bound on the stabilization threshold of $pe$.
\end{restatable}
\begin{myproof}
  Let $\valuation \in \auto(\Valuation)$.
  We first prove that
  \begin{equation}
    \max\braced{C, |\sigma|(Log)} \; \geq \; \sth^\auto_{pe}(\valuation)
  \end{equation}
  for the poly-exponential expression $pe = \sum_{j=1}^{\ell} p_j \cdot n^{a_j} \cdot b_j^n$.

  The cases where $\sigma(p_j) = 0$ for all $1 \leq j \leq \ell$ or where $j=1$ is the only index with $\sigma(p_j) \neq 0$ are handled as in the proof of \Cref{lem:complexity}.

  Otherwise, there is a maximal index $2 \leq j \leq \ell$ with $\sigma(p_j) \neq 0$.
  In \Cref{lem:complexity}, we derived $$\abs{\sum_{i=1}^{j-1} \valuation(p_i) \cdot n^{a_i} \cdot b_i^n} \leq 2 \cdot |\valuation|(\sqcup \{p_1,\ldots,p_{j-1} \})\cdot n^{a_{j-1}}\cdot b_{j-1}^n$$ for $n \geq N_j$, where the inequation is strict if $|\valuation|(\sqcup \{p_1,\ldots,p_{j-1} \}) \neq 0$.

  By assumption, we have $b_\ell > \ldots > b_1$.
  Hence, for $0 < \varepsilon_j < b_j - b_{j - 1}$ we have
  \begin{align*}
     & 2 \cdot |\valuation|(\sqcup \{p_1,\ldots,p_{j-1} \})\cdot n^{a_{j-1}}\cdot b_{j-1}^n \\
     & \; \leq 2 \cdot |\valuation|(\sqcup \{p_1,\ldots,p_{j-1} \})\cdot n^{a_{j}}\cdot (b_{j-1} + \varepsilon_j)^n \tag{for $n \geq M_j'$} \\
     & \; \leq \max\braced{1, 2 \cdot |\valuation|(\sqcup \{p_1,\ldots,p_{j-1} \})}\cdot n^{a_{j}}\cdot (b_{j-1} + \varepsilon_j)^n \\
     & \; = \left(\frac{b_j}{b_{j - 1} + \varepsilon_j}\right)^{\log_{\frac{b_j}{b_{j - 1} + \varepsilon_j}} \left(\max\braced{1, 2 \cdot |\valuation|(\sqcup \{p_1,\ldots,p_{j-1} \}})\right)}\cdot n^{a_{j}}\cdot (b_{j-1} + \varepsilon_j)^n \\
     & \; \leq \left(\frac{b_j}{b_{j - 1} + \varepsilon_j}\right)^n\cdot n^{a_j}\cdot (b_{j-1} + \varepsilon_j)^n \tag{for $n \geq \log_{\frac{b_j}{b_{j - 1} + \varepsilon_j}} \left(\max\braced{1,2 \cdot |\valuation|(\sqcup \{p_1,\ldots,p_{j-1} \}})\right)$} \\
     & \; = n^{a_j} \cdot b_j^n
  \end{align*}
  for $n \geq \max\braced{M_j',\log_{\frac{b_j}{b_{j - 1}+\varepsilon_j}} (\max\braced{1,2 \cdot |\valuation|(\sqcup \{p_1,\ldots,p_{j-1} \})})}$.
  Here, the first inequation is strict if $|\valuation|(\sqcup \{p_1,\ldots,p_{j-1} \}) \neq 0$.

  From this point on, we consider the bound $\log_k(b)$ instead of the application of the logarithm function, i.e., we omit the expression $\max\braced{1,b}$ in the argument of the logarithm in the following.
  In total we have shown that for all \[n \geq \max\braced{M_j',N_j,\log_{\frac{b_j}{b_{j - 1}+\varepsilon_j}} (2 \cdot |\valuation|(\sqcup \{p_1,\ldots,p_{j-1} \}))},\]
  we have
  \[
    \abs{\sum_{i=1}^{j-1} \valuation(p_i) \cdot n^{a_i} \cdot b_i^n} < n^{a_j}\cdot b_j^n.
  \]

  Now, we can proceed as in the proof of \Cref{lem:complexity} and show that either always $\valuation(pe) < 0$ or always $\valuation(pe) > 0$ holds for all \[n \geq \max\{M_j',N_j,\log_{\frac{b_j}{b_{j - 1}+\varepsilon_j}} (2 \cdot |\valuation|(\sqcup \{p_1,\ldots,p_{j-1} \}))\}.\]

  Note that we have $C \geq M_j'$ and $C \geq N_j$ for all $2 \leq j \leq \ell$.
  Moreover, for all $2 \leq j \leq \ell$, we have $\log_{\frac{b_j}{b_{j - 1}+\varepsilon_j}} (2 \cdot |\valuation|(\sqcup \{p_1,\ldots,p_{j-1} \})) \leq |\valuation|(Log)$.
  Hence, we have $\max\braced{C, |\valuation|(Log)} \geq \sth^\auto_{pe}(\valuation)$.
  \qed
\end{myproof}

Note that in \Cref{lem:complexity_logarithmic}, $\varepsilon_j$ can be chosen arbitrarily.
However, the choice influences the resulting bound:
If $\varepsilon_j$ is higher, i.e., closer to $b_j - b_{j - 1}$, then $$\log_{\frac{b_j}{b_{j - 1} + \varepsilon_j}}(\sqcup \braced{p_1,\ldots, p_{j-1}})$$ yields higher values.
On the other hand, if $\varepsilon_j$ is closer to $0$, then the value of $M_j'$ is increasing.
However, the choice of $\varepsilon_j$ (resp.\ the base of the logarithm) has no impact on the asymptotic complexity bound since $\log_a(x) = \frac{\log_b(x)}{\log_b(a)}$ for any $a,b > 1$.
In practice, we choose $\varepsilon_j = \frac{b_j - b_{j - 1}}{2}$.

\begin{example}
  \label{exa:sthLoop}
  We now improve the polynomial bound from $\normalfont{\textsf{PB}}(n^5)$ (i.e., of degree 5) in \Cref{exa:sthPoly} to a logarithmic (sub-linear) bound from $\normalfont{\textsf{LB}}$ by applying \Cref{lem:complexity_logarithmic}.
  Here, we choose $\varepsilon_j = \frac{1}{2}$.
  Thus, we have
  \[
    \begin{array}{rcll}
      M'_2 & = & 1, & \text{as 1 is the 1-monotonicity threshold of $(1.5,0)$ and $(1,0)$}  \\
      M'_3 & = & 1, & \text{as 1 is the 1-monotonicity threshold of $(9.5,0)$ and $(9,0)$}.
    \end{array}
  \]
  Hence, we get $C = \max\{1, M'_2, N_2, M'_3, N_3 \} = 1$.
  So by \Cref{lem:complexity_logarithmic},
  \begin{align*}
     & \max\braced{C,\log_6(2 \cdot \sqcup \{ - x_3^2 + 2\cdot x_3^5\}),\log_{\frac{32}{19}} (2 \cdot \sqcup \{2\cdot x_3^5 - x_3^2, 2\cdot x_1 + x_3^2\})} \\
     & = \max\braced{1,\log_6(2\cdot x_3^2 + 4\cdot x_3^5), \log_{\frac{32}{19}} (4\cdot x_1 + 2\cdot x_3^2 + 4\cdot x_3^5)} \\
     & = \max \braced{1, \tfrac{1}{\log_2\left(\frac{32}{19}\right)}\cdot
    (\log_{2}(2) + \log_{2} ( 2\cdot x_1 + x_3^2 + 2\cdot x_3^5)) } \\
     & \leq \max \braced{1, \tfrac{3}{2} + \tfrac{3}{2}\cdot \log_{2} ( 2\cdot x_1 + x_3^2 + 2\cdot x_3^5)} \\
     & = \tfrac{3}{2} + \tfrac{3}{2}\cdot \log_{2} ( 2\cdot x_1 + x_3^2 + 2\cdot x_3^5) \tag{where $\leq$ is interpreted
      point-wise for every $|\valuation|\in\Valuation$}
  \end{align*}
  is a \emph{logarithmic} bound on the stabilization threshold of \eqref{exa:pe}, i.e., a bound from $\normalfont{\textsf{LB}}$.
\end{example}

Now we use the results on stabilization thresholds of poly-exponential expressions in order to infer runtime bounds for \emph{tnn}-loops.
For any atom $\alpha = (s_1 < s_2)$ (resp.\ $s_2-s_1 > 0$), let $pe_\alpha\in\PPEE(\ZZ,\NN)$ be a poly-exponential expression which results from multiplying $(s_2-s_1)[\vec{x}/{\mbox{\normalfont{$\cl{\vec{x}}$}}}]$ with the least common multiple of all denominators occurring in $(s_2-s_1)[\vec{x}/{\mbox{\normalfont{$\cl{\vec{x}}$}}}]$ or in $\auto(x_1), \ldots, \auto(x_d)$.
This multiplication ensures that $pe_\alpha$ is from $\PPEE(\ZZ,\NN)$ instead of $\PPEE(\QQ,\NN)$, and moreover, it ensures that that $pe_\alpha$ has the form $\sum_{j=1}^{\ell} p_j \cdot n^{a_j} \cdot b_j^n$ where $\valuation(p_j) \in \ZZ$ for all $\valuation\in\auto(\Valuation)$, as required for \Cref{lem:complexity,lem:complexity_logarithmic}.

For any initial state $\valuation \in \auto(\Valuation)$, if $\valuation(pe_\alpha[n / \sth^\auto_{pe_\alpha}(\valuation)])$ is non-positive, then the atom $\alpha$ is violated and it will never be satisfied again after $\sth^\auto_{pe_\alpha}(\valuation)$ loop iterations.
Thus, if we consider the maximum of the stabilization thresholds of all $pe_\alpha$ for $\alpha\in\guard$ for a terminating loop $\IntLoop$, then computing a bound on this function yields a bound on the runtime complexity.

\begin{definition}[Stabilization Threshold for Loops]
  \label{def:Stabilization Threshold for Loops}
  Let $\IntLoop$ be a loop.
  The \emph{stabilization threshold} $\sth^\auto_{\IntLoop} :\auto(\Valuation) \to \NN$ of $\IntLoop$ is defined as $\sth^\auto_{\IntLoop}(\valuation) = \max\braced{\sth^\auto_{pe_{\alpha}}(\valuation)\mid\alpha\in\guard}$ for any $\valuation \in \auto(\Valuation)$.
\end{definition}

\begin{example}
  \label{ex:tnn}
  Consider the following tnn-loop which results from the leading example \eqref{WhileExampleChained} by removing the variables $x_4$ and $x_5$ (which both do not influence the runtime complexity and violate the tnn-property, see also \Cref{runtimeBoundWhileExample}):
  \begin{equation}
    \WhileLoop{x_2^2 - x_3^5 < x_1 \, \wedge \, x_2 \neq 0}{\myvec{x_1 \\ x_2 \\ x_3} \leftarrow \myvec{9\cdot x_1 + 4\cdot x_3^2 \\ 4\cdot x_2 \\ x_3}}
    \label{exa:simplified_tnn}
  \end{equation}

  A closed form (with start value $n_0 = 0$) for this tnn-loop is $\cl{x_1} = (x_1 + \frac{1}{2}\cdot x_3^2) \cdot 9^n - \frac{1}{2}\cdot x_3^2$, $\cl{x_2} = x_2 \cdot 4^n$, and $\cl{x_3} = x_3$.
  Note that this closed form results from substituting $n$ by $2\cdot n$ in the closed form of the unchained loop (see \Cref{exa:closed_form_twn}).
  The guard of the tnn-loop has the atoms $2\cdot(x_1 - x_2^2 + x_3^5) > 0$ (since we multiplied with the least common multiple of all denominators in $\cl{x_1}$ to obtain a poly-exponential expression in $\PPEE(\ZZ,\NN)$), $x_2 > 0$, and $-x_2 > 0$ (since $x_2 \neq 0$ is transformed into $0 < x_2 \lor 0 < -x_2$).
  So by inserting $\cl{\vec{x}}$ into the atoms of the guard, we obtain $(-2\cdot x_2^2) \cdot 16^n + (2\cdot x_1 + x_3^3) \cdot 9^n - x_3^2 + 2\cdot x_3^5 > 0$, $x_2 \cdot 4^n > 0$, and $-x_2 \cdot 4^n > 0$.
  Both $x_2 \cdot 4^n$ and $-x_2 \cdot 4^n$ have a stabilization threshold of $0$ as their sign is always determined by $\pm x_2$.
  Note that \Cref{lem:complexity,lem:complexity_logarithmic} yield 1 as a bound on their stabilization threshold.
  We obtained $\tfrac{3}{2} + \tfrac{3}{2}\cdot \log_{2} ( 2\cdot x_1 + x_3^2 + 2\cdot x_3^5)$ as a bound on the stabilization threshold of the remaining atom in \Cref{exa:sthLoop}.
\end{example}

The following theorem summarizes the approach for \emph{tnn}-loops.
Here, we consider \emph{tnn}-loops with updates of the form $\update(x_i) = c_i \cdot x_i + p_i$ where $c_i\in\NN$ and $p_i\in\QQ[x_{i+1},\dots,x_{d}]$ so that the theorem is applicable for the \emph{tnn}-loops $L_t \chain L_t$ in \Cref{lem:correctness_chaining}.
\begin{restatable}[Computing Runtime Bounds for \emph{tnn}-Loops]{theorem}{RBLoops}
  \label{thm:RBLoops}
  Let $L = \IntLoop$ be a terminating tnn-loop with $\update(x_i) = c_i \cdot x_i + p_i$ where $c_i\in\NN$ and $p_i\in\QQ[x_{i+1},\dots,x_{d}]$.
  For any atom $\alpha$, let $pe_\alpha\in\PPEE(\ZZ,\NN)$ be constructed by using a closed form with start value $n_0$, and let $\bound_\alpha$ be a bound on the stabilization threshold of $pe_\alpha$, i.e., $|\valuation|(\bound_\alpha) \geq \sth^\auto_{pe_{\alpha}}(\sigma)$ for all $\valuation \in \auto(\Valuation)$.
  Then $\run = \max (\{ \bound_\alpha \mid\alpha\in\guard\} \cup \{ n_0 \})$ is a bound on the stabilization threshold of $\IntLoop$ and thus, it is also a runtime bound, i.e., $|\valuation|(\run) \geq \sth^\auto_{\IntLoop}(\sigma) \geq \rc(\valuation)$ for all $\valuation \in \auto(\Valuation)$.
\end{restatable}
\makeproof{thm:RBLoops}{
  \RBLoops*
  \begin{myproof}
    Let $\alpha$ be an atom which occurs in the guard $\guard$.
    For $n \geq n_0$, by the definition of $pe_\alpha$, we have $\valuation(pe_\alpha) \leq 0$ iff $\valuation(\update^{n}(\alpha)) = \false$.
    Thus, we either always have $\valuation(\update^{n}(\alpha)) = \true$ or $\valuation(\update^{n}(\alpha)) = \false$ for all $n \geq \max \{ \sth^\auto_{pe_\alpha}(\sigma), n_0 \}$.
    Hence, for all $n \geq |\sigma|(\run) \geq \max ( \{\sth^\auto_{pe_{\alpha}}(\sigma) \mid \alpha \in \guard \} \cup \{n_0\})$, the value of $\sigma(\update^n(\guard))$ is the same.

    As $L$ is terminating, there is a minimal $n_1 = \rc(\sigma) \in \NN$ such that $\sigma(\update^{n_1}(\guard)) = \false$.
    Assume that $n_1 > |\sigma|(\run)$.
    If for all $n \geq |\sigma|(\run)$ we had $\sigma(\update^n(\guard)) = \true$, then this would contradict $\sigma(\update^{n_1}(\guard)) = \false$.
    If for $n = |\sigma|(\run)$ we had $\sigma(\update^n(\guard)) = \false$, then this would contradict the minimality of $n_1$.
    So the assumption was wrong, and we have $|\sigma|(\run) \geq n_1 = \rc(\sigma)$, i.e., $\run$ is indeed a runtime bound.
    \qed
  \end{myproof}
}
\begin{example}
  \label{exa:tnn_final}
  Finally, we can infer a runtime bound for the leading example.
  Note that \eqref{exa:simplified_tnn} terminates as every iteration increases $x_2^2$ by a factor of $16$ whereas $x_1$ is only increased by a factor of $9$.
  Thus, $x_2^2$ eventually outgrows the value of $x_1$ as $x_2 \neq 0$.
  In practice, we use SMT solvers to prove termination of such loops via the complete approach discussed in the previous subsection.
  Note that this loop does not admit a linear ranking function over $\RR$ \cite{heizmann2015RankingTemplatesLinear}.

  For every atom of its guard, we have shown how to compute a bound on the stabilization threshold.
  For $x_2 > 0$ and $-x_2 > 0$ we obtained the bounds $0$ or $1$ in \Cref{ex:tnn}.
  Furthermore, we inferred the logarithmic bound $b_{x_2^2 - x_3^5 < x_1} = \tfrac{3}{2} + \tfrac{3}{2}\cdot \log_{2} ( 2\cdot x_1 + x_3^2 + 2\cdot x_3^5)$ in \Cref{exa:sthLoop}.
  Thus, $b_{x_2^2 - x_3^5 < x_1}$ is a bound on the runtime complexity of \eqref{exa:simplified_tnn} and also of $\eqref{WhileExample} \chain \eqref{WhileExample} = \eqref{WhileExampleChained}$.
  Hence, the runtime complexity of the loop \eqref{WhileExample}
  is logarithmic in its input, see \Cref{runtimeBoundWhileExample}.
  Note that while in our previous conference paper \cite{lommen2022AutomaticComplexityAnalysis} we used a more complicated technique to infer bounds by considering invariants, it would yield a much worse bound for this example (a polynomial bound of degree 5), since it could not yet infer any logarithmic bounds.
\end{example}

\subsection{Size Bounds for Solvable Loops}
\label{sect:loops_size_bounds}

In this section, we present a technique to compute size bounds for loops by using closed forms and runtime bounds.
Size bounds should be bounds on the values of variables up to the point where the loop guard is not satisfied anymore for the first time.
More precisely, a \emph{size bound} on a variable $v$ is a bound on the absolute value of $v$ after $n$ iterations of the update $\update$, where $n$ is bounded by the runtime complexity.
Note that \Cref{Size Bounds of Loops} requires that size bounds also hold \emph{before} evaluating the loop (this will be needed later in \Cref{sect:Size Bounds for Integer Programs} to handle programs with loops that consist of several transitions).
\begin{definition}[Size Bounds for Loops]
  \label{Size Bounds of Loops}
  The function $\size: \VSet \rightarrow\BoundSet$ is a \emph{size bound} for a loop $\IntLoop$ if for all $v\in\VSet$ and all $\valuation\in\Valuation$, we have $$|\valuation|(\size(v)) \geq \sup\braced{|\valuation(\update^n(v))| \mid n \leq \rc(\valuation)}.$$
\end{definition}

We extend $||\cdot||$ from polynomials to poly-exponential expressions, i.e., $||\cdot||$ is used to transform any poly-exponential expression into a (weakly monotonic) bound from $\BoundSet$.
For any poly-exponential expression $pe = \sum_{j} (\sum_i c_{i,j}\cdot \beta_{i,j})\cdot n^{a_j} \cdot b_j^n$ where $c_{i,j} \in \AA$ and the $\beta_{i,j}$ are normalized monomials of the form $x_1^{e_1} \cdot \ldots \cdot x_d^{e_d}$, $||pe||$ denotes $\sum_{j}\left(\sum_i |c_{i,j}| \cdot \beta_{i,j}\right)\cdot n^{a_j} \cdot |b_j|^n $.

Now we determine size bounds by over-approximating the closed form $\cl{x}$ by the non-negative, weakly monotonically increasing function $||\cl{x}||$.
Then we substitute $n$ by a runtime bound $\run$.
Due to the monotonicity, this results in a bound on the size of $x$ not only at the end of the loop, but also during the iterations of the loop.
Since the closed form is only valid for $n$ iterations with $n \geq n_0$, we ensure that size bounds are also correct for less than $n_0$ iterations by symbolically evaluating the update.
\begin{restatable}[Size Bounds for Loops with Closed Forms]{theorem}{SBClosedForms}
  \label{thm:size_bounds_closed_form}
  Let $\cl{\vec{x}}$ be a closed form for the loop $\IntLoop$ with start value $n_0$ and let $\run\in\BoundSet$ be a runtime bound.
  Then the (absolute) size of $v\in\VSet$ is bounded by
  \[\size(v) =
    \left\{ \begin{array}{ll}
      ||\cl{v}|| \, [n / \run],      & \text{if $n_0 = 0$} \\
      \max\braced{||\cl{v}|| \, [n / \run],
      \; ||\update||^{n_0 - 1}(v) }, & \text{otherwise}
    \end{array} \right.
  \]
  Here, we define $||\update||:\VSet \to \NN[\VSet]$ with $||\update||(v) = ||\update(v)||$ for all $v \in \VSet$.
  Hence, the function $\size$ is a size bound for $\IntLoop$.
\end{restatable}
\makeproof{thm:size_bounds_closed_form}{
  \SBClosedForms*
  \begin{myproof}
    We have to prove that
    \begin{equation}
      \label{eq:size_bounds_sup}
      |\valuation|(\size(v)) \geq \sup\braced{|\valuation(\update^n(v))| \mid n \leq \rc(\valuation)}
    \end{equation}
    holds for all states $\valuation:\VSet \to \ZZ$ and all $v\in\VSet$.
    First, note that the evaluation of every summand of $|\valuation|(\size(v))$ is non-negative for all $\valuation:\VSet \to \ZZ$.
    Let $n \leq \rc(\valuation)$ for a $\valuation:\VSet \to \ZZ$.
    If $n < n_0$, then \eqref{eq:size_bounds_sup} clearly holds as $|\valuation(\update^n(v))| \leq |\valuation|(||\update||^n(v)) \leq |\valuation|(||\update||^{n_0-1}(v))$ and the summand $||\update||^{n_0-1}(v)$ occurs in $\size(v)$.
    Otherwise, consider $n \geq n_0$.
    For all $\valuation:\VSet \to \ZZ$, the function $$ f(n) = |\valuation| (|| \cl{v}||)$$ increases weakly monotonically in $n$.
    Since $\run$ is a bound on the runtime complexity, i.e., $|\valuation|(\run) \geq \rc(\valuation)$, we can conclude that $|\valuation|(\size(v)) \geq |\valuation| (||\cl{v}|| [n/\run]) = f(|\valuation|(\run)) \geq f(\rc(\valuation))\linebreak
      \geq f(n) \geq |\valuation(\cl{v})| = |\valuation(\update^n(v))|$ holds.
    Here, the second and third inequation hold because $f$ is weakly monotonically increasing and $|\valuation|(\run) \geq \rc(\valuation) \geq n$.
    The last inequation holds as $f(n)$ over-approximates $|\valuation(\cl{v})|$.
    So in total we have that $|\valuation|(\size(v)) \geq |\valuation(\update^n(v))|$ holds for all $n \leq \rc(\valuation)$, which proves \eqref{eq:size_bounds_sup}.
    \qed
  \end{myproof}
}
\begin{example}
  \label{exa:sizeboundLoop}
  As mentioned in \Cref{ex:closedFormEx,exa:closed_form_solvable}, for the loop \eqref{WhileExample}, a closed form for $x_4$ with start value $0$ is $\normalfont{\cl{x_4}} =\tfrac{1}{2}\cdot\alpha\cdot (-\im)^n + \tfrac{1}{2}\cdot\compconj{\alpha}\cdot \im^n$ where $\alpha = (1 + 3\im)\cdot x_4 + 2\im\cdot x_5$.
  Hence, $||\cl{x_4}|| = || \tfrac{1}{2}\cdot\alpha\cdot (-\im)^n + \tfrac{1}{2}\cdot\compconj{\alpha}\cdot \im^n|| = (|\tfrac{1+ 3\rm{i}}{2}|\cdot x_4 + |\im|\cdot x_5 )\cdot |-\rm{i}|^n + (|\tfrac{1- 3\im}{2}|\cdot \mbox{$x_4$} + | -\rm{i}|\cdot \mbox{$x_5$} )\cdot |\rm{i}|^n = \sqrt{10}\cdot \mbox{$x_4$} + 2\cdot \mbox{$x_5$}$,
  as $| \tfrac{1+ 3 \rm{i}}{2}| = | \tfrac{1- 3\rm{i}}{2}| = \tfrac{\sqrt{10}}{2}$ and $| \rm{i}| = | -\rm{i}| = 1$.
  So the approach infers \emph{linear} size bounds for $x_4$ and $x_5$ (the similar computations for $x_5$ are omitted) while the incomplete technique of \cite{brockschmidt2016AnalyzingRuntimeSize} infers exponential size bounds in this example.
  As the over-approximation $||\cl{x_4}||$ does not depend on $n$, it directly yields a size bound, i.e., $\size(x_4) = ||\cl{x_4}||$.
  In contrast, in the over-approximation $||\cl{x_1}|| = \left( x_1 + \frac{1}{2} \cdot x_3^2 \right)\cdot 3^n + \frac{1}{2}\cdot x_3^2$, we have to replace $n$ by the runtime bound $\run = 4 + 3\cdot\log_{2}(2\cdot x_1 + x_3^2 + 2\cdot x_3^5)$ (see \Cref{runtimeBoundWhileExample}).
  Thus, we obtain the overall polynomial size bound
  \begin{align*}
    \size(x_1) & = \left( x_1 + \tfrac{1}{2} \cdot x_3^2 \right)\cdot 3^{4 + 3\cdot\log_{2}(2\cdot x_1 + x_3^2 + 2\cdot x_3^5)} + \tfrac{1}{2}\cdot x_3^2 \\
               & = \left( x_1 + \tfrac{1}{2} \cdot x_3^2 \right)\cdot 3^4\cdot 3^{\tfrac{3}{\log_3\left(2\right)}\cdot\log_{3}(2\cdot x_1 + x_3^2 + 2\cdot x_3^5)} + \tfrac{1}{2}\cdot x_3^2 \\
               & = \left( x_1 + \tfrac{1}{2} \cdot x_3^2 \right)\cdot 3^4\cdot (2\cdot x_1 + x_3^2 + 2\cdot x_3^5)^{\tfrac{3}{\log_3\left(2\right)}} + \tfrac{1}{2}\cdot x_3^2 \\
               & \leq \left( x_1 + \tfrac{1}{2} \cdot x_3^2 \right)\cdot 3^4\cdot (2\cdot x_1 + x_3^2 + 2\cdot x_3^5)^5 + \tfrac{1}{2}\cdot x_3^2
    \tag{where $\leq$ is interpreted point-wise w.r.t.\ $|\valuation|\in\Valuation$}.
  \end{align*}
  In particular, this example illustrates that logarithmic runtime bounds are also useful to improve size bounds.
  It shows that due to the logarithmic runtime bound, we obtain a polynomial size bound for $x_1$, whereas the approach of our conference papers \cite{lommen2022AutomaticComplexityAnalysis,lommen2023TargetingCompletenessUsing} would yield an exponential size bound.
\end{example}

Note that the technique of \Cref{thm:size_bounds_closed_form} is not limited to closed forms which are poly-exponential expressions.
More precisely, \Cref{thm:size_bounds_closed_form}
is applicable to all loops where we can compute over-approximating bounds for the closed form and the runtime complexity (if the bound for the closed form is weakly monotonically increasing in $n$).
For example, the update $\update(x) = x^2$ has the closed form $x^{(2^n)}$, but it does not admit a poly-exponential closed form due to $x$'s super-exponential growth.
However, by instantiating $n$ by a runtime bound, we can still compute a size bound for this update.
The reason for focusing on poly-exponential expressions is that they were needed for the technique in \Cref{sect:Runtime Bounds for PRS-Loops} to compute runtime bounds automatically.

\subsection{Completeness}
\label{sect:completenessLoops}

In this section, we summarize the procedure for the computation of runtime and size bounds for loops, and present completeness results.
These observations are captured in the following algorithm.
Since we can detect \emph{prs}-loops and their periods by \Cref{Bound on the Period}, \Cref{lem:correctness_chaining} allows us to compute runtime bounds for all terminating \emph{prs}-loops.
This is illustrated in \Cref{fig:illustration} and \Cref{alg:loops}:
If $L$ is an unsolvable loop, then we first try to transform $L$ into a solvable loop $L_x$ in \Cref{alg:unsolvable}.
In practice, we might apply this preprocessing step multiple times using several different polynomials $q$.
(In contrast, if $L$ is already solvable, then we have $q = x$ and $L_x = L$.)

\begin{figure}[t]
  \centering
  {\scriptsize
    \begin{tikzpicture}[->,>=stealth',shorten >=1pt,auto,node distance=3.1cm,semithick,initial text=$ $]
      \node (r0) {
        \begin{tabular}{c}
          Runtime \\
          Bound:
        \end{tabular}
      };
      \node (r1) [right of=r0,xshift=-1.4cm]{
        \begin{tabular}{c}
          \emph{prs}-loop $L$ \\
          $p\cdot ||\auto||(\run) + p - 1$
        \end{tabular}
      };
      \node (r2) [right of=r1,xshift=.6cm]{
        \begin{tabular}{c}
          $L_p$ \\
          $||\auto||(\run)$
        \end{tabular}
      };
      \node (r3) [right of=r2,xshift=1cm]{
        \begin{tabular}{c}
          $L_t$ with 	$\update_t:\VSet\rightarrow\QQ[\VSet]$ \\
          $\run$ by \Cref{sect:Runtime Bounds for PRS-Loops}
        \end{tabular}
      };
      \node (s0) [below of=r0,yshift=1cm] {
        \begin{tabular}{c}
          Size \\
          Bound:
        \end{tabular}
      };
      \node (s1) [below of=r1,yshift=1cm] {
        \begin{tabular}{c}
          solvable loop $L$ \\
          $\clExp{}{s}$
        \end{tabular}
      };
      \node (s3) [below of=r3,yshift=1cm] {
        \begin{tabular}{c}
          $L_t'$ with	 	$\update_t':\VSet\rightarrow\AA[\VSet]$ \\
          $\clExp{}{t}$ by \cite{frohn2019TerminationTriangularInteger}
        \end{tabular}
      };
      \draw (r1) edge node {chaining} (r2);
      \draw (r1) edge [below,draw=none] node {\cref{lem:correctness_chaining} (a)} (r2);
      \draw (r2) edge node {\mbox{\scriptsize $\auto:\VSet\rightarrow\QQ[\VSet]$}} (r3);
      \draw (r2) edge [below,draw=none] node {\cref{lem:correctness_chaining} (b)} (r3);
      \node (h1) [below of=r1,yshift=2.75cm] {};
      \node (h2) [below of=r2,yshift=2.75cm] {};
      \node (h3) [below of=r3,yshift=2.75cm] {};
      \draw (h2) edge [below,bend left=12,decorate, decoration={snake,amplitude=1.5pt}] node {\cref{lem:correctness_chaining} (c) \& (d)} (h1.center);
      \draw (h3) edge [below,bend left=12,decorate, decoration={snake,amplitude=1.5pt}] node {\cref{lem:correctness_chaining} (c) \& (d)} (h2.center);
      \draw (r0) edge [->,right,loosely dashed] node
        {\cref{thm:size_bounds_closed_form}} (s0);
      \draw (s1) edge node {\cref{lem:transform_solvable} by $\auto':\VSet\rightarrow\AA[\VSet]$} (s3);
      \node (h1a) [below of=s1,yshift=2.75cm] {};
      \node (h3a) [below of=s3,yshift=2.75cm] {};
      \draw (h3a) edge [above,bend left=5,decorate, decoration={snake,amplitude=1.5pt}] node {\cref{thm:closed_form_solvable}} (h1a.center);
    \end{tikzpicture}
  }
  \caption{Computing Runtime and Size Bounds for Solvable Loops via \emph{twn}-Loops}
  \vspace*{-.2cm}
  \label{fig:illustration}
\end{figure}
\begin{figure}[t]
  \begin{algorithm}[H]
    \DontPrintSemicolon
    \caption{Algorithm to Compute Runtime and Size Bounds for Loops}\label{alg_loops}
    \Input{A loop $L = \IntLoop$}
    \Output{A runtime bound $\run \in \BoundSet$ and a size bound $\size: \VSet\to\BoundSet$ for $L$}
    Initialize $\run$ and $\size$: $\run \gets \omega$ and $\size(v) \gets \omega$ for all $v\in\VSet$. \label{alg_loops:initial}\;
    Try to transform $L$ into a solvable loop $L_x$ w.r.t.\ a polynomial $q$ as in \Cref{def:transform_unsolvable}.\label{alg:unsolvable}\;
    \If{$L_x$ is a terminating prs-loop with period $p$}{
      Obtain $L_p$ by chaining $L_x$ $p$-times.\; \label{alg:loops:chain}
      Compute automorphism $\auto:\VSet\to\QQ[\VSet]$ and \emph{twn}-loop $L_t$ as in \Cref{lem:correctness_chaining}.\; \label{alg:loops:auto}
      \If{$L_t$ is a tnn-loop} {
        $\run \gets p\cdot ||\auto||(\text{result of \Cref{thm:RBLoops} on $L_t$ and $\auto$}) + p - 1$\; \label{alg:loops:tnn}
      }
      \Else{
        $\run \gets 2\cdot p\cdot ||\auto||(\text{result of \Cref{thm:RBLoops} on $L_t
              \chain L_t$ and $\auto$})
          + 2\cdot p - 1$\; \label{alg:loops:twn}
      }
    }
    \If{$L$ is a solvable loop}{
      $\size \gets \text{result of \Cref{thm:size_bounds_closed_form} using the runtime bound
        } \run$\; \label{alg:loops:size}
    }
    \return $\run \, [x/||q||]$ and $\size$.
  \end{algorithm}
  \caption{Algorithm Illustrating Runtime and Size Bound Computations}
  \vspace*{-.4cm}
  \label{alg:loops}
\end{figure}

For runtime bounds, $L_x$ is transformed to $L_p$ by chaining in \Cref{alg:loops:chain}, and $L_p$ is transformed further to $L_t$ by an automorphism $\auto$ in \Cref{alg:loops:auto} as in \Cref{lem:correctness_chaining}.
If $L_t$ is not already a \emph{tnn}-loop, then we chain $L_t$ once more in \Cref{alg:loops:twn}.
Now, we can use \Cref{lem:complexity,lem:complexity_logarithmic} together with \Cref{thm:RBLoops} to compute a runtime bound for this loop in \Cref{alg:loops:tnn,alg:loops:twn}.
The runtime bound $\run$ for $L_t$ can then be transformed into a runtime bound for $L_p$ and further into a runtime bound for $L_x$ and $L$.
For size bounds, $L$ is directly transformed into a \emph{twn}-loop $L_t'$ by an automorphism $\auto'$ if $L$ is solvable.
The closed form $\clExp{}{t}$ obtained for $L_t'$ is transformed via the automorphism $\auto'$ into a closed form	$\clExp{}{s}$ for $L$.
Then the runtime bound for $L$ is inserted into this closed form to yield a size bound for $L$ in \Cref{alg:loops:size}.
So in \Cref{fig:illustration}, standard arrows denote transformations of loops and wavy arrows denote transformations of runtime bounds or closed forms.
Note that we are not interested in the size of polynomial combinations of variables.
Hence, the preprocessing from unsolvable loops to solvable loops is not used for size bounds.

As shown by \Cref{thm:completeness}, the technique for the computation of runtime and size bounds is ``complete'' for the class of terminating \emph{prs}-loops, i.e., finite (polynomial) runtime bounds and thus, also finite size bounds are always computable.
For \emph{linear} \emph{prs}-loops, i.e., \emph{prs}-loops with only linear arithmetic in the update and the guard, termination is decidable.

A \emph{prs}-loop is a \emph{unit} \emph{prs}-loop if $|\lambda| \leq 1$ holds for all its eigenvalues $\lambda$.
A solvable loop is called \emph{strict} if the bases $b_j$ of the exponential expressions $b_j^n$ in the closed form of the guards of the resulting \emph{tnn}-loop are pairwise different, i.e., we can order them strictly via $b_\ell > \ldots > b_1$.\footnote{A sufficient condition for strictness which does not refer to the transformation from solvable loops to \emph{tnn}-loops is that $\update(\guard) \to \guard$ holds and that the absolute values of the $b_j$ in the closed form of the guards of the original loop are pairwise different.
  Here, $\update(\guard) \to \guard$ is needed to ensure that the runtime complexity of the loops $(\guard,\update_t\circ\update_t)$ and $L_t\chain L_t$ is the same.
  Hence, we only have to consider $\guard$ (and not the chained guard of $L_t\chain L_t$).
  As the absolute values of the $b_j$'s are pairwise different, the \emph{tnn}-loop $(\guard,\update_t\circ\update_t)$ is strict.}
Note that pairwise different eigenvalues $c_i$ of the \emph{tnn}-loop do not imply strictness, since non-linear combinations of variables in the loop guard might yield other exponential expressions.
For example, the polynomial $x_1\cdot x_2$ and the eigenvalues $c_1 = 2$ and $c_2 = 3$ yield the exponential expression $6^n$ instead of $2^n$ or $3^n$.
The following theorem only considers loops with updates $\update: \VSet\to\ZZ[\VSet]$, and termination and runtime complexity over $\ZZ$.
\begin{restatable}[Completeness Results for Loops]{theorem}{CompletenessLoops}
  \label{thm:completeness}
  \hspace*{0.1cm}
  \begin{itemize}
    \item Termination is decidable for all linear prs-loops.
    \item Polynomial runtime bounds and finite size bounds are computable for all terminating prs-loops.
    \item For terminating \emph{unit} prs-loops, these size bounds are polynomial as well.
    \item For terminating \emph{strict} prs-loops, the runtime bounds are logarithmic and the size bounds are polynomial.
  \end{itemize}
\end{restatable}
\begin{myproof}
  \begin{enumerate}
    \item For a linear \emph{prs}-loop $L$, we proceed as in \cref{lem:correctness_chaining}.
          So we first compute the loop $L_p$ via chaining (\cref{def:chaining}) which is still solvable and linear (\cref{lem:correctness_chaining} (a)).
          Then we transform it via an automorphism $\vartheta$ into a linear \emph{twn}-loop $L_t$ (\cref{lem:transform_solvable}) that is terminating on $\vartheta(\ZZ^d)$ iff the \emph{prs}-loop $L$ is terminating on $\ZZ^d$ (\cref{lem:correctness_chaining} (c)).
          Termination of the linear \emph{twn}-loop $L_t$ can be decided by the approach of \cite{frohn2020TerminationPolynomialLoops}.
          Note that linearity of $L_t$ is important here since it guarantees that the existential arithmetic formula over $\ZZ$ which results from the reduction of \cite{frohn2020TerminationPolynomialLoops}
          is linear as well.
    \item As shown in \Cref{thm:RBLoops}, since $\vartheta$ is a rational automorphism (\cref{lem:correctness_chaining} (b)), one can compute a polynomial runtime bound for $L_t$ on $\vartheta(\ZZ^d)$ and by \cref{lem:correctness_chaining} (d), this yields a polynomial runtime bound for $L$.

          By transforming the \emph{prs}-loop $L$ directly into a \emph{twn}-loop as in \cref{lem:transform_solvable}, from the closed form of the resulting \emph{twn}-update we can obtain a closed form for $L$'s update (\Cref{thm:closed_form_solvable}).
          Thus, when inserting the runtime bound for $L$ into this closed form, \Cref{thm:size_bounds_closed_form} yields a finite size bound for $L$.
    \item Now we show that for terminating unit \emph{prs}-loops, we compute polynomial size bounds.
          Note that when transforming the \emph{prs}-loop $L$ into a \emph{twn}-loop with update $\update_t'$, for all $1 \leq i \leq d$ we have $\update_t'(x_i) = c_i \cdot x_i + p_i$ where $c_i$ is an eigenvalue of $L$.
          The reason is that due to the definition of the transformation in \cref{lem:transform_solvable}, $c_i$ is on the diagonal of the Jordan normal form of $L$'s respective update matrix $A_\SSet$.
          Hence, we have $|c_i| \leq 1$, as $L$ is a unit \emph{prs}-loop.
          We call any \emph{twn}-update of this form \emph{unit}.

          By induction on $i$, we now prove that any unit \emph{twn}-update $\update_t'$ has closed forms $\cl{x_i} = \sum_{j = 1}^m p_j \cdot n^{a_j} \cdot b_j^n$ where $|b_j| \leq 1$, $a_j\in\NN$, and $p_i\in\AA[\VSet]$.
          We call such a poly-exponential expression \emph{unit}.
          \paragraphProof{Induction Base:}
          We have $\update_t'(x_1) = c \cdot x_1 + p$ with $|c| \leq 1$ (as $\update_t'$ is unit) and $p\in\AA$.
          Moreover, we have $(\update_t')^n(x_1) = x_1 \cdot c^n + \sum_{i=1}^n c^{n-i}\cdot p = x_1 \cdot c^n + \frac{c^n - 1}{c - 1}\cdot p = (x_1 + \frac{p}{c - 1})\cdot c^n - \frac{p}{c - 1}$ if $c < 1$ and $(\update_t')^n(x_1) = x_1 + n\cdot p$ if $c = 1$.
          The last expressions are both unit closed forms $\cl{x_1}$.
          \paragraphProof{Induction Step:}
          In the induction step we assume that we have unit closed forms $\cl{x_1},\ldots,\cl{x_k}$ for the variables $x_1,\ldots,x_k$.
          Similar to the induction base, we have $\update_t'(x_{k + 1}) = c \cdot x_{k + 1} + p$ with $|c| \leq 1$ and $p\in\AA[x_{k+2},\ldots,x_d]$.
          So following \cite[Lemma 14]{frohn2019TerminationTriangularInteger}, we have to show that $(\update_t')^n(x_{k+1}) = c^n \cdot x_{k + 1} + \sum_{i = 1}^n c^{n-i} \cdot p[x_j / \cl{x_j} \mid j \in \braced{1,\ldots,k}][n / i - 1]$ can be transformed into a unit closed form.

          By the induction hypothesis, the closed forms $\cl{x_1},\ldots,\cl{x_k}$ are unit.
          We insert them into $p$ such that we obtain the expression $\sum_{j = 1}^m p_j \cdot n^{a_j} \cdot b_j^n$ (see \eqref{thm:completeness_label1} below).
          This expression is unit, because being unit is closed under addition and multiplication, i.e., if $p,q$ are unit poly-exponential expressions, then $p + q$ and $p \cdot q$ are unit poly-exponential expressions as well.
          \begin{align}
             & c^n \cdot x_{k + 1} + \sum_{i = 1}^n c^{n-i} \cdot p[x_j / \cl{x_j} \mid j \in \braced{1,\ldots,k}][n / i - 1] \nonumber \\
             & = c^n \cdot x_{k + 1} + \sum_{i = 1}^n c^{n-i} \cdot \underbrace{\left(\sum_{j = 1}^m p_j \cdot n^{a_j} \cdot b_j^n\right)}_{\coloneqq\; p[x_j / \cl{x_j} \mid j \in \braced{1,\ldots,k}]}[n / i - 1] \label{thm:completeness_label1}
            \\
             & = c^n \cdot x_{k + 1} + \sum_{j = 1}^m \underbrace{\sum_{i = 1}^n c^{n - i} \cdotp_j \cdot (i - 1)^{a_j} \cdot b_j^{i - 1}}_{\eqqcolon\; q_j} \label{thm:completeness_label2}
          \end{align}

          So, after commuting both sums in \eqref{thm:completeness_label2}, it remains to show that $q_j$ can be transformed into a unit poly-exponential expression.
          In \cite[Thm.\ 12]{frohn2019TerminationTriangularInteger} it is shown how to transform $q_j$ into a poly-exponential expression.
          Note that only $c^{n - i}$ or $b_j^{i - 1}$ lead to exponential factors $c^n$ or $b_j^n$, respectively.
          However, we have both $|c| \leq 1$ (since $\update_t'$ is unit) and $|b_j| \leq 1$ (by the induction hypothesis).

          To summarize, we have proven that \cref{Closed Forms for TWN Updates}
          yields unit closed forms $\sum_{j = 1}^m p_j \cdot n^{a_j} \cdot b_j^n$ for the \emph{twn}-loop resulting from transforming $L$.
          Remember that we obtain closed forms for solvable loops by reverting the automorphism (see \Cref{lem:transform_solvable}).
          This automorphism only affects the polynomials $p_j \in \AA[\VSet]$, but it does not modify the exponential factors $b_j^n$.
          Thus, the closed forms for the original \emph{prs}-loop $L$ have the form $\cl{x} = \sum_{j = 1}^m p_j' \cdot n^{a_j} \cdot b_j^n$ with $|b_j| \leq 1$, i.e., they are unit.
          Hence, we have $||\cl{x}|| = \sum_{j = 1}^m ||p_j'|| \cdot n^{a_j}
            \cdot |b_j|^n \leq \sum_{j = 1}^{m} ||p_j'|| \cdot n^{a_j}$.
          Applying \Cref{thm:size_bounds_closed_form}, i.e., replacing $n$ by the polynomial runtime bound, finally yields a polynomial size bound.
    \item Finally, consider a strict \emph{prs}-loop with period $p$
          and exponential expressions $|b_1| < \dots < |b_\ell|$.
          Here, we obtain logarithmic runtime bounds for $L_t\chain L_t$ by \Cref{lem:complexity_logarithmic} and thus also for $L_t$, $L_p$, and $L$.
          In particular, the size bounds for strict loops are polynomial as all exponential expressions $b^n[n/\run]$ in the construction of \Cref{thm:size_bounds_closed_form} yield polynomials.
          \qed
  \end{enumerate}
\end{myproof}

\section{Runtime and Size Bounds for Integer Programs}
\label{sect:global_integer_programs}

Up to now, we focused on \emph{isolated} loops.
In particular, we introduced techniques to infer runtime and size bounds for \emph{prs}-loops.
However, in practical applications, programs can have consecutive or nested loops.
In \cite{brockschmidt2016AnalyzingRuntimeSize,giesl2022ImprovingAutomaticComplexity} we presented an \emph{incomplete} modular technique for complexity analysis of \emph{integer programs} for which most questions regarding termination or complexity are undecidable.
The original incomplete approach used individual ranking functions for different subprograms.
Based on this, we now introduce a modular approach to automatically infer runtime bounds for programs possibly consisting of multiple consecutive or nested loops by handling some subprograms as \emph{prs}-loops (such that we can apply the \emph{complete} techniques of \Cref{sect:loops}
to them) and by using ranking functions for others.
Since the approach via (linear) ranking functions is incomplete, of course our modular approach for larger programs with possibly nested loops is incomplete as well.

\begin{example}
  \label{ex:leadingPseudo}
  In \Cref{fig:pseudocode}, we integrate the loop \eqref{WhileExample} into a larger program.
  First, the loop \eqref{WhileExample} is called within another loop, i.e., we have \emph{nested} loops.
  The runtime of the outer loop depends on the non-deterministic choice of the temporary variable $y$.
  In this paper, we are interested in worst-case executions which in this example corresponds to always choosing $y = 1$.
  Furthermore, we extended the outer loop body with reset statements:
  So for example, $x_1$ is set to $x_6$.
  We also added a second, \emph{consecutive} loop whose runtime depends on the size of $x_4$.
  To analyze it, we will need size bounds of the prs-loop \eqref{WhileExample}.
  Thus, in this section, we demonstrate how to analyze programs with arbitrary control-flow automatically.
  \begin{figure}[t]
    \begin{align*}
       & \textbf{while $0 < y \leq x_6$ do} \\
       & \qquad (x_1,x_2,x_3,x_4,x_5,x_6) \gets (x_6,x_3,2,x_6,x_6,x_6 - y) \\
       & \qquad \text{execute Loop \eqref{WhileExample}} \\
       & x_6 \gets x_4 \\
       & \textbf{while $0 < x_6$ do} \\
       & \qquad x_6 \gets x_6 - 1
    \end{align*}
    \caption{A Pseudocode Program with the \emph{prs}-Loop \eqref{WhileExample}}
    \label{fig:pseudocode}
  \end{figure}
\end{example}

In order to compute runtime bounds, we analyze subprograms in topological order, i.e., in case of multiple consecutive loops, we start with the first loop and propagate knowledge about the resulting values of variables to subsequent loops.
By inferring runtime and size bounds for one subprogram after the other, in the end we obtain a bound on the runtime complexity of the whole program.

We first try to compute runtime bounds for subprograms by multiphase linear ranking functions ($\MRFs$, see \cite{giesl2022ImprovingAutomaticComplexity,ben-amram2017MultiphaseLinearRankingFunctions,heizmann2015RankingTemplatesLinear,ben-amram2019MultiphaseLinearRankingFunctions}), since they can also handle larger subprograms that do not correspond to single-path loops as in \Cref{sect:loops}.
If $\MRFs$ do not yield a finite runtime bound for the respective subprogram, then if possible, we use the \emph{prs}-technique from \Cref{sect:loops}
on the unsolved parts of the subprogram.
So for the first time, ``complete'' complexity analysis techniques like the ones of \Cref{sect:loops} for subclasses of programs with \emph{non-linear} arithmetic are combined with incomplete techniques based on (linear) ranking functions like \cite{brockschmidt2016AnalyzingRuntimeSize,giesl2022ImprovingAutomaticComplexity}.
In contrast to \cite{brockschmidt2016AnalyzingRuntimeSize,giesl2022ImprovingAutomaticComplexity}, the modular approach works for any method of bound computation (not only for ranking functions for runtime bounds and for local size bounds as in \cite{brockschmidt2016AnalyzingRuntimeSize}).

We start by introducing integer programs in \Cref{sect:integer_programs} and \emph{global} runtime and size bounds in \Cref{sect:global_bounds}.
In \Cref{sect:Runtime Bounds for Integer Programs}, we show how to compute global runtime bounds for a program and how to use the complete technique for the inference of runtime bounds of loops from \Cref{sect:Runtime Bounds for PRS-Loops} for subprograms that correspond to \emph{prs}-loops.
Similarly, in \Cref{sect:Size Bounds for Integer Programs} we present a technique to compute global size bounds for general programs, using the complete technique for size bounds of loops from \Cref{sect:loops_size_bounds} for subprograms that correspond to solvable loops.
Finally, similar to \Cref{sect:completenessLoops} where we discussed the completeness of the techniques for loops, in \Cref{sect:Completeness} we discuss the completeness of the approach for general programs.

\subsection{Integer Programs}
\label{sect:integer_programs}
For integer programs, we use a formalism based on transitions, which also allows us to represent while-loops like \eqref{WhileExample} easily (see \Cref{sect:conclusion} for a discussion of related work).
An integer program may have two kinds of non-determinism.
Non-deterministic branching is realized by multiple transitions with the same start location whose guards are non-exclusive.
Non-deterministic sampling is modeled by temporary variables $\TVSet$ (which can be restricted in the guard of a transition).
Temporary variables are updated arbitrarily in each evaluation step, only restricted by the transition's guard.
Intuitively, these variables can be set by an adversary trying to ``sabotage'' the execution of the program in order to obtain long runtimes.
When inferring runtime bounds, we can only handle temporary variables by ranking functions, since the
\emph{twn}-technique to characterize the dynamics of the loop via the ``dominating summand'' is not capable of handling temporary variables.

\begin{definition}[Integer Program]
  \label{Integer Program}
  $\IntProgram$ is an \emph{integer program} where
  \begin{itemize}
    \item $\VSet$ is a finite set of \emph{program variables} and $\TVSet$ is a finite set of \emph{temporary variables}
          with $\TVSet \cap \VSet = \emptyset$.
    \item $\LSet$ is a finite set of \emph{locations} with an \emph{initial location}
          $\location_0\in\LSet$.
    \item $\TSet$ is a finite set of \emph{transitions}.
          A transition is a 4-tuple $(\location,\guard,\update,\location')$ with a \emph{start location} $\location\in\LSet$, \emph{target location} $\location'\in\LSet\setminus\braced{\location_0}$, \emph{guard} $\guard\in\FormulaSet(\VSet\cup\TVSet)$, and \emph{update function} $\update: \VSet\rightarrow\ZZ[\VSet\cup\TVSet]$ mapping program variables to update polynomials.
  \end{itemize}
\end{definition}
Transitions $(\location_0,\_,\_,\_)$ are called \emph{initial} and $\TSet_0 \subseteq \TSet$ denotes the set of all initial transitions.
Note that $\location_0$ has no incoming transitions.
\begin{example}
  In the integer program of \cref{fig:ITS} which corresponds to the pseudocode of \cref{fig:pseudocode}, we omitted identity updates $\update(v) = v$ and guards where $\guard$ is $\true$.
  Here, $\VSet = \braced{x_1,\ldots,x_6}$, $\TVSet = \braced{y}$, $\LSet = \{\location_0, \dots, \location_3\}$, where $\location_0$ is the initial location, and $\TSet = \{ t_0, \ldots, t_5 \}$, where $t_3$ has non-linear arithmetic in its update and guard.
  Note that the loop in \eqref{WhileExample} corresponds to transition $t_3$.
  \begin{figure}[t]
    \centering
    \begin{tikzpicture}[->,>=stealth',shorten >=1pt,auto,node distance=3.5cm,semithick,initial text=$ $]
      \hspace*{-0.2cm}
      \node[state,initial] (q0) {$\location_0$};
      \node[state] (q1) [right of=q0,xshift=-1.75cm]{$\location_1$};
      \node[state] (q2) [right of=q1, node distance=4.5cm]{$\location_2$};
      \node[state] (q3) [below of=q1, node distance=2cm]{$\location_3$};
      \draw (q0) edge node [text width=3.5cm,align=center,above] {\footnotesize $t_0$} (q1);
      \draw (q2) edge[bend left=-25] node [text width=4cm,align=center,above]
        {$t_2$}
      (q1);
      \draw (q2) edge [loop right] node [text width=2.8cm,align=center]
        {
          \footnotesize $t_3:\guard = (x_2^2 - x_3^5 < x_1) \wedge x_2 \neq 0$ \\
          $
            \begin{array}{rcl}
              \update(x_1) & = & 3\cdot x_1 + x_3^2       \\
              \update(x_2) & = & -2 \cdot x_2             \\
              \update(x_3) & = & x_3                      \\
              \update(x_4) & = & 3\cdot x_4 + 2\cdot x_5  \\
              \update(x_5) & = & -5\cdot x_4 - 3\cdot x_5 \\
            \end{array}
          $
        }
      (q2);
      \draw (q1) edge[bend left=-10] node [text width=3cm, below, align=center]
        {
          \footnotesize \mbox{$t_1 : \guard = (0 < y \leq x_6)$}
          $
            \begin{array}{rclcrl}
              \update(x_1) & = & x_6 \quad & \update(x_2) & = & x_3     \\
              \update(x_3) & = & 2 \quad   & \update(x_4) & = & x_6     \\
              \update(x_5) & = & x_6 \quad & \update(x_6) & = & x_6 - y \\
            \end{array}
          $
        }
      (q2);
      \draw (q1) edge node [left, align=center]
        {
          \footnotesize $t_4: x_6 \leq 0$ \\
          $\update(x_6) = x_4$
        }
      (q3);
      \draw (q3) edge [loop left] node [text width=2.4cm,align=center]
        {
          $t_5: \guard = (x_6 > 0)$\\
          $\update(x_6) = x_6 - 1$
        }
      (q3);
    \end{tikzpicture}
    \caption{An Integer Program with Non-Linear Arithmetic}\label{fig:ITS}
  \end{figure}
\end{example}

To define the semantics of integer programs, an evaluation step moves from one configuration $(\location,\valuation)\in\LSet\times\Valuation$ to another configuration $(\location',\valuation')$ via a transition $(\location, \guard, \update, \location')$ where $\valuation(\guard)$ holds.
Here, we extend states by temporary variables, i.e., we consider states $\valuation,\valuation': \VSet\cup\TVSet\to\ZZ$, and $\valuation'$ is obtained by applying the update $\update$ on $\valuation$.
From now on, we fix a program $\IntProgram$.

\begin{definition}[Evaluation of Integer Programs]
  \label{def:Evaluation of Integer Programs}
  A \emph{configuration} is an element of $\LSet \times \Valuation$.
  For two configurations $(\location,\valuation)$ and $(\location',\valuation')$, and a transition $t = (\location_t,\guard,\update,\location_{t}')\in\TSet$, $(\location,\valuation)\rightarrow_t(\location',\valuation')$ is an \emph{evaluation}
  step by $t$ if
  \begin{itemize}
    \item $\location = \location_t$ and $\location' = \location_{t}'$,
    \item $\valuation(\guard) = \normalfont{\true}$, and
    \item for every program variable $v\in\VSet$ we have $\valuation(\update(v)) = \valuation'(v)$.
  \end{itemize}
  We denote the union of all relations $\to_t$ for $t \in \TSet$ by $\to_{\TSet}$.
  Whenever it is clear from the context, we omit the transition $t$ resp.\ the set $\TSet$ in the index.
  We also abbreviate $(\location_0,\valuation_0)\rightarrow(\location_1,\valuation_1) \cdots \rightarrow(\location_k,\valuation_k)$ by $(\location_0,\valuation_0)\rightarrow^k(\location_k,\valuation_k)$ and write $(\location,\valuation)\rightarrow^*(\location',\valuation')$ if $(\location,\valuation)\rightarrow^k(\location',\valuation')$ for some $k \geq 0$.
\end{definition}
Similar to \Cref{run example}, we denote states $\valuation\in\Valuation$ again as tuples $(\valuation(x_1),\ldots,\linebreak
  \valuation(x_6), \valuation(y)) \in \ZZ^7$.
Then for the program in \cref{fig:ITS}, we have the evaluation $(\location_0,(1,6,1,6,8,10,0)) \to_{t_0} (\location_1,(1,6,1,6,8,10,10)) \!\to_{t_1}\! (\location_2,(10,1,2,10,10,0,\linebreak
  10)) \to^9_{t_3}\!
  (\location_2,(12\cdot 3^9-2,(-2)^9,2,50,-80,0,0)) \to_{t_2} \circ \to_{t_4} (\location_3,(12\cdot 3^9-2,(-2)^9,2, 50,-80,50,0)) \to^{50}_{t_5} (\location_3,(12\cdot 3^9-2,(-2)^9,2,50,-80,0,0))$.
So temporary variables like $y$ can be updated arbitrarily in each step.
Note that this is not the longest evaluation as $\valuation(y) = 10$ was chosen non-deterministically such that $t_1$ could only be evaluated once.

\subsection{Runtime and Size Bounds for Integer Programs}
\label{sect:global_bounds}

In this subsection, we recapitulate preliminaries on (global) runtime and size bounds of integer programs from \cite{brockschmidt2016AnalyzingRuntimeSize,giesl2022ImprovingAutomaticComplexity}.
While up to now we defined runtime and size bounds for loops in \Cref{def:runtimeLoop,Size Bounds of Loops}, we now recapitulate the definition of these notions for general integer programs.
In \cref{sect:Runtime Bounds for Integer Programs,sect:Size Bounds for Integer Programs} we will show how previously computed runtime and size bounds of loops can be used for the computation of such bounds for general programs.
The \emph{runtime complexity} $\rc(\initial)$ of a program $\Program = \IntProgram$ corresponds to the length of a longest evaluation starting in the initial state $\initial$.
We call a program $\Program$ \emph{terminating} if the runtime complexity $\rc(\initial)$ is finite for all initial states $\initial$, i.e., if there is no infinite evaluation from any initial configuration $(\location_0,\initial)$.
\begin{definition}[Runtime Complexity]
  \label{def:runtimeComplexityIntegerProg}
  The \emph{runtime complexity} is $\rc\!:\Valuation\rightarrow\NNC$ with $\rc(\initial)=\sup\braced{k\in\NN\mid\exists (\location',\valuation').\, (\location_0,\initial)\rightarrow^k_\TSet(\location',\valuation')}$.
\end{definition}

Again, we measure the size of variables by their absolute values.
So if $\initial$ denotes the initial state, then $|\initial|$ maps every variable to its initial ``size'', i.e., its initial absolute value.
$\glo: \TSet \rightarrow \BoundSet$ is a \emph{global runtime bound} if for each transition $t$ and initial state $\initial\in \Valuation$, $\glo(t)$ evaluated in the state $|\initial|$ over-approximates the number of evaluation steps with $t$ in any run starting in the configuration $(\location_0,\valuation_0)$.
Let $\rightarrow^*_{\TSet} \circ \rightarrow_t$ denote the relation where arbitrary many evaluation steps are followed by a step with $t$.
\begin{definition}[Global Runtime Bound]
  \label{def:gloTimeBound}
  The function $\glo: \TSet \rightarrow \BoundSet$ is a \emph{global runtime bound}
  if for all $t \in \TSet$ and all states $\initial \in \Valuation$ we have $|\initial|(\glo(t)) \; \geq \; \sup \braced{ k \in \NN \mid \exists \, (\location', \valuation').\; (\location_0, \valuation_0) \; (\rightarrow^*_{\TSet} \circ \rightarrow_t)^k \; (\location', \valuation') }$.
\end{definition}
For the example in \Cref{fig:ITS}, a global runtime bound for $t_0$ and $t_4$ is given by $\glo(t_0) = \glo(t_4) = 1$.
Moreover, we have $\glo(t_1) = \glo(t_2) = x_6$, as $x_6$ is bounded from below by $t_1$'s guard $0 < x_6$, the value of $x_6$ decreases by the non-deterministic, positive value $y$ in $t_1$, and no transition increases $x_6$.
By adding the bounds for all transitions, a global runtime bound $\glo$ yields an upper bound on the program's runtime complexity.
So for all $\initial\in\Valuation$ we have $|\initial|(\sum_{t\in\TSet}\glo(t)) \geq \rc(\initial).$

Now we define size bounds for program variables $v$ after evaluating a transition $t$:
The bound $\Size(t,v)$ is a \emph{size bound} for $v\in\VSet$ w.r.t.\ $t\in\TSet$ if for any run starting in $\initial\in\Valuation$, $|\initial|(\Size(t, v))$ is greater or equal to the largest absolute value of $v$ after evaluating $t$.
\begin{definition}[Size Bound]
  \label{def:gloSizeBound}
  The function $\Size: (\TSet \times \VSet) \rightarrow \BoundSet$ is a (global) \emph{size bound}
  if for all $(t, v) \in \TSet \times \VSet$ and all states $\initial \in \Valuation$ we have $$|\initial|(\Size(t, v)) \geq \sup\braced{
      |\valuation'(v)| \mid \exists\, \location' \in \LSet.
      \; (\location_0, \valuation_0) \; (\rightarrow^*_\TSet \circ \rightarrow_t) \; (\location', \valuation')}.$$
\end{definition}

\begin{example}
  \label{ex:SizeBounds}
  As an example, we give size bounds for some of the variables w.r.t.\ the transitions $t_0, t_1, t_2$, and $t_4$ in \cref{fig:ITS} which can be inferred automatically by the incomplete approach \cite{brockschmidt2016AnalyzingRuntimeSize}.
  In \Cref{exa:SBLifting}, we will use the new technique based on the complete method for size bounds of loops to compute polynomial size bounds for $t_3$ as well.
  Since $t_0$ does not change any program variables, a size bound is $\Size(t_0, x_i) = x_i$ for all $1 \leq i \leq 6$.
  Note that the value of $x_6$ is never increased and is bounded from below by $0$ in any run through the program.
  Thus, $\Size(t_1, x_i) = x_6$ for $i\in\braced{1,4,5,6}$, $\Size(t_1, x_3) = 2$, and $\Size(t_1, x_2) = \max\braced{x_3,2}$.
  Later in \Cref{exa:SBLifting}, we will see that $\Size(t_3,x_4) = (2 + \sqrt{10})\cdot x_6$.
  As $t_2$ does not change any program variables, we have $\Size(t_2,x_4) = \max\braced{\Size(t_1,x_4),\Size(t_3,x_4)} = (2 + \sqrt{10})\cdot x_6$.
  Similarly, we have $\Size(t_4,x_6) = \max\braced{\Size(t_0,x_4),\Size(t_2,x_4)} = \max\braced{x_4, (2 + \sqrt{10})\cdot x_6}$.
\end{example}

\subsection{Lifting Runtime Bounds for Integer Programs}
\label{sect:Runtime Bounds for Integer Programs}

To infer global runtime and size bounds, we lift \emph{local} bounds (i.e., bounds which only hold for a subprogram with transitions $\TSet'\subseteq\TSet\setminus\TSet_0$) to global bounds for the \emph{full} program.
For the subprogram, one considers runs which start after evaluating an \emph{entry transition} of $\TSet'$.

\begin{definition}[Entry Transitions]
  Let $\emptyset\neq\TSet' \subseteq \TSet\setminus \TSet_0$.
  Its \emph{entry transitions} are $ \entry_{\TSet'} = \braced{t \mid t=(\location',\guard,\update,\location)\in\TSet\setminus\TSet'\wedge \text{ there is a transition } (\location,\wildcard,\wildcard,\wildcard)\in\TSet'}$.
\end{definition}
\noindent
So in \cref{fig:ITS}, we have $\entry_{\{t_3\}}	= \{ t_1 \}$ and $\entry_{\braced{t_5}}	= \{ t_4 \}$.

The approach is \emph{modular} since it computes local bounds for program parts separately.
In contrast to global runtime bounds, a \emph{local} runtime bound $\locsoldtwo{\TSet'_>}{\TSet'}: \entry_{\TSet'}
  \to \BoundSet$ only takes a subset $\TSet'$ into account.
A \emph{local run} is started by an entry transition $\pret\in\entry_{\TSet'} $ followed by transitions from $\TSet'$.
A \emph{local runtime bound} considers a subset $\TSet'_>\subseteq \TSet'$ and over-approximates the number of evaluation steps of any transition from $\TSet'_>$ in an arbitrary local run of the subprogram with the transitions $\TSet'$.
More precisely, for every $t \in \TSet'_>$, $\locsold{\TSet'_>}{\TSet'}{r}$ over-approximates the number of applications of $t$ in any run of $\TSet'$, if $\TSet'$ is entered via $\pret \in \entry_{\TSet'}$.
To indicate that these runtime bounds are ``local'' (i.e., that they can regard arbitrary subprograms $\TSet'$), we add the subscript ``loc''.
So a \emph{local runtime bound} is a function $\locsoldtwo{\TSet'_>}{\TSet'}:\entry_{\TSet'}\to\BoundSet$ whereas a global runtime bound is a function $\glo: \TSet\to\BoundSet$.
In the case where $\TSet' = \TSet$, $\sum_{t_0 \in \TSet_0}\locsold{\TSet'_>}{\TSet}{t_0}$ corresponds to a global runtime bound $\glo(t)$ for any $t\in\TSet'_>$.
To highlight that $\locsold{\TSet'_>}{\TSet'}{\pret}$ is a bound on the number of evaluation steps of transitions from $\TSet'_>$ after entering the subprogram $\TSet'$ via the transition $\pret$, we often write $\locs{\TSet'_>}{\TSet'}{\pret}$ instead of $\locsold{\TSet'_>}{\TSet'}{\pret}$.
In contrast to \cite{brockschmidt2016AnalyzingRuntimeSize,giesl2022ImprovingAutomaticComplexity}, we now allow different local bounds for different entry transitions $r$.\footnote{As shown in \cite{lommen2022AutomaticComplexityAnalysis}, this can indeed lead to a smaller asymptotic bound for the whole program.
  Moreover, different local runtime bounds for different entry transitions will also be needed to handle simple cycles which can be entered via different entry transitions, see \Cref{entryTransitionArgument}.}
However, local runtime bounds do not consider how often an entry transition from $\entry_{\TSet'}$ is evaluated or how large a variable is when we evaluate an entry transition.
\begin{definition}[Local Runtime Bound]
  \label{def:locUpperTimeBound}
  Let $\emptyset\neq\TSet'_>\subseteq \TSet'\subseteq\TSet\setminus \TSet_0$.
  The function $\locs{\TSet'_>}{\TSet'}{}: \entry_{\TSet'} \to \BoundSet$ is a \emph{local runtime bound} for $\TSet'_>$ w.r.t.\ $\TSet'$ if for all $t \in \TSet'_>$, all $\pret\in\entry_{\TSet'}$ with $r = (\location', \wildcard,\wildcard,\location)$, and all $\valuation \in \Valuation$ we have
  \begin{align*}
    |\valuation|( & \locs{\TSet'_>}{\TSet'}{\pret}) \geq \\
                  & \sup \braced{ k \in \NN \mid \exists\, \valuation',
      (\location'', \valuation''). \; (\location', \valuation') \to_{\pret} \, (\location, \valuation) \; (\to_{\TSet'}^* \circ \to_t)^k \; (\location'', \valuation'')}.
  \end{align*}

\end{definition}
To simplify the presentation, in the definition above, we consider every possible configuration $(\location',\valuation')$ instead of only configurations that can be reached from the initial location $\location_0$.

\begin{example}
  \label{ex:local runtime bounds}
  In \cref{fig:ITS}, a local runtime bound for $\TSet'_> = \TSet' = \braced{t_5}$ is $\locs{\braced{t_5}}{\braced{t_5}}{t_4} = x_6$.
\end{example}

Local runtime bounds can often be inferred automatically by approaches based on ranking functions (see, e.g., \cite{brockschmidt2016AnalyzingRuntimeSize}) or by the complete technique for terminating \emph{prs}-loops from \Cref{thm:completeness}.
The latter is applicable whenever a transition of the program \emph{corresponds} to a \emph{prs}-loop.

\begin{definition}[Correspondence between Loops and Transitions]
  \label{def:correspondence_loops_transitions}
  Let $t = (\location, \guard, \update,\location)$ be a transition with $\guard\in\FormulaSet(\VSet')$ for some program variables $\VSet'\subseteq\VSet$ such that $\update(v) \in \ZZ[\VSet']$ for all $v \in \VSet'$.
  The transition $t$ \emph{corresponds} to a loop $(\guard', \update')$ with $\guard'\in\FormulaSet(\braced{x_1,\ldots,x_d})$ and $\update': \braced{x_1,\ldots,x_d}\to\ZZ[x_1,\ldots,x_d]$ via the variable renaming $\pi: \braced{x_1,\ldots,x_d}\to \VSet'$ if $\guard$ is $\pi(\guard')$ and for all $1\leq i \leq d$ we have $\update(\pi(x_i)) = \pi(\update'(x_i))$.
\end{definition}

\begin{example}
  \label{loopAsTransition}
  Consider the loop \eqref{WhileExample} again, i.e., the loop $(\guard',\update')$ where $\guard'$ is $x_2^2 - x_3^5 < x_1 \land x_2 \neq 0$ and $\update'(x_1) = 3\cdot x_1 + x_3^2$, $\update'(x_2) = -2 \cdot x_2$, $\update'(x_3) = x_3$, $\update'(x_4) = 3\cdot x_4 + 2\cdot x_5$, and $\update'(x_5) = -5\cdot x_4 -3\cdot x_5$.
  The transition $t_3$ from the program of \Cref{fig:ITS} corresponds to this loop using the variable renaming with $\pi(x_i) = x_i$ for all $1 \leq i \leq 5$.
\end{example}

Note that \Cref{def:correspondence_loops_transitions} allows us to consider a subset $\VSet'$ of the program variables $\VSet$.
This is useful to remove variables which would otherwise destroy properties like the \emph{twn}-property.
For example, if we only consider the variables $\braced{x_1,x_2,x_3}$ instead of $\braced{x_1,\dots,x_5}$ as in \Cref{loopAsTransition}, then we directly obtain a \emph{twn}-loop instead of the \emph{prs}-loop \eqref{WhileExample}.

The following corollary shows how to obtain local runtime bounds from runtime bounds of loops.
\begin{corollary}[Local Bounds from Runtime Bounds of Loops]
  \label{lem:complexityLoops}
  Let $L$ be a loop with the runtime bound $\run$ (as in \Cref{def:runtimeLoop}) and let the transition $t_L$ correspond to $L$ via the variable renaming $\pi$.
  Then $\locs{\braced{t_L}}{\braced{t_L}}{r} = \pi(\run)$ for all $\pret \in \entry_{\TSet'}$ is a local runtime bound for $\braced{t_L} = \TSet'_>= \TSet'$ in the program $\Program$ (as in \Cref{def:locUpperTimeBound}).
\end{corollary}

\begin{example}
  \label{exa:LocalBoundsLoops}
  A runtime bound for the loop $L = (\guard',\update')$ from \eqref{WhileExample} is $\run = 4 + 3\cdot\log_{2}(2\cdot x_1 + x_3^2 + 2\cdot x_3^5)$, see \Cref{runtimeBoundWhileExample}.
  Hence, by \Cref{lem:complexityLoops}, we can use $\run$ as a local runtime bound for $\TSet'_> = \TSet' = \braced{t_3}$ in the program of \Cref{fig:ITS}, i.e., we obtain $$\locs{\braced{t_3}}{\braced{t_3}}{t_1} = 4 + 3\cdot\log_{2}(2\cdot x_1 + x_3^2 + 2\cdot x_3^5).$$
\end{example}

As mentioned, to improve size and runtime bounds repeatedly, we treat the strongly connected components (SCCs)\footnote{As usual, a graph is \emph{strongly connected} if there is a path from every node to every other node.
  A \emph{strongly connected component} is a maximal strongly connected subgraph.
}
of the program in topological order such that improved bounds for previous transitions are already available when handling the next SCC.
We first try to infer local runtime bounds by multiphase-linear ranking functions (see \cite{giesl2022ImprovingAutomaticComplexity} which also contains a heuristic for choosing $\TSet'_>$ and $\TSet'$ when using ranking functions).
If ranking functions do not yield finite local bounds for all transitions of the SCC, then we apply the complete technique from \Cref{sect:Runtime Bounds for PRS-Loops} on the remaining unbounded transitions, if possible.
(At the end of the current subsection, we will show how our implementation chooses $\TSet'_>$ and $\TSet'$ in that case, see \Cref{alg:twncycleSym,lem:twncycleSym}.)

Afterwards, we use size bounds $\Size(\pret,v)$ to lift previously computed local runtime bounds $\locs{\TSet'_>}{\TSet'}{\pret}$ to global runtime bounds.
To this end, the following \cref{thm:time-bound} generalizes the approach of \cite{brockschmidt2016AnalyzingRuntimeSize,giesl2022ImprovingAutomaticComplexity}.
Each local run is started by an entry transition $\pret$.
Hence, we can use an already computed global runtime bound $\glo(\pret)$ to over-approximate the number of times that such a local run is started.
To over-approximate the size of each variable $v$ when entering the local run, we instantiate it by the size bound $\Size(\pret,v)$.
So size bounds on previous transitions are needed to compute runtime bounds, and similarly, runtime bounds are needed to compute size bounds (both in the technique of \Cref{thm:size_bounds_closed_form} for size bounds of loops, and in the incomplete approach of \cite{brockschmidt2016AnalyzingRuntimeSize} for size bounds of integer programs, see \Cref{sect:Size Bounds for Integer Programs}).
Here, ``$b \; [v/ \Size(\pret,v) \mid v \in \VSet]$'' denotes the expression that results from the bound $b$ by replacing every program variable $v$ by $\Size(\pret,v)$.
The weak monotonic increase of $b$ ensures that the over-approximation of the variables $v$ in $b$ by $\Size(\pret,v)$ indeed also leads to an over-approximation of $b$.
The analysis starts with an \emph{initial} runtime bound $\glo$ and an \emph{initial} size bound $\Size$ which map all arguments to $\omega$, except for the transitions $t$ which do not occur in cycles of $\TSet$, where $\glo(t) = 1$.
Afterwards, $\glo$ and $\Size$ are refined repeatedly, where we alternate between computing runtime and size bounds.

\begin{restatable}[Computing Global Runtime Bounds]{theorem}{GlobalRB}
  \label{thm:time-bound}
  Let $\glo$ be a global runtime bound, $\Size$ be a size bound, and
  $\emptyset\neq \TSet'_> \subseteq \TSet' \subseteq \TSet \setminus \TSet_0$.
  Moreover, let $\locsoldtwo{\TSet'_>}{\TSet}$ be a local runtime bound for $\TSet'_>$ w.r.t.\ $\TSet'$.
  Then $\glopr$ is also a global runtime bound, where for all $t \in \TSet$ we define:
  \[
    \glopr(t)\!=\!		\left\{
    \begin{array}{ll}
      \!\glo(t),                                                                                                                              & \hspace*{-.5cm}\text{if $t\!\in\!\TSet\!\setminus\!\TSet'_>$} \\
      \!\sum_{\pret \in \entry_{\TSet'}} \glo(\pret)\cdot (\locs{\TSet'_>}{\TSet'}{\pret} \left[v/\Size(\pret,v) \mid v\!\in\!\VSet \right]), & \text{if $t\!\in\!\TSet'_>$}
    \end{array}
    \right.
  \]
\end{restatable}
\makeproof{thm:time-bound}{
  \GlobalRB*
  \begin{myproof}
    We show that for all $t\in\TSet$ and all $\valuation_0 \in \Valuation$ we have
    \begin{align*}
      |\initial|(\glopr(t)) \geq \sup \braced{ k \in \NN \mid \exists \, (\location', \valuation').\; (\location_0, \valuation_0) \; (\rightarrow^*_{\TSet} \circ \rightarrow_t)^k \; (\location', \valuation') }.\label{eq:sound_timebound}
    \end{align*}
    The case $t \notin \TSet'_>$ is trivial, since then we have $\glopr(t) = \glo(t)$ and $\glo$ is a global runtime bound.
    For $t \in \TSet'_>$, let $(\location_0, \valuation_0) \, (\to^*_{\TSet} \circ \to_{t})^k \, (\location', \valuation')$ and we have to show $|\initial|(\glopr(t)) \geq k$.

    If $k = 0$, then we clearly have $|\initial|(\glopr(t)) \geq 0 = k$.
    Hence, we consider $k > 0$.
    We represent the evaluation as follows for numbers $\tilde{k}_i \geq 0$ and $k_i' \geq 1$:
    \begin{alignat*}{3}
      (\prel_0, \prestate_0)            & (\to^{\tilde{k}_0}_{\TSet \setminus \TSet'} \circ \to_{\pret_1})          &  & (\actl_1, \actstate_1)        &  & \to^{k_1'}_{\TSet'} \\
      (\prel_1, \prestate_1)            & (\to^{\tilde{k}_1}_{\TSet \setminus \TSet'} \circ \to_{\pret_2})          &  & (\actl_2, \actstate_2)        &  & \to^{k_2'}_{\TSet'} \\
                                        & \hspace*{1.5cm}\vdots                                                     &  &                               &  & \\
      (\prel_{m-1}, \prestate_{m-1}) \, & (\to^{\tilde{k}_{m-1}}_{\TSet \setminus \TSet'} \circ \to_{\pret_{m}}) \, &  & (\actl_{m}, \actstate_{m}) \, &  & \to^{k_{m}'}_{\TSet'} (\prel_m, \prestate_m)
    \end{alignat*}
    So for the evaluation steps from $(\prel_i, \prestate_i)$ to $(\actl_{i+1}, \actstate_{i+1})$ we only use transitions from $\TSet\setminus\TSet'$, and for the evaluation steps from $(\actl_i, \actstate_i)$ to $(\prel_i, \prestate_i)$ we only use transitions from $\TSet'$.
    Thus, $t$ can only occur in the following finite sequences of evaluation steps:
    \begin{equation}
      \label{SubsetEvaluation}
      (\actl_i, \actstate_i) \to_{\TSet'} (\actl_{i,1}, \actstate_{i,1}) \to_{\TSet'} \ldots \to_{\TSet'} (\actl_{i,k_i'-1}, \actstate_{i,k_i'-1}) \to_{\TSet'} (\prel_i, \prestate_i).
    \end{equation}
    For every $1 \leq i \leq m$, let $k_i \leq k_i'$ be the number of times that $t$ is used in the evaluation \eqref{SubsetEvaluation}.
    Clearly, we have
    \begin{equation}
      \label{SumDecreasingTransition}
      \sum_{i=1}^{m} k_i = k.
    \end{equation}

    As $\Size$ is a size bound, we have $|\initial| (\Size(\pret_i, v)) \geq |\actstate_i(v)|$ for all $v\in\VSet$.
    Hence, by the definition of local runtime bounds and as bounds are weakly monotonically increasing functions, we can conclude that
    \begin{equation}
      \label{gloThmHelp}
      |\initial| (\locs{\TSet'_>}{\TSet'}{\pret_i} \left[v/\Size(\pret_i,v) \mid v \in
        \VSet \right]) \; \geq \; |\actstate_i|(\locs{\TSet'_>}{\TSet'}{\pret_i})
      \; \geq \; k_i.
    \end{equation}

    Finally, we need to analyze how often such evaluations $(\actl_i, \actstate_i) \to^*_{\TSet'} (\prel_i, \prestate_i)$ can occur.
    Every entry transition $\pret_i\in\entry_{\TSet'}$ can occur at most $|\initial| (\glo(\pret_i))$ times in the complete evaluation, as $\glo$ is a global runtime bound.
    Thus, we have
    \begin{align*}
      |\initial| (\glopr(t))
      {} = {}
                 & \sum_{\pret \in \entry_{\TSet'}}|\initial|(\glo(\pret))\cdot
      |\initial|(\locs{\TSet'_>}{\TSet'}{\pret} \left[v/\Size(\pret,v) \mid v \in \VSet \right]) \\
      {} \geq {} & \sum_{i=1}^m |\initial| (\locs{\TSet'_>}{\TSet'}{\pret_i}
      \left[v/\Size(\pret_i,v) \mid v \in \VSet \right] ) \\
      {} \geq {} & \sum_{i=1}^m k_i \tag{by \eqref{gloThmHelp}} \\
      {} = {}    & k \tag{by \eqref{SumDecreasingTransition}}
    \end{align*} \qed
  \end{myproof}
}

Note that in \Cref{thm:time-bound} we cannot replace the sum over all $\pret \in \entry_{\TSet'}$ by taking the maximum over all $\pret \in \entry_{\TSet'}$.
The reason is that during a run of the overall program, the subprogram $\TSet'$ may be entered several times via different transitions from $\entry_{\TSet'}$, and $\glo(t)$ has to over-approximate the sum of all applications of transitions from $\TSet'_>$ during the full run.

\begin{example}
  \label{ex:global_runtime_bound}
  Based on the local runtime bounds in \Cref{ex:local runtime bounds,exa:LocalBoundsLoops}, we can compute the remaining global runtime bounds for the example.
  See \Cref{ex:SizeBounds} for the size bounds of $t_1$ and $t_4$.
  By using $\TSet' = \TSet'_{>} = \{ t_3\}$, we obtain
  \begin{align*}
    \glo(t_3) & = \glo(t_1)\cdot	(4 + 3\cdot\log_{2}(2\cdot x_1 + x_3^2 + 2\cdot x_3^5)\left[v/\Size(t_1,v) \mid v\in\VSet \right]) \\
              & = x_6 \cdot	(4 + 3\cdot\log_{2}(2\cdot x_6 + 68))
  \end{align*}
  as $\entry_{\{t_3\}}	= \{ t_1 \}$, $\glo(t_1) = x_6$, $\Size(t_1,x_3) = 2$, and $\Size(t_1,x_1) = x_6$.
  Similarly, when using $\TSet' = \TSet'_{>} = \{ t_5\}$, we have
  \begin{align*}
    \glo(t_5) & = \glo(t_4)\cdot(x_6\left[v/\Size(t_4,v) \mid v\in\VSet \right]) \\
              & = \max\braced{x_4, (2 + \sqrt{10})\cdot x_6}
  \end{align*}
  since $\entry_{\{t_5\}}	= \{ t_4 \}$, $\glo(t_4) = 1$, and $\Size(t_4,x_6) = \max\braced{x_4, (2 + \sqrt{10})\cdot x_6}$.
  Thus, overall	we have a linearithmic runtime bound $\sum_{1 \leq i \leq 5}\glo(t_i)$, i.e., a bound $\bound$ with $\bound[v/n \mid v \in \VSet] \in \mathcal{O}(n \cdot \log(n))$.
  Note that it is due to the size and runtime bound technique from \Cref{sect:loops} that we obtain a linearithmic runtime bound in this example in contrast to our conference papers \cite{lommen2022AutomaticComplexityAnalysis,lommen2023TargetingCompletenessUsing} which would infer a quadratic runtime bound.
  Furthermore, to the best of our knowledge, all other state-of-the-art tools fail to infer polynomial size or finite runtime bounds for this example.
  Similarly, if one modifies $t_4$ such that $x_6$ is set to $x_1$ instead of $x_4$, then the approach yields a polynomial runtime bound (see \Cref{exa:SBLifting}), whereas \cite{lommen2022AutomaticComplexityAnalysis,lommen2023TargetingCompletenessUsing} infers an exponential runtime bound for the full program.
  The reason is that \cite{lommen2022AutomaticComplexityAnalysis} cannot infer logarithmic runtime bounds for the transitions.
  Hence, \cite{lommen2023TargetingCompletenessUsing} does not obtain polynomial size bounds for $x_1$.
\end{example}

\paragraph{Local Runtime Bounds via Simple Cycles:}
To increase its applicability, we now show that the technique for the computation of local runtime bounds via \emph{prs}-loops can also be applied to infer runtime bounds for larger cycles in programs instead of just self-loops.
To this end, we have to extend the notion of \emph{chaining} from loops (see \Cref{def:chaining}) to transitions that occur in a cycle.\footnote{The chaining of a loop $L$ in \Cref{def:chaining} corresponds to $p-1$ chaining steps of a transition $t$ with itself via \Cref{Chaining Transitions}, i.e., to $t \chain \ldots \chain t$.}

\begin{definition}[Chaining Transitions]
  \label{Chaining Transitions}
  Let $t_1,\ldots,t_n \in \TSet$ be transitions with updates $\update_i:\VSet \to \ZZ[\VSet]$ (i.e., their updates do not introduce temporary variables), where $t_i = (\location_i, \guard_i, \update_i, \location_{i+1})$ for all $1 \leq i \leq n$.
  So the target location of $t_i$ is the start location of $t_{i+1}$.
  Then the transition $t_1 \chain \ldots \chain t_n = (\location_1, \guard, \update, \location_{n+1})$ results from \emph{chaining} $t_1,\ldots,t_n$ where
  \[
    \begin{array}{rcl}
      \guard     & = & \guard_1 \, \land \, \update_1(\guard_2) \, \land \, \update_1(\update_2(\guard_3)) \, \land \, \ldots \, \land \, \update_{1}(\ldots\update_{n-1}(\guard_n)\ldots) \\
      \update(v) & = & \update_1(\ldots\update_n(v)\ldots) \text{ for all $v \in \VSet$, i.e., $\update = \update_1 \circ \ldots \circ \update_n$.}
    \end{array}
  \]
\end{definition}

To determine whether a cycle of transitions corresponds to a \emph{prs}-loop, for every entry transition of the cycle, we \emph{chain} the transitions of the cycle, starting with the transition which follows the entry transition.
In this way, we obtain loops consisting of a single transition.
If the chained transition corresponds to a terminating \emph{prs}-loop, we can apply \cref{thm:completeness,lem:complexityLoops} to compute a local runtime bound.
Any local bound on the chained transition is also a bound on each of the original transitions.\footnote{\label{entryTransitionArgument}
  This is sufficient for the improved definition of local bounds in \Cref{def:locUpperTimeBound} where in contrast to \cite{brockschmidt2016AnalyzingRuntimeSize,giesl2022ImprovingAutomaticComplexity} we do not require a bound on the \emph{sum} but only on \emph{each} transition in the considered set $\TSet'$.
  Moreover, here we benefit from the extension to compute individual local bounds for different entry transitions.}

Note that this replacement of a cycle by a self-loop which results from chaining its transitions is only sound for \emph{simple} cycles.
For such cycles, each iteration through the cycle can only be done in a unique way.
So the cycle must not have any subcycles and there also must not be any indeterminisms concerning the next transition to be taken.
The following definition introduces simple cycles formally.

\begin{definition}[Simple Cycle]
  \label{def:simpleCycle}
  $\mathcal{C}= \{t_1,\ldots,t_n\}\subset\TSet$ is a \emph{simple cycle with the sequence} $t_1,\ldots,t_n$ if the updates of the transitions in $\mathcal{C}$ only map program variables to polynomials without temporary variables, and there are pairwise different locations $\location_1,\ldots,\location_n$ such that $t_i = (\location_i, \wildcard, \wildcard, \location_{i+1})$ for $1 \leq i \leq n-1$ and $t_n = (\location_n, \wildcard, \wildcard, \location_1)$.
\end{definition}
\noindent
The above definition ensures that if there is an evaluation with $\to_{t_i} \circ \to^*_{\mathcal{C}\setminus\{t_i\}} \circ \to_{t_i}$, then the steps with $\to^*_{\mathcal{C}\setminus\{t_i\}}$ have the form $\to_{t_{i+1}} \circ \ldots \circ \to_{t_n} \circ \to_{t_1} \circ \ldots \circ \to_{t_{i-1}}$.

By \cref{lem:complexityLoops}, we obtain a bound on the number of evaluations of the \emph{complete cycle}.
However, we also have to consider a \emph{partial execution} which stops before traversing the full cycle.
Therefore, we have to increase every local runtime bound by 1.
To see this, consider a simple cycle $\{t_1,\ldots,t_n\}$ where $t_1 \star \dots \star t_n$ corresponds to the loop $L$.
Now, an evaluation of the simple cycle looks like this for suitable configurations $c_i$:
\[
  \begin{array}{l@{\qquad}ll}
    \underbrace{c_0 \to_{t_1} \dots \to_{t_n} c_{n}}_{\text{1.\ iteration of
    $L$}}                                                &
    \underbrace{\to_{t_1} \dots \to_{t_n} c_{2\cdot n}}_{\text{2.\ iteration of
    $L$}}                                                & \dots \\[.7cm]
    \underbrace{c_{m\cdot n} \to_{t_1} \dots \to_{t_n} c_{(m + 1)\cdot
    n}}_{\text{$L$ terminates after $m + 1$ iterations}} &
    \underbrace{\to_{t_1} \dots
      \to_{t_i}
      c_{(m+1) \cdot n + i}}_{\text{only partial evaluation of cycle (i.e., $i < n$)}}
  \end{array}
\]
If we want to count the number of executions of a transition $t_j$ in such an
evaluation, then we can use a runtime bound on $L$ (which is $\geq m + 1$) and increase this by
one for the potential application of $t_j$ in the end.

\begin{figure}[t]
  \begin{algorithm}[H]
    \DontPrintSemicolon
    \caption{Algorithm to Compute Local Runtime Bounds for Simple Cycles}\label{alg:twncycleSym}
    \Input{A program $\IntProgram$ and a simple cycle $\mathcal{C}=
        \{t_1,\ldots,t_n\}\subset\TSet$}
    \Output{A local runtime bound $\locsoldtwo{\mathcal{C}}{\mathcal{C}}$ for $\mathcal{C} = \TSet'_>= \TSet'$}
    Initialize $\locsoldtwo{\mathcal{C}}{\mathcal{C}}$: $\locs{\mathcal{C}}{\mathcal{C}}{\pret} \gets \omega$ for all	$\pret\in\entry_{\mathcal{C}}$. \label{alg:initial}\;
    \ForAll{$\pret \in\entry_{\mathcal{C}}$ \label{alg:forloop}} {
      Let $i \in \{1,\ldots,n\}$ such that $\pret$'s target location is the start location $\location_i$ of $t_i$. \label{alg:startLocation}\;
      Let $t = t_i \chain \ldots \chain t_n \chain t_1 \chain \ldots \chain t_{i-1}$. \label{alg:chaining} \;
      \If{$t$ corresponds to a loop with runtime bound $\run$ via a variable renaming $\pi$
        \label{alg:if}
      }
      {
        Set	$\locs{\mathcal{C}}{\mathcal{C}}{\pret} \gets
          1 + \pi(\run)$
        \label{alg:then}\;
      }
    }
    \return local runtime bound $\locsoldtwo{\mathcal{C}}{\mathcal{C}}$
  \end{algorithm}
  \caption{Algorithm Computing Local Runtime Bounds for Simple Cycles\label{fig:twncycleSym}}
  \vspace*{-.4cm}
\end{figure}

\Cref{alg:twncycleSym} in \Cref{fig:twncycleSym} describes how to compute a local runtime bound for a simple cycle $\mathcal{C} = \{ t_1, \ldots, t_n \}$ as above.
In the loop of \Cref{alg:forloop}, we iterate over all entry transitions $\pret$ of $\mathcal{C}$.
If $\pret$ reaches the transition $t_i$, then in \Cref{alg:startLocation,alg:chaining} we chain $t_i \chain \ldots \chain t_n \chain t_1 \chain \ldots \chain t_{i-1}$ which corresponds to one iteration of the cycle starting in $t_i$.
If a suitable renaming (and thus also reordering) of the variables turns the chained transition into a \emph{prs}-loop, then we use \cref{thm:completeness,lem:complexityLoops} to compute a local runtime bound $\locs{\mathcal{C}}{\mathcal{C}}{\pret}$ in \Cref{alg:if,alg:then}.
In practice, to use the complete technique for runtime bounds of \emph{prs}-loops for a transition $t$ in a program, the tool \KoAT{} searches for those simple cycles that contain $t$ and where the chained cycle is a \emph{prs}-loop or can be transformed into a \emph{prs}-loop by the preprocessing step for unsolvable loops.
Among those cycles it chooses the one with the smallest runtime bounds for its entry transitions.
\begin{restatable}[Correctness of \Cref{alg:twncycleSym}]{lemma}{twncycleSym}
  \label{lem:twncycleSym}
  Let $\mathcal{C}\subset\TSet$ be a simple cycle in the program $\Program$.
  Then the result $\locsoldtwo{\mathcal{C}}{\mathcal{C}}: \entry_{\mathcal{C}} \to \BoundSet$ of \Cref{alg:twncycleSym} is a local runtime bound for $\mathcal{C} = \TSet'_> = \TSet'$.
\end{restatable}
\makeproof{lem:twncycleSym}{
  \twncycleSym*
  \begin{myproof}
    Let $\pret\in\entry_{\mathcal{C}}$.
    If $\locs{\mathcal{C}}{\mathcal{C}}{\pret} = \omega$, then the claim is trivial.
    Otherwise, let $\location_i$ be the target location of $\pret$, let $1 \leq j \leq n$, and let $\valuation\in\Valuation$.
    By \cref{lem:complexityLoops}, $\pi(\run)$ is a local runtime bound for $\{ t \}
      = \TSet_{>}' = \TSet'$ when using the entry transition $r$. Hence, we obtain
    \[
      \begin{array}{cl}
             & |\valuation|(\locs{\mathcal{C}}{\mathcal{C}}{\pret})                                                                                                                                                                                                                    \\
        =    & 1 + |\valuation|(\pi(\run))                                                                                                                                                                                                                                             \\
        \geq & 1 + \sup \{ k \in \NN \mid \exists\, \initial, \valuation'. \; (\location_0, \initial) \rightarrow_{\TSet}^* \circ \rightarrow_{\pret} (\location_i, \valuation) \to^k_{t_i \chain \ldots \chain t_n \chain t_1 \chain \ldots \chain t_{i-1}}
        (\location_i, \valuation') \}                                                                                                                                                                                                                                                  \\
        =    & \sup \{ k +1 \mid \exists\, \initial, \valuation'. \;                                                                                                                                                                                                                   \\
             & \;(\location_0, \initial) \rightarrow_{\TSet}^* \circ \rightarrow_{\pret} \; (\location_i, \valuation) \; (\to_{t_i} \circ ... \circ \to_{t_{j-1}} \circ \to_{t_j} \circ \to_{t_{j+1}} \circ ... \to_{t_{i-1}})^k \; (\location_i, \valuation') \}                      \\
        \geq & \sup \{ k \in \NN \mid \exists\,\initial, (\location',\valuation'). \; (\location_0, \initial) \rightarrow_{\TSet}^* \circ \rightarrow_{\pret} \; (\location_i, \valuation) \; (\to^*_{\mathcal{C}\setminus\{t_j\}} \circ \to_{t_j})^k \; (\location', \valuation') \},
      \end{array}
    \]
    since $\mathcal{C}$ is a simple cycle. \qed
  \end{myproof}
}

\begin{example}
  \label{exa:simpleCycleTime}
  Consider the integer program which results from replacing $t_3 =
    (\location_2,\guard,\update,\location_2)$ in \Cref{fig:ITS} by two new transitions
  $t_{3a} = (\location_2, \guard,\update_{a}, \location_2')$ and $t_{3b} = (\location_2',
    \true, \update_{b}, \location_2)$ and the new location $\location_2'$ where
  $\update_{a}(v) = \update(v)$ for $v \in \{ x_4, x_5 \}$, $\update_{b}(v) = \update(v)$
  for $v \in \{ x_1,
    x_2, x_3, x_6 \}$, and $\update_{a}$ resp.\ $\update_{b}$ are the identity on the remaining variables.
  If we now consider the simple cycle $\mathcal{C} = \braced{t_{3a},t_{3b}}$, then
  \Cref{lem:twncycleSym} yields a local time bound $\locs{\mathcal{C}}{\mathcal{C}}{t_1} =
    \locs{\braced{t_3}}{\braced{t_3}}{t_1} + 1 = 5 + 3\cdot\log_{2}(2\cdot x_1 + x_3^2 +
    2\cdot x_3^5)$ (see \Cref{exa:LocalBoundsLoops}) for $\mathcal{C} = \TSet'_> = \TSet'$
  since $t_{3a} \chain t_{3b} = t_{3}$.
  Similarly as in \Cref{ex:global_runtime_bound}, we can now lift this local runtime bound to a global runtime bound.
\end{example}

\subsection{Lifting Size Bounds for Integer Programs}
\label{sect:Size Bounds for Integer Programs}

In this section, we incorporate the technique to compute size bounds based on closed forms from \Cref{sect:loops_size_bounds} into the incomplete setting of general integer programs.
To this end, we introduce \emph{local size bounds}
$\locsize{t',\TSet'}(v)$ which give a bound on the size of the program variable $v$ after evaluating the transition $t'$ in the subprogram $\TSet'$.
Again, to simplify the formalism, in the following definition we consider every possible configuration $(\location,\valuation)$ and not only configurations which are reachable from the initial location $\location_0$.

\begin{definition}[Local Size Bounds]
  \label{def:local_size_bounds}
  Let $\emptyset\neq\TSet'\subseteq\TSet\setminus\TSet_0$ and $t'\in\TSet'$.
  $\locsize{t',\TSet'}:\VSet\to\BoundSet$ is a \emph{local size bound} for $t'$ w.r.t.\ $\TSet'$ if for all $v\in\VSet$ and all $\valuation\in\Valuation$:
  \begin{align*}
    |\valuation| (\locsize{t',\TSet'}(v)) \geq \sup\{ & |\valuation'(v)|\mid \exists \location' \in \LSet, (\wildcard,\wildcard,\wildcard,\location)\in\entry_{\TSet'}. \\
                                                      & (\location,\valuation)\; (\to^*_{\TSet'}\circ \to_{t'}) \; (\location',\valuation')\}.
  \end{align*}
\end{definition}

\Cref{thm:lift_size_bounds} below yields a \emph{modular} procedure to infer (global) size bounds from previously computed local size bounds.
A local size bound for a transition $t'$ w.r.t.\ a subprogram $\TSet'\subseteq\TSet\setminus\TSet_0$ is lifted to a global one by inserting size bounds for all entry transitions.
Again, this is possible because we only use weakly monotonically increasing functions as bounds.
As before, ``$b\left[v/p_v \mid v\in\VSet \right]$'' denotes the bound which results from replacing every program variable $v$ by $p_v$ in the bound $b$.

\begin{restatable}[Computing Global Size Bounds]{theorem}{GlobalSB}
  \label{thm:lift_size_bounds}
  Let $\emptyset\neq\TSet'\subseteq\TSet\setminus\TSet_0$, let $\locsize{t',\TSet'}$ be a local size bound for a transition $t'$ w.r.t.\ $\TSet'$ and let $\Size: (\TSet\times\VSet)\to\BoundSet$ be a size bound for $\Program$.
  Let \[\Size'(t',x) = 	\max \braced{\locsize{t',\TSet'}(x) \left[v/\Size(\pret,v) \mid v\in\VSet \right]
      \mid
      \pret \in \entry_{\TSet'}
    }
  \] and 	$\Size'(t,x) = \Size(t,x)$ for all $t' \neq t$.
  Then $\Size'$ is also a size bound for $\Program$.
\end{restatable}
\makeproof{thm:lift_size_bounds}{
  \GlobalSB*
  \begin{myproof}
    We show that for all $(t,x)\in\TSet\times\VSet$ and all $\initial\in\Valuation$ we
    have \[|\initial|(\Size'(t,x)) \geq \sup\	\braced{ |\valuation'(x)| \mid \exists\,
        \location' \in \LSet.\;(\location_0, \valuation_0) \; (\to^*_\TSet \circ \to_t) \;
        (\location', \valuation')}.\]

    The case $t' \neq t$ is trivial, as we have $\Size'(t,x) = \Size(t,x)$ for all $x\in\VSet$ and $\Size$ is a size bound for the program $\Program$.
    Otherwise, any evaluation with $\to^*_\TSet \circ \to_{t'}$ starting in an initial configuration $(\location_0, \valuation_0)$ has the following form: $$(\location_0,\valuation_0)\, \to^*_\TSet \circ \to_{\pret'}\, (\location,\valuation)\,\to_{\TSet'}^* \circ\to_{t'} \,(\location', \valuation') \quad \text{for some }
      \pret'\in\entry_{\TSet'}.$$

    Thus, we have to prove for all such evaluations and all program variables $x\in\VSet$ that \[|\valuation_0|(\Size'(t',x)) \geq |\valuation'(x)|\] holds:
    \[
      \begin{array}{rcl}
        |\valuation_0|(\Size'(t',x)) & =    & |\valuation_0|\left(\max_{\pret \in
          \entry_{\TSet'}} \braced{\locsize{t',\TSet'}(x) \left[v/\Size(\pret,v)
        \mid v\in\VSet \right]}\right)                                                                                                                                                                                                                                                   \\
                                     & \geq & |\valuation_0|\left(\locsize{t',\TSet'}(x) \left[v/\Size(\pret',v) \mid v\in\VSet \right]\right)                                                                                                                                           \\
                                     & \geq & \locsize{t',\TSet'}(x) \left[v/|\valuation(v)| \mid v\in\VSet \right]                                                                                                                                                                      \\
                                     &      & \qquad\parbox{8cm}{(as $|\valuation_0|(\Size(\pret', v)) \geq |\valuation(v)|$ for all $v\in\VSet$ since $\Size$ is a size bound for $\Program$ and $\locsize{t',\TSet'}(x)$ is a weakly monotonically increasing bound from $\BoundSet$)} \\
                                     & =    & |\valuation|(\locsize{t',\TSet'}(x))                                                                                                                                                                                                       \\
                                     & \geq & |\valuation'(x)|                                                                                                                                                                                                                           \\
                                     &      & \qquad \parbox{8.5cm}{(by \Cref{def:local_size_bounds}, since $\locsize{t',\TSet'}$ is a local size bound for transition $t'$)}
      \end{array}
    \] \qed
  \end{myproof}
}

To obtain local size bounds which can then be lifted via \Cref{thm:lift_size_bounds}, we look for transitions $t_L$ that correspond to a loop $L$ and then we compute a size bound for $L$ as in \Cref{sect:loops_size_bounds}.
The following lemma shows that size bounds for loops as in \Cref{Size Bounds of Loops}
indeed yield local size bounds for the corresponding transitions.\footnote{\label{size bounds loops n 0}
  Local or global size bounds for transitions only have to hold if the transition is indeed taken.
  In contrast, size bounds for loops also have to hold if there is no loop iteration (see \Cref{Size Bounds of Loops}).
  This will be needed in \Cref{lem:cyclesSB} to compute local size bounds for simple cycles.}

\begin{restatable}[Local Size Bounds via Loops]{lemma}{LocalSBLoops}
  \label{lem:lift_size_bounds_loops}
  Let	$\size_L$ be a size bound for a loop $L$ (as in \cref{Size Bounds of Loops}) and let the transition $t_L$ correspond to the loop $L$ via a variable renaming $\pi$.
  Then $\pi \circ \size_L \circ \pi^{-1}$ is a local size bound for $t_L$ w.r.t.\ $\{t_L\}$ (as in \cref{def:local_size_bounds}).
\end{restatable}
\makeproof{lem:lift_size_bounds_loops}{
  \LocalSBLoops*
  \begin{myproof}
    Let the transition $t = (\location, \guard, \update, \location)$ correspond to the loop $L = (\guard',\update')$ via the variable renaming $\pi$.
    By \cref{def:local_size_bounds}, we have to show that for all $v\in\VSet$ and all $\valuation\in\Valuation$ we have
    \[
      \mbox{\small $ |\valuation|(\pi(\size_L(\pi^{-1}(v)))) \geq \sup\{ |\valuation'(v)|\mid \exists \,(\wildcard,\wildcard,\wildcard,\location)\in\entry_{\braced{t_L}}.\; (\location,\valuation) \, (\to^*_{t_L}\circ \to_{t_L}) \, (\location,\valuation')\}.$}
    \]

    \noindent
    Recall that $\rc: \valuation\to\overline{\NN}$ denotes the runtime complexity of $L$.
    Thus, we have
    \begin{align*}
      \;      & |\valuation|(\pi(\size_L(\pi^{-1}(v)))) \\
      = \;    & |\valuation \circ \pi|(\size_L(\pi^{-1}(v))) \\
      \geq \; & \sup\braced{|\valuation(\pi(\update'^n(\pi^{-1}(v))))| \mid n \leq \rc(\valuation \circ \pi)}
      \tag*{(by \cref{Size Bounds of Loops})} \\
      = \;    & \sup\braced{|\valuation(\update^n(\pi(\pi^{-1}(v))))| \mid n \leq \rc(\valuation \circ \pi)}
      \tag*{(by the correspondence between $L$ and $t_L$)} \\
      = \;    & \sup\braced{|\valuation(\update^n(v))| \mid n \leq \rc(\valuation \circ \pi)} \\
      =	\;    & \sup\{ |\valuation'(v)|\mid \forall\, 0 \leq n \leq \rc(\valuation\circ \pi). \;(\location,\valuation)\to^n_{t_L} (\location,\valuation')\} \\
      \geq	\; & \sup\{ |\valuation'(v)|\mid \forall\, 1 \leq n \leq \rc(\valuation\circ \pi). \;(\location,\valuation)\to^n_{t_L} (\location,\valuation')\} \tag*{(by omitting $n = 0$ in the supremum)} \\
      = \;    & \sup\{ |\valuation'(v)|\mid (\location,\valuation) \; (\to^*_{t_L}\circ \to_{t_L}) \; (\location,\valuation')\}\tag*{(by the definition of $\rc$ and the correspondence between $L$ and $t_L$)} \\
      \geq \; & \sup\{ |\valuation'(v)|\mid \exists \, (\wildcard,\wildcard,\wildcard,\location)\in\entry_{\braced{t_L}}.\; (\location,\valuation) \; (\to^*_{t_L}\circ \to_{t_L}) \; (\location,\valuation')\}
    \end{align*} \qed
  \end{myproof}
}
\begin{example}
  \label{exa:SBLifting}
  The bound $\size_L(x_1) = \left( x_1 + \tfrac{1}{2} \cdot x_3^2\right)\cdot 3^4\cdot (2\cdot x_1 + x_3^2 + 2\cdot x_3^5)^5 + \tfrac{1}{2}\cdot x_3^2$ is a size bound for $x_1$ in the loop \eqref{WhileExample}, see \cref{exa:sizeboundLoop}.
  The transition $t_3$ in the program of \cref{fig:ITS} corresponds to this loop $L$ via the identity $\pi$.
  Since $\entry_{\braced{t_3}} = \{t_1\}$, \cref{thm:lift_size_bounds,lem:lift_size_bounds_loops} yield the following (non-linear) size bound for $x_1$ in the full program of \cref{fig:ITS} (see \cref{ex:SizeBounds} for $\Size(t_1,v)$):
  \begin{align}
    \Size(t_3,x_1) & =
    \size_L(x_1) \left[v/\Size(t_1,v) \mid v\in\VSet \right] \nonumber \\
                   & =
    \left( \tfrac{1}{2} \cdot 2^2 + x_6 \right)\cdot 3^4\cdot (2^2 + 2\cdot 2^5 + 2\cdot
    x_6)^5 + \tfrac{1}{2}\cdot 2^2 \nonumber \\
                   & =
    \left( 2 + x_6 \right)\cdot 3^4\cdot (2 \cdot x_6 + 68)^5 + 2 \label{newSizebound}
  \end{align}
  Recall that $\size_L(x_1)$ and thus also $\Size(t_3,x_1)$ are polynomial due to the new technique to infer logarithmic runtime bounds.
  In contrast, our conference papers \cite{lommen2022AutomaticComplexityAnalysis,lommen2023TargetingCompletenessUsing} can only infer an exponential size bound $\size_L(x_1)$ and thus, the resulting size bound $\Size(t_3,x_1)$ would also be exponential.
  To the best of our knowledge, all other existing approaches fail to infer a finite size bound $\Size(t_3,x_1)$.

  For that reason, they would also fail to infer a finite runtime bound if afterwards $x_6$ is set to $x_1$ instead of $x_4$, i.e., if transition $t_5$ is executed $x_1$ times.
  For this modified program, our conference papers \cite{lommen2022AutomaticComplexityAnalysis,lommen2023TargetingCompletenessUsing} would obtain an exponential runtime bound, but the new approach would infer a polynomial runtime bound.
  Here we benefit from the construction which takes the different roles of variables in bounds into account.
  To this end, it is crucial that \Cref{lem:complexity,lem:complexity_logarithmic} consider each variable individually when computing runtime bounds for loops, in contrast to the technique of \cite{hark2020PolynomialLoopsTermination} which uses a coarser over-approximation that does not distinguish the effects of different variables.
  Thus, instead of a polynomial of degree 27, in our approach the bound $\Size(t_3,x_1)$ in \eqref{newSizebound} is only a polynomial of degree 6, since all occurrences of $x_3$ in $\size_L(x_1)$ are instantiated by the constant value $\Size(t_1,x_3) = 2$.

  We infer the remaining size bounds in an analogous way.
  For instance, $\size_L(x_4) =\sqrt{10}\cdot x_4 + 2\cdot x_5$ is a size bound for $x_4$ in the loop \eqref{WhileExample}, see again \cref{exa:sizeboundLoop}.
  Hence, we obtain $\Size(t_3,x_4) =\size_L(x_4) \left[v/\Size(t_1,v) \mid v\in\VSet \right] = (\sqrt{10}\cdot x_4 + 2\cdot x_5)\left[v/\Size(t_1,v) \mid v\in\VSet \right] = (\sqrt{10} + 2)\cdot x_6$.
  Here, the original incomplete technique of \cite{brockschmidt2016AnalyzingRuntimeSize} would yield an exponential size bound instead of the linear size bound.
\end{example}

The approach alternates between improving size and runtime bounds for individual transitions.
We start with $\Size(t_0,v) = ||\update(v)||$ for initial transitions $t_0\in\TSet_0$ where $\update$ is $t_0$'s update, and $\Size(t,\wildcard) = \omega$ for $t\in\TSet\setminus\TSet_0$.
To improve the size bounds of transitions that correspond to (possibly non-linear) solvable loops, we can use closed forms (\cref{thm:size_bounds_closed_form}) and the lifting via \cref{thm:lift_size_bounds}.
Otherwise, we use the incomplete technique of \cite{brockschmidt2016AnalyzingRuntimeSize} to improve size bounds (where \cite{brockschmidt2016AnalyzingRuntimeSize} essentially only succeeds for updates without non-linear arithmetic).
As mentioned, both the technique from \cite{brockschmidt2016AnalyzingRuntimeSize}
and the approach from \cref{thm:size_bounds_closed_form}
rely on runtime bounds to compute size bounds.
On the other hand, as shown in \Cref{sect:Runtime Bounds for Integer Programs}, size bounds for ``previous'' transitions are needed to infer (global) runtime bounds for transitions in a program via \Cref{thm:time-bound}.
For that reason, the alternated computation resp.\ improvement of global size and runtime bounds for the transitions is repeated until all bounds are finite or no bound could be improved in the previous iteration.

\paragraph{Local Size Bounds via Simple Cycles:}
In \Cref{def:correspondence_loops_transitions,lem:lift_size_bounds_loops} we considered transitions with the same start and target location that directly correspond to loops.
To increase the applicability of the technique for size bounds via closed forms of loops, we now proceed similarly as in the previous section and again consider simple cycles, where iterations through the cycle can only be done in a unique way.

Now we want to compute a \emph{local} size bound for the transition $t_n$ w.r.t.\ a simple cycle $\mathcal{C} = \{t_1, \ldots, t_n\}$ where $t_1\chain \ldots \chain t_n$ corresponds to a loop $L$ via a variable renaming $\pi$.
Then a size bound $\size_L$ for the loop $L$ yields the size bound $\pi \circ \size_L \circ \pi^{-1}$ for $t_n$ regarding evaluations through $\mathcal{C}$ starting in $t_1$.
However, to obtain a local size bound $\locsize{t_n,\mathcal{C}}$, we have to consider evaluations starting after any entry transition $(\wildcard,\wildcard,\wildcard,\location_i)\in\entry_{\mathcal{C}}$.
Hence, we use $|| \, \update_i(\ldots\update_n(\pi(\size_L(\pi^{-1}(v))))\ldots) \, ||$ for any $(\wildcard,\wildcard,\wildcard,\location_i)\in\entry_{\mathcal{C}}$ where $\update_j$ is the update of $t_j$.
In this way, we also capture evaluations starting in $\location_i$, i.e., without evaluating the full cycle.
Note that we do not have to consider such partial evaluations in the special case $\location_i = \location_1$.

\begin{restatable}[Local Size Bounds for Simple Cycles]{lemma}{SBCycles}
  \label{lem:cyclesSB}
  Let $\mathcal{C} = \braced{t_1,\ldots,t_n}\subset\TSet$ be a simple cycle and
  let $\size_L$ be a size bound for a loop $L$ where $t_1\chain \ldots \chain t_n$ corresponds to $L$ via a variable renaming $\pi$.
  For each $1 \leq j \leq n$, let $\update_j$ be the update of $t_j$.
  Then a \emph{local size bound} $\locsize{t_n,\mathcal{C}}(v)$ for $t_n$ w.r.t.\ $\mathcal{C}$ is
  \[\begin{array}{l@{}l}
      \locsize{t_n,\mathcal{C}}(v) = \max ( & \braced{\pi(\size_L(\pi^{-1}(v))) \mid (\wildcard,\wildcard,\wildcard,\location_1)\in\entry_{\mathcal{C}}
      } \cup                                                                                                                                            \\
                                            & \braced{|| \,
          \update_i(...\update_n(\pi(\size_L(\pi^{-1}(v))))...) \,||\mid 1\!<\!i\!\leq\!n,
          (\wildcard,\wildcard,\wildcard,\location_i)\!\in\!\entry_{\mathcal{C}}
        }).
    \end{array}\]
\end{restatable}
\makeproof{lem:cyclesSB}{
  \SBCycles*
  \begin{myproof}
    We have to show that for all $v\in\VSet$, \Cref{def:local_size_bounds} holds for $\locsize{t_n,\mathcal{C}}(v)$, i.e., for all $\valuation\in\Valuation$ we have
    \[
      |\valuation|(\locsize{t_n,\mathcal{C}}(v)) \geq \sup\{ |\valuation'(v)|\mid \exists \location' \in \LSet, (\wildcard,\wildcard,\wildcard,\location)\in\entry_{\mathcal{C}}.\; (\location,\valuation) \; (\to^*_{\mathcal{C}}\circ \to_{t_n}) \; (\location',\valuation')\}.
    \]
    Here, the evaluation always has the following form for an $\pret = (\wildcard,\wildcard,\wildcard,\location_j)\in\entry_{\mathcal{C}}$:
    \[
      (\location_j,\valuation)\; (\to_{t_j}\circ\ldots\circ \to_{t_n}) \; (\location_1,\valuation')\; (\to_{t_1}\circ \ldots\circ\to_{t_n})^*\;(\location_1, \valuation'')
    \]
    If we are in the first case, i.e., $j = 1$, then we only have to consider evaluations $(\location_1,\valuation')\; (\to_{t_1}\circ \ldots\circ\to_{t_n})^*\;(\location_1, \valuation'')$.
    Hence, in that case there is no partial evaluation $(\location_j,\valuation)\; (\to_{t_j}\circ\ldots\circ \to_{t_n}) \; (\location_1,\valuation')$ of the simple cycle and we have $(\location_j,\valuation) = (\location_1,\valuation')$.

    We now prove that $|\valuation|(\locsize{t_n,\mathcal{C}}(v))$ is a bound on $|\valuation'(v)|$ and on all $|\valuation''(v)|$.
    If $j > 1$, then we have
    \begin{align*}
      |\valuation|(\locsize{t_n,\mathcal{C}}(v)) & \geq |\valuation|\left(\max\braced{ || \,
          \update_i(\ldots\update_n(\pi(\size_L(\pi^{-1}(v))))\ldots) \, || \; \mid 1 <
          i \leq n, \;
      (\wildcard,\wildcard,\wildcard,\location_i)\in\entry_{\mathcal{C}}}\, \right) \\
                                                 & \geq |\valuation| \left(\, || \, \update_j(\ldots\update_n(\pi(\size_L(\pi^{-1}(v))))\ldots) \, || \, \right)
    \end{align*}
    Note that $|\valuation|(||\,(\update_j\circ ... \circ\update_n\circ\pi)(x_i)\,||) \geq |\valuation \circ \update_j\circ \dots\circ\update_n\circ\pi| (x_i)$ for all variables $x_i \in \{x_1,\dots,x_d\}$ of the loop $L$.
    Moreover, $\size_L(\pi^{-1}(v)) \in \BoundSet$ for all $v \in \VSet$.
    Hence, we have
    \begin{align*}
      |\valuation| \left(\, || \, \update_j(\ldots\update_n(\pi(\size_L(\pi^{-1}(v))))\ldots) \, || \, \right) & \geq |\valuation \circ \update_j\circ\ldots\circ\update_n\circ \pi| (\size_L(\pi^{-1}(v))) \\
                                                                                                               & = |\valuation'\circ \pi| (\size_L(\pi^{-1}(v))).
    \end{align*}
    Similarly, if $j =1$, then we have $\valuation = \valuation'$ which implies
    \begin{align*}
      |\valuation|(\locsize{t_n,\mathcal{C}}(v)) & \geq
      |\valuation|\left(\pi(\size_L(\pi^{-1}(v)))\right) \\
                                                 & = |\valuation'|\left(\pi(\size_L(\pi^{-1}(v)))\right) \\
                                                 & = |\valuation' \circ \pi|\left(\size_L(\pi^{-1}(v))\right).
    \end{align*}
    So for all $1 \leq j \leq n$, we obtain $|\valuation|(\locsize{t_n,\mathcal{C}}(v)) \geq |\valuation' \circ \pi|\left(\size_L(\pi^{-1}(v))\right)$.
    Since $\size_L$ is a size bound for the loop $L$, by \cref{Size Bounds of Loops} we have $|\valuation'\circ \pi| (\size_L(\pi^{-1}(v))) \geq |\valuation'\circ \pi|(\pi^{-1}(v)) =|\valuation'| (\pi(\pi^{-1}(v))) = |\valuation'|(v)$.
    Hence, this proves $|\valuation|(\locsize{t_n,\mathcal{C}}(v)) \geq |\valuation'|(v)$.

    Now we prove that $|\valuation|(\locsize{t_n,\mathcal{C}}(v))$ is a bound on $|\valuation''(v)|$ if $(\location_1,\valuation') \; (\to_{t_1}\circ \ldots\circ\linebreak
      \to_{t_n})^+\;(\location_1, \valuation'')$ where ``$+$'' denotes the transitive closure.
    By a similar argument as in \cref{lem:lift_size_bounds_loops}, $\pi \circ \size_L \circ \pi^{-1}$ is a local size bound for $t_1\chain \ldots \chain t_n$ w.r.t.\ $\{ t_1\chain \ldots \chain t_n \}$, i.e., we have
    \begin{align*}
      |\valuation'| (\pi(\size_L(\pi^{-1}(v)))) & \geq \sup\braced{|\widetilde{\valuation}(v)|\mid (\location_1,\valuation') \; (\to_{t_1}\circ \ldots\circ\to_{t_n})^+ \; (\location_1,\widetilde{\valuation})} \\
                                                & \geq |\valuation''|(v).
    \end{align*}
    So in total we have $|\valuation|(\locsize{t_n,\mathcal{C}}(v)) \geq |\valuation'| (\pi(\size_L(\pi^{-1}(v)))) \geq |\valuation''|(v)$.
    \qed
  \end{myproof}
}
\begin{example}
  \label{ex:simpleCycle}
  As in \Cref{exa:simpleCycleTime}, we again replace $t_3 = (\location_2, \guard,\update, \location_2)$ by the same transitions $t_{3a} = (\location_2, \guard, \update_{a}, \location_2')$ and $t_{3b} = (\location_2', \true, \update_{b}, \location_2)$ with the new location $\location_2'$ in the program of \Cref{fig:ITS}, where $\update_{a}(v) = \update(v)$ for $v \in \{ x_4, x_5 \}$, $\update_{b}(v) = \update(v)$ for $v \in \{
    x_1, x_2, x_3, x_6 \}$, and $\update_{a}$ resp.\ $\update_{b}$ are the identity on the remaining variables.
  Then $\braced{t_{3a},t_{3b}}$ forms a simple cycle and \Cref{lem:cyclesSB} allows us to compute local size bounds	$\locsize{t_{3a},\braced{t_{3a},t_{3b}}}$ and $\locsize{t_{3b},\braced{t_{3a},t_{3b}}}$ w.r.t.\ $\braced{t_{3a},t_{3b}}$, because the chained transitions $t_{3a} \chain t_{3b} = t_3$ and $t_{3b} \chain t_{3a}$ both correspond to the loop \eqref{WhileExample}.
  They can then be lifted to global size bounds as in \Cref{exa:SBLifting} using size bounds for the entry transitions $\entry_{\braced{t_{3a},t_{3b}}} = \{t_1\}$.
\end{example}

This shows how we choose $t'$ and $\TSet'$ when lifting local size bounds to global ones with \Cref{thm:lift_size_bounds}:
For a transition $t'$ we search for a simple cycle $\TSet'$ such that chaining the cycle results in a solvable loop for which we can infer a finite runtime bound (by \Cref{lem:correctness_chaining} or by $\MRFs$ \cite{giesl2022ImprovingAutomaticComplexity}) and have entry transitions $\entry_{\TSet'}$ with finite size bounds.
For all other transitions, we compute size bounds as in \cite{brockschmidt2016AnalyzingRuntimeSize}.

So our heuristic to choose cycles for \emph{size bounds} is slightly different from the heuristic to choose cycles for \emph{runtime bounds}, since for size bounds, we are not restricted to \emph{prs}-loops.
Furthermore, we do not have to consider runtime bounds for the entry transitions in the size bound computations.

\paragraph{Size Bounds for Commuting Programs:}
Up to now our approach to infer size bounds either uses the technique of \cite{brockschmidt2016AnalyzingRuntimeSize,giesl2022ImprovingAutomaticComplexity} or is restricted to simple cycles.
In this subsection, we present a new extension to infer size bounds for sets of loops or simple cycles that are \emph{commuting}.
Two transitions (resp.\ their updates $\update$ and $\update'$) are called \emph{commuting} if $\update \circ \update' = \update' \circ \update$ holds.
Sufficient criteria for commuting loops are if the updates only change disjoint variables or if both updates are linear.
A set of simple cycles with a joint entry location $\location$ is commuting if all pairs of transitions are commuting which result from chaining the simple cycles when starting in $\location$.
\begin{definition}[Commuting Integer Programs]
  \label{def:commuting_integer_programs}
  Let $\mathcal{C}_1,\dots,\mathcal{C}_m$ be disjoint simple cycles which share a common location $\location$ such that all entry transitions $\entry_{\TSet'}$ of $\TSet' = \mathcal{C}_1\cup \dots\cup\mathcal{C}_m$ end in $\location$.
  For all $1 \leq i \leq m$, if $\mathcal{C}_i$ is a simple cycle with the sequence $t_1,\dots,t_n$ such that the start location of $t_1$ is $\location$, then we define $c_i = t_1\chain\dots\chain t_n$.
  We say that $\mathcal{C}_1,\dots,\mathcal{C}_m$ are \emph{commuting} if $c_i$ and $c_j$ are commuting for all $i,j\in\braced{1,\dots,m}$.
  The integer program $(\VSet,\TVSet,\braced{\location_0,\location},\location_0,\braced{c_1,\dots,c_m,t})$ where $t = (\location_0,\true,\identity,\location)$ is the corresponding \emph{commutator program}, denoted as $[c_1,\dots,c_m]$.
\end{definition}

\begin{figure}[t]
  \begin{subfigure}[t]{.6\textwidth}
    \centering
    \begin{tikzpicture}[->,>=stealth',shorten >=1pt,auto,semithick,initial text=$ $]

      \node[state,initial] (q0) {$\location_0$};
      \node[state] (q1) [above right of=q0,node distance=1.5cm,draw=none]{};
      \node[state] (q2) [right of=q0,node distance=1.5cm,draw=none]{$\dots$};
      \node[state] (q3) [below right of=q0,node distance=1.5cm,draw=none]{};
      \node[state] (q7) [right of=q2,node distance=1.5cm]{$\location$};
      \node[state] (q4) [below left of=q7,node distance=1.5cm,draw=none]{};
      \node[state] (q5) [above left of=q7,node distance=1.5cm,draw=none]{};

      \node[state] (c1) [above right of=q7,node distance=1.5cm,minimum size=0.25cm]{};

      \node[state] (c2a) [right of=q7,node distance=1.5cm,minimum size=0.25cm]{};
      \node[state] (c2b) [below right of=q7,node distance=1.5cm,minimum size=0.25cm]{};

      \node[state] (c1_label) [above right of=c2a,node distance=0.8cm,draw=none]{$\mathcal{C}_1$};
      \node[state] (c2_label) [below right of=c2a,node distance=0.8cm,draw=none]{$\mathcal{C}_2$};

      \draw (q0) edge node {} (q1);
      \draw (q0) edge node {} (q3);
      \draw (q4) edge node {} (q7);
      \draw (q5) edge node {} (q7);
      \draw (q2) edge node {} (q7);

      \draw (q7) edge [bend left=20] node {} (c1);
      \draw (c1) edge [bend left=20] node {} (q7);

      \draw (q7) edge node {} (c2a);
      \draw (c2a) edge [bend left=20,right] node {} (c2b);
      \draw (c2b) edge node {} (q7);
    \end{tikzpicture}
    \caption{Original Program}
    \label{fig:originalProgram}
  \end{subfigure}
  \hfill
  \begin{subfigure}[t]{.4\textwidth}
    \centering
    \begin{tikzpicture}[->,>=stealth',shorten >=1pt,auto,semithick,initial text=$ $]
      \node[state,initial] (q0) {$\location_0$};
      \node[state] (q1) [right of=q0,node distance=1.5cm]{$\location$};

      \node[state] (dummy1) [above right of=q0,node distance=1.5cm,draw=none]{};
      \node[state] (dummy2) [below right of=q0,node distance=1.5cm,draw=none]{};

      \draw (q0) edge node {$t$} (q1);

      \draw (q1) edge [in=30,out=60,loop] node {$c_1$} (q1);
      \draw (q1) edge [out=330,in=300,loop] node {$c_2$} (q1);
    \end{tikzpicture}
    \caption{Commutator Program $[c_1,c_2]$}
    \label{fig:commutatorProgram}
  \end{subfigure}
  \caption{Illustration of \Cref{def:commuting_integer_programs}}
  \label{fig:commutatorProgramIllustration}
\end{figure}
The previous definition is illustrated in \Cref{fig:originalProgram,fig:commutatorProgram}.
The original program in \Cref{fig:originalProgram} has two cycles $\mathcal{C}_1$ and $\mathcal{C}_2$.
These cycles are chained into the transitions $c_1$ and $c_2$ in the commutator program $[c_1,c_2]$ in \Cref{fig:commutatorProgram}.
We can compute closed forms for both $c_1$ and $c_2$ and as in \Cref{thm:size_bounds_closed_form}, we can use these closed forms to infer size bounds for $c_1$ and $c_2$ by considering runtime bounds of $[c_1,c_2]$.
As the order of execution of $\mathcal{C}_1$ and $\mathcal{C}_2$ (resp.\ $c_1$ and $c_2$) is irrelevant since they are commuting, we can combine their size bounds into a single size bound which is sound for both $c_1$ and $c_2$.

\begin{example}
  \label{exa:commuting}
  As an example, consider the program in \Cref{fig:commutingITS}.
  Here, we have two commuting transitions $t_{2a}$ and $t_{2b}$, i.e., $\update_{2a} \circ \update_{2b} = \update_{2b} \circ \update_{2a}$, which form
  two simple cycles $\mathcal{C}_1 = \braced{t_{2a}}$ and $\mathcal{C}_2 = \braced{t_{2b}}$.
  Currently, we cannot infer global size bounds for $t_{2a}$ and $t_{2b}$, as $t_{2b}$ is an entry transition of $t_{2a}$ and vice versa.
  Thus, we do not have a ``starting point'' for the size bound analysis.
  Hence, we cannot infer a finite runtime bound for this program as the runtime of $t_4$ depends on the size of $x_2$ after the transitions $t_{2a}$ and $t_{2b}$.
  To ease the illustration of our approach, as in the example of \Cref{fig:ITS}, we kept the non-linear monomial $x_3^2$ constant in order to simplify the closed form of $x_2$.
  However, our approach also works if $x_3$ is non-constant.
\end{example}

\begin{figure}
  \center
  \begin{tikzpicture}[->,>=stealth',shorten >=1pt,auto,semithick,initial text=$ $]
    \node[state,initial] (q0) {$\location_0$};
    \node[state] (q1) [right of=q0,node distance=1.5cm]{$\location_1$};
    \node[state] (q2) [right of=q1,node distance=1.5cm]{$\location_2$};

    \draw (q0) edge node {$t_1$} (q1);
    \draw (q1) edge node {$t_3$} (q2);

    \draw (q1) edge [loop above] node [xshift=1.5cm,yshift=-0.3cm] {
        \footnotesize $
          \begin{array}{rcl}
            t_{2a}:\guard     & = & (x_1 > 0)   \\
            \update_{2a}(x_1) & = & x_1 - 1     \\
            \update_{2a}(x_2) & = & x_2 + x_3^2 \\
          \end{array}
        $
      } (q1);
    \draw (q1) edge [loop below] node [xshift=-1.3cm,yshift=0.3cm] {
        \footnotesize $
          \begin{array}{rcl}
            t_{2b}:\guard     & = & (x_1 > 0)       \\
            \update_{2b}(x_1) & = & x_1 - 1         \\
            \update_{2b}(x_2) & = & x_2 + x_3^2 + 1 \\
          \end{array}
        $
      } (q1);
    \draw (q2) edge [out=330,in=300,loop] node {
        \footnotesize $
          \begin{array}{rcl}
            t_4:\guard   & = & (x_2 > 0) \\
            \update(x_2) & = & x_2 - 1   \\
          \end{array}
        $
      } (q2);

  \end{tikzpicture}
  \caption{Example of a Commuting Integer Program}
  \label{fig:commutingITS}
\end{figure}

Let $\mathcal{C}_1,\dots,\mathcal{C}_m$ be commuting as in \Cref{def:commuting_integer_programs} and $[c_1,\dots,c_m]$ be the respective commutator program with a global runtime bound $\glo$.
In \Cref{lem:commuting_loops} we will introduce a technique to infer local size bounds for all transitions in $\TSet' = \mathcal{C}_1\cup\dots\cup\mathcal{C}_m$ w.r.t\ $\TSet'$.
Then, we can lift a local size bound for a transition $t_j\in\mathcal{C}_1$ w.r.t.\ $\TSet'$ to a global one by only considering the entry transitions of $\TSet'$ instead of the entry transitions of $\mathcal{C}_1$ which include transitions of $\mathcal{C}_2,\dots,\mathcal{C}_m$ whose current (global) size bound may be $\omega$ (we always have $\entry_{\TSet'} \subseteq \entry_{\mathcal{C}_1}$).

For all $i\in\braced{1,\dots,m}$, let $L_i = (\guard_i,\update_i)$ be the loop which corresponds to $c_i$ via a variable renaming $\pi_i$.
Moreover, let $\clExp{\vec{x}}{i}$ be a closed form (with start value $n_{0,i}$) for $L_i$.
Then similar to \Cref{thm:size_bounds_closed_form}, we obtain the following size bound $\size_i$:
\[ \size_i(v) =
  \left\{ \begin{array}{ll}
    ||\clExp{v}{i}|| \, [n / \pi_i^{-1}(\glo(c_i))], & \text{if $n_0 = 0$} \\
    \max\braced{||\clExp{v}{i}|| \, [n / \pi_i^{-1}(\glo(c_i))],
    \; ||\update_i||^{n_{0,i} - 1}(v) },             & \text{otherwise}
  \end{array}
  \right.
\]
Here, we use the global runtime bound $\glo(c_i)$ of the commutator program instead of the runtime bound of $L_i$.
The reason is that we have to consider arbitrary runs in the commutator program.
So $\size_i(v)$ corresponds to a bound on the size of $v$ after the transition $c_i$ in the commutator program $[c_1,\ldots,c_m]$.

Next, we ``merge'' the local size bounds $\pi_i \circ \size_i \circ \pi_i^{-1}$ of all simple cycles which yields
\begin{equation}
  \label{size-comm}
  \size_{[c_1,\ldots,c_m]}(v) = \pi_m ( \size_m(\pi^{-1}_m( \ldots
  (\pi_1 ( \size_1(\pi^{-1}_1(v)))) \ldots))).
\end{equation}
So $\size_{[c_1,\ldots,c_m]}(v)$ is a bound on the size of $v$ after \emph{any} transition in the commutator program $[c_1,\ldots,c_m]$.
The implementation uses an arbitrary order of the expressions $\size_1,\dots,\size_m$, as this does not significantly affect performance in practice.
However, this approach could be refined as follows:
Instead of directly replacing $n$ with the runtime bound in the definition of $\texttt{sb}_i$ (which ``destroys'' the commutativity of $\size_1,\dots,\size_m$), one could introduce a unique loop counter $n_i$ for each loop and substitute $n_i$ with the runtime bound of the corresponding transition $c_i$ \emph{after} combining all $\texttt{sb}_i$.
This has the advantage that all $\texttt{sb}_i$ remain commutative, i.e., then the choice of permutation becomes irrelevant.

Finally, to obtain a local size bound $\locsize{t_j,\TSet'}(v)$ for $t_j$ w.r.t.\ $\TSet'$, we take $\size_{[c_1,\ldots,c_m]}(v)$, but we have to apply the remaining partial evaluation of the cycle $\mathcal{C}_1 = \{ t_1, \ldots, t_n \}$ from $t_1$ to $t_j$ in the end.
Let $\update_i'$ be the update of $t_i$, i.e., after multiple traversals of the cycles $\mathcal{C}_1,\ldots,\mathcal{C}_m$, there is a final run via $t_1,\ldots,t_j$ which uses the update $\update_1' \circ \ldots \circ \update_j'$.
To take the size after the multiple traversals of the cycles $\mathcal{C}_1,\ldots,\mathcal{C}_m$ into account, in this update the variables have to be instantiated according to $\size_{[c_1,\ldots,c_m]}$, i.e., we obtain $\size_{[c_1,\ldots,c_m]} \circ \update_1' \circ \ldots \circ \update_j'$.\footnote{\label{OtherDirection}
  Note that here the partial cycle $t_1,\ldots,t_j$ is executed \emph{after} several full iterations whereas in \Cref{lem:cyclesSB} the partial cycle $t_i,\ldots,t_n$ is executed \emph{before} several full iterations.}
\begin{restatable}[Size Bounds for Commuting Simple Cycles]{lemma}{SBCommuting}
  \label{lem:commuting_loops}
  Consider commuting simple cycles $\mathcal{C}_1, \ldots,\linebreak[2]
    \mathcal{C}_m$ as in \Cref{def:commuting_integer_programs}.
  So in particular, $\mathcal{C}_1$ is a simple cycle with the sequence $c_1 = t_1 \chain \dots \chain t_n$ where the start location of $t_1$ is $\location$ and the update of $t_i$ is $\update_i'$.
  Let $\size_{[c_1,\ldots,c_m]}$ be defined as in \eqref{size-comm} and let $1 \leq j \leq n$.
  Then a \emph{local size bound} for $t_j$ w.r.t.\ $\TSet' = \mathcal{C}_1 \cup \ldots \cup \mathcal{C}_m$ is $$ \locsize{t_j,\TSet'}(v) = \left\{
    \begin{array}{l@{\quad}l}
      || \, \size_{[c_1,\ldots,c_m]}(\update_1'(\ldots\update_j'(v)\ldots)) \,||, & \text{if $j < n$} \\
      \phantom{||} \, \size_{[c_1,\ldots,c_m]}(v),                                & \text{if $j = n$}
    \end{array}
    \right.
  $$
\end{restatable}

\makeproof{lem:commuting_loops}{
  \SBCommuting*
  \begin{myproof}
    We have to show that for all $v\in\VSet$, \Cref{def:local_size_bounds} holds for $\locsize{t_j,\TSet'}(v)$, i.e., for all $\valuation\in\Valuation$ we have
    \[
      |\valuation|(\locsize{t_j,\TSet'}(v)) \geq \sup\{ |\valuation'(v)|\mid \exists
      \location' \in \LSet, (\wildcard,\wildcard,\wildcard,\location)\in\entry_{\TSet'}.\;
      (\location,\valuation) \; (\to^*_{\TSet'}\circ \to_{t_j}) \;
      (\location',\valuation')\}.
    \]
    The evaluation always has the following form for any state $\valuation\in\Valuation$:
    \[
      (\location,\valuation)\; \to_{\TSet'}^* \; (\location,\bar{\valuation})\;
      (\to_{t_1}\circ \ldots\circ\to_{t_j}) \;(\location', \valuation')
    \]
    This evaluation can be split into an evaluation $(\location,\valuation)\; \to_{\braced{c_1,\dots,c_m}}^* \; (\location,\bar{\valuation})$ in the commutator program $[c_1,\dots,c_m]$ and into its second part $(\location,\bar{\valuation})\; (\to_{t_1}\circ \ldots\circ\to_{t_j}) \;(\location', \valuation')$ in the original program, whenever $j < n$.
    For $j = n$, we directly have $(\location,\valuation)\; \to_{\braced{c_1,\dots,c_m}}^* \; (\location,\bar{\valuation}) = (\location', \valuation')$.

    For the evaluation $(\location,\valuation)\; \to_{\TSet'}^* \; (\location,\bar{\valuation})$ we can re-order the application of the transitions $c_1, \ldots, c_m$ such that $c_m$ is applied (repeatedly) first and $c_1$ is applied (repeatedly) last, i.e., $(\location,\valuation) = (\location,\valuation_m) \to_{c_m}^* \dots \to_{c_2}^* (\location,\valuation_1) \to_{c_1}^* (\location,\bar{\valuation})$ where the guards of the transitions are ignored.
    This is possible as all transitions $\braced{c_1,\dots,c_m}$ commute pairwise.\footnote{However, the truth value of the guards is not necessarily preserved by commutation during an evaluation.
      This is one of the reasons why it is not trivial to apply a similar technique to compute runtime bounds for such commuting loops.
    }
    Let $v\in\VSet$.
    By induction on $k\in\braced{1,\dots,m}$ we show that $\abs{\bar{\valuation}(v)} \leq \abs{\valuation_{k}}(\size_{[c_1,\ldots,c_k]}(v))$ where
    \[
      \size_{[c_1,\ldots,c_k]}(v) = \pi_k ( \size_k(\pi^{-1}_k( \ldots
      (\pi_1 ( \size_1(\pi^{-1}_1(v)))) \ldots))).\]
    \paragraphProof{Induction Base:}
    If $k = 1$, then we have $\abs{\bar{\valuation}(v)} \leq \abs{\valuation_{1}}(\size_{[c_1]}(v)) = \abs{\valuation_1}(\pi_1(\size_1(\pi^{-1}_1(v))))$ by \Cref{thm:size_bounds_closed_form}
    and \Cref{lem:lift_size_bounds_loops}.\footnote{Note that while $\size_1$ uses $\glo(c_i)$ instead of the runtime of the corresponding loop, the runtime complexity of the loop is always smaller or equal to the number of applications of the corresponding transition in the commutator program.
    }
    \paragraphProof{Induction Step:}
    In the induction step, in a similar way we obtain
    \begin{equation}
      \label{proof:commuting1}
      \abs{\valuation_{k - 1}(v)} \leq \abs{\valuation_{k}}(\pi_{k}(\size_{k}(\pi^{-1}_{k}(v))))
    \end{equation}
    by \Cref{thm:size_bounds_closed_form} and \Cref{lem:lift_size_bounds_loops} for $m \geq k > 1$.
    Thus, the following inequations hold:
    \begin{align}
      \abs{\bar{\valuation}(v)} & \leq \abs{\valuation_{k - 1}}(\size_{[c_1,\ldots,c_{k-1}]}(v))
      \tag{by the induction hypothesis}
      \nonumber \\
                                & = \abs{\valuation_{k - 1}}(\pi_{k-1} ( \size_{k-1}(\pi^{-1}_{k-1}( \ldots
      (\pi_1 ( \size_1(\pi^{-1}_1(v)))) \ldots))))\nonumber \\
                                & \leq
      \abs{\valuation_{k}}(\pi_{k}(\size_{k}(\pi^{-1}_{k}(\pi_{k-1} ( \size_{k-1}(\pi^{-1}_{k-1}( \ldots
      (\pi_1 ( \size_1(\pi^{-1}_1(v)))) \ldots))))\label{proof:commuting2} \\
                                & = \abs{\valuation_{k}}(\size_{[c_1,\ldots,c_k]}(v))\nonumber
    \end{align}
    Here, \eqref{proof:commuting2} holds by monotonicity of the occurring bounds and \eqref{proof:commuting1}.

    Hence, we have $\abs{\bar{\valuation}(v)} \leq \abs{\valuation_{m}}(\size_{[c_1,\ldots,c_m]}(v)) = \abs{\valuation}(\size_{[c_1,\ldots,c_m]}(v))$.
    This proves the lemma for the case $j = n$.
    Otherwise, we have
    \[|\valuation'(v)| = |\bar{\valuation}(\update_1'(\dots\update_j'(v)\dots))| \leq |\bar{\valuation}|(|| \update_1'(\dots\update_j'(v)\dots) ||).\]
    Combing both results we can conclude that
    \[ |\valuation'(v)| \; \leq \;
      |\bar{\valuation}|(|| \update_1'(\dots\update_j'(v)\dots) ||)
      \; \leq \;
      |\valuation|( \underbrace{||
      \size_{[c_1,\ldots,c_m]}(\update_1'(\dots\update_j'(v)\dots)) ||}_{=\;
        \locsize{t_j,\TSet'}(v)}) \]
    \qed
  \end{myproof}
}

\begin{example}
  Reconsider \Cref{exa:commuting} and the commutator program $[t_{2a},t_{2b}]$.
  Then the global runtime bound $\glo(t_{2a}) = \glo(t_{2b}) = x_1$ for this commutator program
  can be automatically computed by ranking functions.
  Let $L_1$ and $L_2$ be the loops corresponding to $t_{2a}$ and $t_{2b}$ with closed forms $\clExp{x_1}{1} = \clExp{x_1}{2} = x_1 - n$, $\clExp{x_2}{1} = x_2 + n \cdot x_3^2$, $\clExp{x_2}{2} = x_2 + n \cdot x_3^2 + n$, and $\clExp{x_3}{1} = \clExp{x_3}{2} = x_3$.
  Then, for example, we have $\size_1(x_1) = \size_2(x_1) = 2\cdot x_1$, $\size_1(x_2) = x_2 + x_1\cdot x_3^2$, $\size_2(x_2) = x_2 + x_1\cdot x_3^2 + x_1$, and $\size_1(x_3) = \size_2(x_3) = x_3$.
  Hence, \[
    \begin{array}{rcl}
      \size_{[c_1,c_2]}(x_2) & = &
      \size_2(\size_1(x_2))                                                                 \\
                             & = &
      \size_2(x_2) + \size_2(x_1)\cdot \size_2(x_3)^2                                       \\
                             & = & (x_1 + x_2 + x_1\cdot x_3^2) + (2 \cdot x_1) \cdot x_3^2 \\
                             & = & x_1 + x_2 + 3\cdot x_1\cdot x_3^2.
    \end{array}
  \]
  In the example, $\locsize{t_{2a},\braced{t_{2a},t_{2b}}}$ and $\locsize{t_{2b},\braced{t_{2a},t_{2b}}}$ are already global size bounds and thus, $\Size(t_3,x_2) = x_1 + x_2 + 3\cdot x_1\cdot x_3^2$.
  Thus, overall we obtain a polynomial runtime bound for the whole program.
  In contrast, neither the techniques from our conference papers \cite{lommen2022AutomaticComplexityAnalysis,lommen2023TargetingCompletenessUsing} nor any other tool (to the best of our knowledge) can infer a finite runtime bound for this example.
\end{example}

\subsection{Completeness}
\label{sect:Completeness}

For individual loops, we showed in \Cref{thm:completeness} that polynomial runtime bounds and finite size bounds are computable for all terminating \emph{prs}-loops.
In this section, we discuss completeness of the approach to compute runtime and size bounds for general integer programs.
We show that for a large class of programs consisting of consecutive\footnote{\label{CommuteNotComplete}
  We do not obtain a corresponding completeness result for commuting loops, since we do not have a complete technique to compute runtime bounds for the commutator programs.}
\emph{prs}-loops, in case of termination we can always infer finite runtime and size bounds.

For a set $\mathcal{C} \subseteq \TSet$ and $\location, \location' \in \LSet$, let $\location \rightsquigarrow_\mathcal{C} \location'$ hold iff there is a transition $(\location, \wildcard, \wildcard, \location') \in \mathcal{C}$.
We say that $\mathcal{C}$ is a \emph{component} if we have $\location \rightsquigarrow_\mathcal{C}^+ \location'$ for all locations $\location, \location'$ occurring in $\mathcal{C}$, where $\rightsquigarrow_\mathcal{C}^+$ is the transitive closure of $\rightsquigarrow_\mathcal{C}$.
So in particular, we must also have $\location \rightsquigarrow_\mathcal{C}^+ \location$ for all locations $\location$ in the transitions of $\mathcal{C}$.
We call an integer program \emph{simple} if every component is a simple cycle that is reachable from any initial configuration $(\location_0, \valuation_0)$ with $\valuation_0 \in \Sigma$.\footnote{While the requirement that every component is reachable from any initial configuration may seem restrictive, obtaining completeness results for a larger class of programs would be challenging, since for every component, we would have to be able to exactly characterize the state space after its termination.
  Furthermore, we would need to decide termination of every component for initial values from such a state space.
  In particular, if the state space consists of only a single point, then this is equivalent to solving the (non-universal) halting problem (i.e., the positivity problem).
  However,
  to the best of our knowledge, the positivity problem is only known to be decidable for simple recurrences of order up to 9 \cite{OuaknineW14}.}

\begin{definition}[Simple Integer Program]
  \label{def:simple_integer_program}
  An integer program $(\VSet,\TVSet,\LSet,\location_0,\linebreak
    \TSet)$ is \emph{simple} if every component $\mathcal{C} \subseteq \TSet$ is a simple cycle, and for every entry transition $(\wildcard,\wildcard,\wildcard,\location)\in\entry_{\mathcal{C}}$ and every $\initial\in\Valuation$, there exists an evaluation $(\location_0,\initial) \rightarrow^*_{\TSet} (\location,\initial)$.
\end{definition}
In \cref{fig:ITS}, $\braced{t_1,t_2,t_3}$ is a component that is not a simple cycle.
However, if we remove $t_2$ and replace the guards of $t_1$ and $t_4$ by $\true$ and their updates by the identity, then the resulting program $\mathcal{P}'$ is simple (but it still contains non-linear arithmetic).
A simple program terminates iff each of its isolated simple cycles terminates.
Thus, if we can prove termination for every simple cycle, then the overall program terminates.
Hence, if after chaining, every simple cycle corresponds to a linear, unit \emph{prs}-loop, then we can decide termination and infer polynomial runtime and size bounds for the overall integer program.
For terminating, non-unit \emph{prs}-loops, runtime bounds are still polynomial but size bounds can be exponential.
Hence, then the global runtime bounds can be exponential as well.
These results are summarized in \Cref{thm:completeness_integer_programs}.

Note that it is not decidable whether an integer program is simple.
The reason is that one has to check the reachability property required in \Cref{def:simple_integer_program}, i.e., for every component $\mathcal{C}$, every entry transition $(\wildcard,\wildcard,\wildcard,\location)\in\entry_{\mathcal{C}}$, and every $\initial\in\Valuation$, there must be an evaluation $(\location_0,\initial) \rightarrow^*_{\TSet} (\location,\initial)$.
If one strengthened this by requiring that one can reach $\location$ from $\location_0$ using only transitions whose guard is $\true$ and whose update is the identity, then the class of programs in \Cref{thm:completeness_integer_programs} (a) would become decidable (there are only $n$ ways to chain a simple cycle with $n$ transitions and checking whether a loop is a \emph{prs}-loop is decidable by \Cref{Bound on the Period}).

\begin{restatable}[Completeness Results for Integer Programs]{theorem}{CompletenessIP}
  \label{thm:completeness_integer_programs}
  \hspace*{0.1cm}
  \begin{enumerate}
    \item Termination is decidable for all simple \emph{linear} integer programs where after chaining, all simple cycles correspond to prs-loops.
    \item Finite runtime and size bounds are computable for all simple integer programs where after chaining, all simple cycles correspond to \emph{terminating}
          prs-loops.
    \item If in addition to (b), all simple cycles correspond to \emph{unit} prs-loops, then the runtime and size bounds are \emph{polynomial}.
    \item If in addition to (b), all simple cycles correspond to \emph{unit} and \emph{strict} prs-loops, then the runtime bounds are logarithmic.
  \end{enumerate}
\end{restatable}
\begin{myproof}
  \begin{enumerate}
    \item A simple integer program terminates iff every loop corresponding to a simple cycle terminates, where we only consider loops starting from a target location of an entry transition of the simple cycle.
          In the following, we prove this claim.
          We define $\mathfrak{S}$ to be the set consisting of all loops that correspond to $t_1 \chain \ldots \chain t_n$ for a simple cycle $\mathcal{C} = \braced{t_1,\ldots,t_n}$ in $\Program$ where there exists an entry transition from $\entry_{\mathcal{C}}$ ending in the start location of $t_1$.

          If $\initial\in\Valuation$ is a non-terminating initial state for $\Program$, then as $\LSet$ is finite, there exists an $\location\in\LSet$ such that for all $m\in\NN$ there is an $m' \geq m$ and an evaluation $(\location_0,\initial) \rightarrow^*_\TSet (\location,\valuation) \rightarrow_{\mathcal{C}}^{m'} (\location,\valuation')$ for a component $\mathcal{C}$.
          Thus, the loop $L\in\mathfrak{S}$ which corresponds to $\mathcal{C}$ does not terminate on $\valuation\in\Valuation$.

          For the other direction, if $\valuation\in\Valuation$ is a non-terminating initial state for $L\in\mathfrak{S}$, and $L$ corresponds to $t_1 \chain \ldots \chain t_n$ for a simple cycle $\mathcal{C} = \{t_1,\ldots,t_n\}$ where the start location of $t_1$ is the target location of an entry transition from $\entry_{\mathcal{C}}$, then there exists an evaluation $(\location_0,\valuation) \rightarrow^* (\location,\valuation)$ by \Cref{def:simple_integer_program}.
          Hence, $\Program$ does not terminate for the initial state $\valuation$.

          Thus, we can prove termination for every loop in $\mathfrak{S}$ by \Cref{lem:correctness_chaining}(b) and \cite{frohn2020TerminationPolynomialLoops} in order to decide termination of $\Program$.

          \medskip

    \item We can partition the transitions of a simple integer program into the following three sets: initial transitions $\TSet_0$, transitions $\TSet_{\mathrm{cyc}}$ which are part of a simple cycle, and transitions $\TSet_1 = \TSet\setminus(\TSet_0\uplus\TSet_{\mathrm{cyc}})$ which connect simple cycles.
          \medskip
          \paragraphProof{Size and Runtime Bounds for $\TSet_0\uplus\TSet_1$:}
          Note that $\glo(t) = 1$ is a runtime bound for all $t\in\TSet_0\uplus\TSet_1$.
          A size bound can be obtained by $\Size(t,x) = ||\update(x)||$ for all $t = (\location_0,\wildcard,\update,\wildcard)\in\TSet_0$.
          Moreover, we can obtain a size bound for all $t = (\wildcard,\wildcard,\update,\wildcard)\in\TSet_1$ by $\Size(t,x) = \max \{ ||\update(x)|| \left[v/\Size(\pret,v) \mid v\!\in\!\VSet \right] \; \mid \; \pret \in \entry_{\braced{t}} \}$.
          Note that the set $\entry_{\braced{t}}$ might contain transitions from $\TSet_{\mathrm{cyc}}$.
          Due to this, we handle transitions in topological order, starting with $\TSet_0$.
          In this way, we have already inferred finite size and runtime bounds for every entry transition $\pret\in\entry_{\{t\}}$.
          \medskip
          \paragraphProof{Size and Runtime Bounds for $\TSet_{\mathrm{cyc}}$:}
          Let $\mathcal{C}\subseteq\TSet_{\mathrm{cyc}}$ be a simple cycle.
          Assume that we already inferred finite size and runtime bounds for the entry transitions $\entry_{\mathcal{C}}$ of the component $\mathcal{C}$.
          Then $\glo(t) = \sum_{\pret \in \entry_{\mathcal{C}}}\glo(\pret)\cdot (\locs{\{t \}}{\mathcal{C}}{r} \left[v/\Size(\pret,v) \mid v\!\in\!\VSet \right])$ yields a finite (global) runtime bound for $t\in\mathcal{C}$.
          Note that we can compute a polynomial runtime bound\linebreak
          $\locs{\{t \}}{\mathcal{C}}{r}$ w.r.t.\ $\mathcal{C}$ if $\Program$ and hence the loop corresponding to $\mathcal{C}$ terminates (see \Cref{lem:correctness_chaining}(d)).
          Similarly, we can infer finite local size bounds via \Cref{lem:cyclesSB} and then lift them to global bounds $\Size(t,v)$ via \Cref{thm:lift_size_bounds}.
          \medskip

    \item As we only consider unit \emph{prs}-loops, all local size bounds are polynomial (see \Cref{thm:completeness}).
          By \Cref{thm:completeness} also all local runtime bounds are polynomial.
          Note that polynomial bounds are closed under addition, multiplication, and substitution.
          Thus, all bounds which are constructed in the previous step (b) are polynomial.

          \medskip

    \item All size bounds are polynomial by (c).
          Furthermore, all local runtime bounds are logarithmic by \Cref{thm:completeness}.
          Inserting these polynomial size bounds (as in \Cref{thm:time-bound}) into the logarithmic runtime bounds yields logarithmic bounds again.
          \qed
  \end{enumerate}
\end{myproof}

Note that in the example program $\mathcal{P}'$ above, the eigenvalues of the update matrix of $t_5$ have absolute value $1$, i.e., $t_5$ corresponds to a unit \emph{prs}-loop.
In contrast, $t_3$ does not correspond to a unit loop.
However, $\mathcal{P}'$ still has polynomial runtime and size bounds although \Cref{thm:completeness_integer_programs} is not applicable.
The reason is that the only non-unit eigenvalues of $t_3$ are ``compensated'' by its logarithmic runtime.

\section{Evaluation, Implementation, Related Work, and Conclusion}
\label{sect:conclusion}

In this paper, we developed techniques to infer runtime and size bounds automatically and to use them in order to obtain bounds on the runtime complexity of programs.
In particular, we presented complete techniques for important classes of loops and showed how to integrate them into an (incomplete) modular approach for general integer programs.

The resulting procedure to prove termination and to infer runtime and size bounds is complete for certain classes of programs.
Moreover, as shown below by our experiments with the implementation in the tool \KoAT{}, it improves the power of automatic complexity analysis on integer programs significantly.
In particular, for the first time this also allows to perform automatic runtime analysis for programs with non-linear arithmetic.

Thus, we showed that complete techniques for termination and complexity analysis for subclasses of loops are not only of theoretical interest, but they are also important in practice.
To our knowledge, our approach is the first to integrate such results into an incomplete approach for automated complexity analysis like \cite{brockschmidt2016AnalyzingRuntimeSize,giesl2022ImprovingAutomaticComplexity}.
For this integration, we developed several novel contributions which extend and improve the previous approaches in \cite{hark2020PolynomialLoopsTermination,brockschmidt2016AnalyzingRuntimeSize,giesl2022ImprovingAutomaticComplexity}
substantially:
\begin{enumerate}
  \item In \Cref{sect:Runtime Bounds for PRS-Loops}, we extended the technique for the computation of runtime bounds from \emph{twn}-loops to \emph{prs}-loops, and improved it such that these bounds now take the roles of the different variables into account.
        Moreover, we can now infer logarithmic runtime bounds and handle unsolvable loops.
        In \Cref{sect:loops_size_bounds}, we showed how to use closed forms in order to infer size bounds for solvable loops with possibly non-linear arithmetic.
        We proved completeness of our approach for terminating \emph{prs}-loops in \Cref{sect:completenessLoops}.
  \item We embedded our approach for loops into the setting of general integer programs in \Cref{sect:global_integer_programs}
        by lifting local runtime and size bounds for subprograms to global bounds for the full program.
        In particular, local runtime bounds can now either be obtained via the technique for \emph{prs}-loops from \Cref{sect:Runtime Bounds for PRS-Loops}
        or via ranking functions, and local size bounds can either be obtained via the approach of \Cref{sect:loops_size_bounds} for solvable loops or via the incomplete technique of \cite{brockschmidt2016AnalyzingRuntimeSize}.
        We extended our approach from loops to simple cycles and to size bounds for commuting simple cycles.
        Finally, we showed completeness of our approach for simple integer programs with only terminating \emph{prs}-loops in \Cref{sect:Completeness}.
  \item We integrated our approach into our re-implementation of the tool \KoAT, written in \tool{OCaml}.
\end{enumerate}

To infer local runtime bounds, \KoAT{} first applies multiphase linear ranking functions \cite{giesl2022ImprovingAutomaticComplexity,ben-amram2017MultiphaseLinearRankingFunctions,ben-amram2019MultiphaseLinearRankingFunctions,heizmann2015RankingTemplatesLinear}, which can be done very efficiently.
For terminating \emph{twn}-loops where no finite bound was found, it then uses the complete technique from \Cref{sect:Runtime Bounds for PRS-Loops}
for the inference of runtime bounds.
In the future, we plan to extend the implementation of the complete technique for runtime bound computations to all terminating \emph{prs}-loops.
When computing size bounds, \KoAT{} first applies the technique of \cite{brockschmidt2016AnalyzingRuntimeSize} for reasons of efficiency and in case of exponential or infinite size bounds, it tries to compute size bounds via closed forms for solvable loops as in \Cref{sect:loops_size_bounds}.\footnote{So most of the techniques described in the paper (including the complete techniques for subclasses of loops) are implemented in \textsf{KoAT}.
  In particular, the complete technique for size bounds is implemented not only for all \emph{prs}-loops, but even for all solvable loops.
  The only part which is not yet implemented is the transformation from \emph{prs}-loops to \emph{twn}-loops for runtime bounds via rational automorphisms as in \Cref{lem:correctness_chaining}.
  For that reason, the complete techniques for termination and runtime bounds can only be used for \emph{twn}-loops by \textsf{KoAT}, but not for \emph{prs}-loops that are not yet in \emph{twn}-form.
  However, even without the transformation of \Cref{sect:Reducing Runtime Bounds of PRS-Loops to TNN-Loops}, sometimes the removal of variables from a transition as in \Cref{def:correspondence_loops_transitions} and \Cref{loopAsTransition} already suffices to transform a \emph{prs}- into a \emph{twn}-loop.}

Here, the algorithms of \cite{thiemann2016AlgebraicNumbersIsabelle,thiemann2016FormalizingJordanNormal} are used to handle algebraic numbers and to compute Jordan normal forms for the transformation from solvable to \emph{twn}-loops as in \Cref{lem:transform_solvable}.
Moreover, \KoAT{} applies a local control-flow refinement technique \cite{giesl2022ImprovingAutomaticComplexity,lommen2024ControlFlowRefinementProbabilistic}, and it preprocesses the program in the beginning, e.g., by extending the guards of transitions with invariants inferred by \tool{Apron} \cite{jeannet2009ApronLibraryNumerical}.
For all SMT problems (including the termination proofs of \emph{twn}-loops), \KoAT{} uses \tool{Z3} \cite{moura2008Z3SMTSolver}.

\paragraph{Related Work:}

As mentioned in the introduction, there are many approaches to analyze complexity of programs automatically mostly based on (linear) ranking functions, e.g., \cite{ben-amram2017MultiphaseLinearRankingFunctions,albert2019ResourceAnalysisDriven,sinn2017ComplexityResourceBound,brockschmidt2016AnalyzingRuntimeSize,flores-montoya2016UpperLowerAmortized,giesl2022ImprovingAutomaticComplexity,hoffmann2017AutomaticResourceBound,lopez18IntervalBasedResource,carbonneaux2015CompositionalCertifiedResource,albert2012CostAnalysisObjectoriented,lommen2024ControlFlowRefinementProbabilistic,pham2024RobustResourceBounds,HoffmannJ22}.
However, there also exist many complete techniques to decide termination, analyze runtime complexity, or study memory consumption for certain classes of programs, e.g., \cite{tiwari04,braverman06,frohn2019TerminationTriangularInteger,ben-amram2019TightWorstCaseBounds,hosseini2019TerminationLinearLoops,frohn2020TerminationPolynomialLoops,hark2020PolynomialLoopsTermination,ben-amram2016FlowchartProgramsRegular,ben-amram2008LinearPolynomialExponential,xuSymbolicTerminationAnalysis2013,neumann2020RankingFunctionSynthesis}.

Instead of representing integer programs via transitions, there are also techniques based on \emph{cost equation systems}, e.g., \cite{flores-montoya2014ResourceAnalysisComplex,flores-montoya2016UpperLowerAmortized}, implemented in the tool \tool{CoFloCo}.
This approach analyzes program parts independently, using a set of constraints to measure the sizes of variables w.r.t.\ their initial and final values, and it uses linear invariants to compose the results.
So it differs significantly from our approach which can also infer non-linear size bounds.
Similarly, in the tool \tool{PUBS} \cite{albert2008AutomaticInferenceUpper,albert2012CostAnalysisObjectoriented}, \emph{cost relations}
are analyzed which are a system of recursive equations that capture the cost of the program.
Apart from the new version of \tool{KoAT}, \tool{PUBS} is the only tool which can compute logarithmic runtime bounds.
To this end, \tool{PUBS} searches for a ranking function $f: \Valuation\to\NN$ and a constant $k\in\NN$ such that $f(\valuation) \geq k \cdot f(\valuation') + 1$ holds for all evaluation steps from $\valuation$ to $\valuation'$.
To measure the sizes of variables, \tool{PUBS} uses recurrence relations.
An approach for automatic complexity analysis of \tool{OCaml} programs is implemented in the tool \tool{RAML}
\cite{hoffmann2012MultivariateAmortizedResource,hoffmann2017AutomaticResourceBound}.
\tool{RAML} infers worst-case resource bounds for higher-order polymorphic programs by constructing suitable linear optimization problems which are then solved by linear program solvers.
Moreover, the resource consumption of \tool{Liquid Haskell} programs is analyzed in \cite{handley2019LiquidateYourAssets}.
Furthermore, there also exist tools to infer \emph{lower}
bounds on the worst-case runtime complexity, e.g., \textsf{LoAT} \cite{frohn2022ProvingNonTerminationLower} and \textsf{LOBER} \cite{albert2021LowerBoundSynthesisUsing}.
In \cite{meyer2021InferringExpectedRuntimes,lommen2024ControlFlowRefinementProbabilistic}, \tool{KoAT} was extended to probabilistic programs.

There also exist tools which analyze the runtime complexity of \tool{C}-code, e.g., the tool \tool{Loopus} \cite{sinn2017ComplexityResourceBound}, or \tool{MaxCore} \cite{albert2019ResourceAnalysisDriven} with the tools \tool{CoFloCo} or \tool{PUBS} in the backend.
These tools also rely on variants of (linear) ranking functions and size bounds:
\tool{MaxCore}'s size bound computations build upon \cite{DBLP:conf/popl/CousotH78} and \tool{Loopus} considers suitable bounding invariants to infer size bounds.
\tool{KoAT} can be applied to \tool{C}-code as well, by using \tool{Clang} \cite{clang} and \tool{llvm2kittel}
\cite{falke2011TerminationAnalysisPrograms} to transform pointer-free \tool{C}
programs into integer programs.
In order to handle more general \tool{C} programs,\linebreak
we developed the framework \tool{AProVE (KoAT + LoAT)} \cite{lommen2025AProVEKoATLoAT}, which also participates in the annual \emph{Software Verification Competition} (\emph{SV-COMP}) \cite{svcomp}.

Concerning related work on complete techniques, in \cite{tiwari04}, \cite{braverman06}, and \cite{hosseini2019TerminationLinearLoops}, decidability of termination for linear loops over $\RR$, $\QQ$, and $\ZZ$, respectively, was proven.
However, we are not aware of any implementation of these approaches.
Note that for such loops, decidability of the halting problem (i.e., termination for given inputs) is still open.
In fact, deciding this problem reduces to the positivity problem, a well-known open problem which requires breakthroughs in number theory \cite{OuaknineW14}.
Furthermore, \cite{xuSymbolicTerminationAnalysis2013} and \cite{frohn2020TerminationPolynomialLoops,frohn2019TerminationTriangularInteger} present decidability results on termination of solvable and \emph{twn}-loops, respectively.
Our approach builds upon \cite{frohn2020TerminationPolynomialLoops,frohn2019TerminationTriangularInteger} and \cite{hark2020PolynomialLoopsTermination}, which introduced a procedure to infer polynomial runtime bounds for \emph{twn}-loops.

Probably the most popular technique to analyze runtime complexity of programs is the use of ranking functions, see, e.g., \cite{ben-amram2014RankingFunctionsLinearConstraint,heizmann2015RankingTemplatesLinear}.
A complete procedure to infer linear ranking functions -- based on Farkas' quantifier elimination technique -- was introduced in \cite{podelski2004CompleteMethodSynthesis}.
In \cite{neumann2020RankingFunctionSynthesis}, it was proven that polynomial ranking functions are complete for loops over the reals with a continuous semi-algebraic update and compact semi-algebraic guard.
Moreover, \cite{ben-amram2017MultiphaseLinearRankingFunctions,ben-amram2019MultiphaseLinearRankingFunctions}
proved that all linear loops with a (nested) multiphase-linear ranking function have linear runtime.
However, \cite{heizmann2015RankingTemplatesLinear} showed that loops like \eqref{WhileExample} do not admit such a multiphase-linear ranking function.
Regarding size bounds, \cite{ben-amram2008LinearPolynomialExponential,ben-amram2019TightWorstCaseBounds}
introduced a complete approach to compute tight polynomial size bounds for a programming language consisting of for-loops with a fixed number of iterations.
Based on this, in the future, it might be interesting to extend our approach for \emph{prs}-loops in order to obtain a guarantee on the tightness of the bounds.

\paragraph{Evaluation:}
To evaluate the new contributions, we tested \tool{KoAT} on the 519 benchmarks for \emph{Complexity of} \tool{C} \emph{Integer Programs} (\tool{CINT}) from the \emph{Termination Problems Data Base} \cite{tpdb} which is used in the annual \emph{Termination and Complexity Competition (TermComp)}
\cite{giesl2019TerminationComplexityCompetition}.\footnote{In addition, we also evaluated \tool{KoAT} on the 816 examples for \emph{Complexity of Integer Transitions Systems} (\tool{CITS}).
  See \cite{KoATwebpage} for the (similar) results of this evaluation.}

Here, all variables are interpreted as integers over $\ZZ$ (i.e., without overflows).
To distinguish the original version of \KoAT{} \cite{brockschmidt2016AnalyzingRuntimeSize} from the re-implemen\-ta\-tion, we refer to them as \tool{KoAT1} and \tool{KoAT2}, respectively.
We used the following configurations of \tool{KoAT2}, which apply different techniques to infer runtime and size bounds:
\begin{itemize}
  \item \tool{KoAT2} only uses $\MRFs$ of depth $5$ for runtime bounds and the original technique from \cite{brockschmidt2016AnalyzingRuntimeSize} to infer size bounds.
        Thus, this version does not apply any of the new contributions.
  \item \tool{KoAT2(T)} uses the approach of \Cref{lem:complexity} to compute polynomial\linebreak
        runtime bounds for \emph{twn}-loops in \Cref{thm:RBLoops}, \Cref{lem:complexityLoops}, \Cref{thm:time-bound}, and\linebreak
        \Cref{lem:twncycleSym} in addition to \tool{KoAT2}.
  \item \tool{KoAT2(S)} uses the approach of \Cref{thm:size_bounds_closed_form} for size bounds for solvable loops in \Cref{thm:lift_size_bounds}, \Cref{lem:lift_size_bounds_loops}, and \Cref{lem:cyclesSB} in addition to \tool{KoAT2}.
  \item \tool{KoAT2(TS)} combines both \tool{KoAT2(T)} and \tool{KoAT2(S)}.
  \item \tool{KoAT2(F)} fully uses the new techniques of \Cref{lem:unsolvable} (on runtime bounds for unsolvable loops), \Cref{lem:complexity_logarithmic} (on logarithmic runtime bounds for loops), and \Cref{lem:commuting_loops} (on size bounds for commuting simple cycles) in addition to \tool{KoAT2(T)} and \tool{KoAT2(S)}.
        So in contrast to the other versions of \tool{KoAT2}, this variant also contains the contributions of the current paper that are new compared to its conference versions \cite{lommen2022AutomaticComplexityAnalysis,lommen2023TargetingCompletenessUsing}.
\end{itemize}

\medskip

The \tool{CINT} collection consists mainly of examples with linear arithmetic and the existing tools can already solve most of its benchmarks which are not known to be non-terminating.\footnote{The tool \tool{iRankFinder} \cite{domenech2018IRankFinder} proves non-termination for 119 programs in \tool{CINT}.
  \tool{KoAT2} (without the contributions of this paper) already infers finite runtimes for 335 of the remaining $519-119 = 400$ examples in \tool{CINT}.}
While most complexity analyzers are essentially restricted to programs with linear arithmetic, the new approach also succeeds on programs with \emph{non-linear} arithmetic.
The approach of the current paper increases \KoAT's power substantially not only for (non-linear) \emph{twn}-loops but also for programs (possibly with non-linear arithmetic) where the values of variables computed in ``earlier'' loops influence the runtime of ``later'' loops (e.g., the modification of the example from \Cref{fig:ITS} where $t_4$ sets $x_6$ to $x_1$ instead of $x_4$, see the end of \Cref{ex:global_runtime_bound}).

Therefore, we extended \tool{CINT} by 20 new typical benchmarks including the programs in \eqref{WhileExample}, \Cref{fig:ITS}, and the modification of \Cref{fig:ITS} discussed above, as well as all the introduced variations of the leading example, resulting in the collection \tool{CINT${}^+$}.
As mentioned, for \tool{KoAT2} and \tool{KoAT1}, we used \tool{Clang} \cite{clang} and \tool{llvm2kittel} \cite{falke2011TerminationAnalysisPrograms} to transform \tool{C}
programs into integer programs as in \Cref{sect:global_integer_programs}.
We compare the different configurations of \tool{KoAT2} with \tool{KoAT1}, \tool{MaxCore} with the tools \tool{CoFloCo} or \tool{PUBS} in the backend (denoted by \tool{MaxCore(C)} and \tool{MaxCore(P)}, respectively), and \tool{Loopus}.

\begin{table}[t]
  \begin{center}
    \scalebox{0.83}{
      \makebox[\textwidth][c]{
        \setlength{\tabcolsep}{2pt}
        \begin{tabular}{l|c|cc|cc|cc|cc|cc|c|c}
                                   & $\textsf{PB}(1)$ & \multicolumn{2}{c|}{$\textsf{LB}\setminus\textsf{PB}(1)$} & \multicolumn{2}{c|}{$\textsf{PB}(n)\setminus\textsf{LB}$} & \multicolumn{2}{c|}{$\textsf{PB}(n^2)\setminus\textsf{PB}(n)$} & \multicolumn{2}{c|}{$\textsf{PB}(n^{>2})\setminus\textsf{PB}(n^2)$} & \multicolumn{2}{c|}{$< \omega$} & $\mathrm{AVG^+(s)}$ & $\mathrm{AVG(s)}$                                   \\
          \hline \tool{KoAT2(F)}   & 24               & 10                                                        & (6)                                                       & 238                                                            & (2)                                                                 & 73                              & (2)                 & 18                &     & 368 & (12) & 4.71 & 17.47 \\
          \hline \tool{KoAT2(TS)}  & 24               & 0                                                         &                                                           & 237                                                            & (2)                                                                 & 72                              & (2)                 & 25                & (2) & 363 & (9)  & 5.39 & 18.88 \\
          \hline \tool{KoAT2(T)}   & 24               & 0                                                         &                                                           & 236                                                            & (2)                                                                 & 69                              & (1)                 & 19                & (2) & 356 & (8)  & 5.87 & 17.96 \\
          \hline \tool{KoAT2(S)}   & 24               & 0                                                         &                                                           & 232                                                            &                                                                     & 70                              & (1)                 & 13                &     & 342 & (2)  & 5.01 & 14.52 \\
          \hline \tool{KoAT2}      & 24               & 0                                                         &                                                           & 231                                                            &                                                                     & 66                              &                     & 10                &     & 336 & (1)  & 4.78 & 14.11 \\
          \hline \tool{MaxCore(C)} & 23               & 0                                                         &                                                           & 221                                                            & (1)                                                                 & 67                              &                     & 7                 &     & 318 & (1)  & 1.88 & 5.46  \\
          \hline \tool{KoAT1}      & 25               & 0                                                         &                                                           & 170                                                            &                                                                     & 74                              &                     & 12                &     & 291 & (1)  & 1.83 & 2.83  \\
          \hline \tool{MaxCore(P)} & 12               & 5                                                         &                                                           & 164                                                            &                                                                     & 88                              &                     & 3                 &     & 272 &      & 0.85 & 6.16  \\
          \hline \tool{Loopus}     & 17               & 0                                                         &                                                           & 171                                                            &                                                                     & 50                              &                     & 7                 & (1) & 245 & (1)  & 0.53 & 0.53  \\
        \end{tabular}
      }
    }
    \caption{Evaluation on the Collection \tool{CINT${}^+$}}
    \label{fig:CINT}
  \end{center}
\end{table}
\Cref{fig:CINT} gives the results of the evaluation, where as in \emph{TermComp}, we used a timeout of 5 minutes per example.
All tools were run inside an Ubuntu Docker container on a machine with an AMD Ryzen 7 3700X octa-core CPU and $48 \, \mathrm{GB}$ of RAM.
The first entry in every cell denotes the number of benchmarks from \tool{CINT${}^+$} for which the tool inferred the respective bound.
The number in brackets only considers the 20 new examples.
The runtime bounds inferred by the tools are compared asymptotically as functions which depend on the largest initial absolute value $n$ of all program variables (see \Cref{def:bounds} for the formal definition of $\textsf{PB}(n^d)$ and \textsf{LB}).
So for example, \tool{KoAT2(F)} proved an (at most) linear runtime bound for $24 + 10 + 238 = 272$ benchmarks, i.e., for these examples it inferred a runtime bound $b \in\textsf{PB}(n)$ with $|\valuation|(b) \geq \rc(\valuation)$ for all states $\valuation \in \Valuation$.
Overall, this configuration succeeds on $368$ examples, i.e., ``$< \omega$'' is the number of examples where a finite bound on the runtime complexity could be computed by the tool within the time limit.
Moreover, the best configuration is able to prove termination for $377$ benchmarks.
So the termination proof succeeds for 9 additional examples since we do not have to construct actual runtime bounds and do not have to consider size bounds.
``$\mathrm{AVG^+(s)}$'' denotes the average runtime of successful runs in seconds, whereas ``$\mathrm{AVG(s)}$'' is the average runtime of all runs.

To determine how often the complete techniques might potentially be applicable to subprograms in the benchmark collection, note that all but two of the 539 benchmarks in the extended collection \tool{CINT${}^+$} contain cycles after the translation to the \KoAT{} input format (i.e., they are no straightline programs).
In 529 of these 537 examples, \KoAT{} detects at least one solvable loop after chaining simple cycles, and in 503 of them, at least one of these loops is even a \emph{twn}-loop.
Moreover, 104 examples contain at least two commuting simple cycles where after chaining, both correspond to solvable loops.
In total, 50 of the 539 programs from the collection \tool{CINT${}^+$} make use of non-linear arithmetic.

Our experiments show that already on the original benchmarks \tool{CINT}, integrating the new techniques leads to the most powerful approach for runtime complexity analysis.
The effect of these new techniques becomes even clearer when also considering our new examples which contain non-linear arithmetic and loops whose runtime depends on the results of earlier loops in the program.
Thus, the new contributions of the paper are crucial in order to extend automated complexity analysis to larger programs with non-linear arithmetic.

\KoAT's source code, a binary, and a Docker image are available at:
\[\mbox{\url{https://koat.verify.rwth-aachen.de/twn-journal}}\]
This website also has details on the experiments (including the results on the \tool{CITS}
collection), a list and description of the new examples, and \emph{web interfaces} to run different configurations of \KoAT{} directly online.

\bibliographystyle{spbasic} 
\bibliography{bib}
\clearpage \section*{\huge\appendixname}
\appendix

\end{document}